\pdfoutput=1
\documentclass[a4paper,11pt]{article}
\usepackage{amsmath}
\usepackage{amssymb}
\usepackage{graphicx}
\graphicspath{{figures/}} 
\usepackage{xfrac}
\usepackage{todonotes}

\usepackage[utf8]{inputenc}

\usepackage[square,numbers,comma,sort&compress]{natbib}
\usepackage[colorlinks=true,citecolor=blue,linkcolor=purple,urlcolor=purple]{hyperref}
\usepackage{geometry}
\usepackage{booktabs}
\usepackage{slashed}

\def\varabstract{ }
\def\varkeywords{ }
\def\vararxivnumber{ }
\def\vartitle{ }
\def\varsubtitle{ }
\renewcommand{\title}[1]{\gdef\vartitle{#1}}

\renewcommand{\abstract}[1]{\gdef\varabstract{#1}}
\newcommand{\keywords}[1]{\gdef\varkeywords{#1}}

\newtoks\authtoks
\renewcommand{\author}[2][]{%
	\authtoks=\expandafter{\the\authtoks#2$^{#1}$\ }%
}
\newtoks\affiltoks
\newcommand{\affiliation}[2][]{%
    \affiltoks=\expandafter{\the\affiltoks{\item[$^{#1}$]#2}}%
}
\newtoks\emailtoks\newcounter{emailcounter}%
\newcommand{\emailAdd}[1]{%
\ifnum\theemailcounter>0\emailtoks=\expandafter{\the\emailtoks, \typeemail{#1}}%
\else\emailtoks=\expandafter{\typeemail{#1}}%
\fi
\stepcounter{emailcounter}}
\newcommand{\typeemail}[1]{\href{mailto:#1}{\tt #1}}

\renewcommand\maketitle{
	\newgeometry{margin=2cm}
	\pagestyle{empty}\setcounter{page}{0}
	\vspace*{-0.5cm}\hspace*{-0.2cm}\mbox{\!\!\!\includegraphics[width=.14\textwidth]{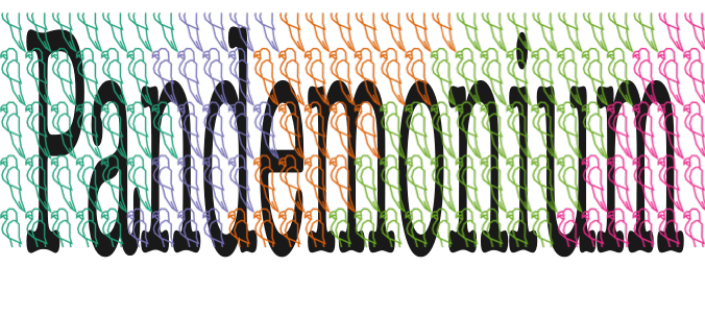}}	{\huge\flushleft\sffamily\bfseries\vartitle\\\Large\varsubtitle\par}
\vskip6ex
{\large\bfseries\raggedright\sffamily\the\authtoks\par}
\vskip2ex
\begin{list}{}{%
\setlength{\leftmargin}{0.28cm}%
\setlength{\labelsep}{0pt}%
\setlength{\itemsep}{-3pt}%
\setlength{\topsep}{-\parskip}}
\itshape\small%
\the\affiltoks
\end{list}
\vskip2ex
\noindent\hspace{0.28cm}\begin{minipage}[l]{.9\textwidth}
\begin{flushleft}
\textit{E-mail:} \the\emailtoks
\end{flushleft}
\end{minipage}
\vskip5ex
\noindent{\renewcommand\baselinestretch{.9}\textsc{Abstract:}}\ \varabstract
\vskip5ex
\if!\varkeywords!\else\noindent{\textsc{Keywords:}}\ \varkeywords \vskip2ex\fi
\if!\vararxivnumber!\else\noindent{\textsc{ArXiv ePrint:}} \href{http://arxiv.org/abs/\vararxivnumber}{\vararxivnumber}\vskip2ex\fi

\newpage
\restoregeometry
\pagestyle{plain}

\setcounter{footnote}{0}
}

\usepackage[square,numbers,comma,sort&compress]{natbib}

\definecolor{MS}{rgb}{0,0,1}

\usepackage{ifthen}

	\newcommand{\barlimc}[7]{
  \pgfmathparse{\mypos+0.3}
  \edef\mypos{\pgfmathresult}
		\node[left,scale=0.6] at (0,\mypos) {#1};
		\pgfmathparse{#3 > 5 ? 1 : 0}
		\ifthenelse{\pgfmathresult=1}{
			\fill[#2] ($(0,\mypos)+(0,-0.1)$) rectangle +(5,0.2);
			\fill[white] ($(0,\mypos)+(3.5,-0.1)$) rectangle +(0.3,0.2);
			\draw[decoration={zigzag},decorate,#2,very thick] (3.4,\mypos) to +(0.5,0);
			\node[left,scale=0.6] at (5,\mypos) {#3};
			}{
			\fill[#2] ($(0,\mypos)+(0,-0.1)$) rectangle +(#3,0.2);
			\node[left,scale=0.6] at (#3,\mypos) {#3};
		}
		\fill[#4] ($(0,\mypos)+(0,-0.1)$) rectangle +(#5,0.2);
		\node[left,scale=0.6] at (#5,\mypos) {#5};
		\fill[#6] ($(0,\mypos)+(0,-0.1)$) rectangle +(#7,0.2);
		\pgfmathparse{#7 <0.3 ? 1 : 0}
		\ifthenelse{\pgfmathresult=1}{
			\node[right,scale=0.6] at (0,\mypos) {#7};
		}{
		\node[left,scale=0.6] at (#7,\mypos) {#7};
	}
}

\title{{\tt Pandemonium}: a clustering tool to partition parameter space - application to the B anomalies}

\author[1,2,3]{Ursula Laa,}\emailAdd{ursula.laa@boku.ac.at}
\author[1]{and German Valencia}\emailAdd{german.valencia@monash.edu}

\affiliation[1]{School of Physics and Astronomy,  Monash University, Wellington Road, Clayton, VIC-3800, Australia}

\affiliation[2]{School of Econometrics and Business Statistics,  Monash University, Wellington Road, Clayton, VIC-3800, Australia}

\affiliation[3]{University of Natural Resources and Life Sciences, Vienna, Department of Landscape, Spatial and Infrastructure Sciences, Institute of Statistics, Peter-Jordan-Straße 82/I, 1190 Vienna, Austria}

\abstract{
We introduce the interactive tool {\tt pandemonium} to cluster model predictions that depend on a set of parameters. The model predictions are used to define the coordinates in observable space which go into the clustering. The results of this partitioning are then visualized in both observable and parameter space to study correlations between them. The tool offers multiple choices for coordinates, distance functions and linkage methods within hierarchical clustering. It provides a set of diagnostic statistics and visualization methods to study the clustering results in order to interpret the outcome. The methods are most useful in an interactive environment that enables exploration, and we have implemented them with a graphical user interface in R. We demonstrate the concepts with an application to phenomenological studies in flavor physics in the context of the so-called B anomalies, exploring the tension between $P_5^\prime$ and $R_K$ and quantifying the resolution in parameter space that can be provided by a given observable set.

}


\begin{document}

\maketitle

{\hypersetup{linkcolor=black}
  \tableofcontents}

\newpage

\section{Introduction}

A common task in physics consists of comparing the predictions of models; be it to available experimental data, or among themselves in the absence of such data. It is also common for the models to differ only in the values of some parameters. When measurements exist, a global fit may be used to determine the preferred values of these parameters. The data, however, contains information beyond that which can be summarized by the preferred parameter set, including confidence intervals as well as tensions between different measurements. When the models are compared to each other, different goals can be achieved by mapping regions of parameter space onto regions of observable space.  These goals may include finding benchmark points for future experimental study; investigating the level at which the models can be probed; or simply understanding the correlations between parameters that are implied by the full set of observables.

This general task can be described as a partitioning of parameter space based on the similarity of the model predictions, and clustering algorithms are a useful tool for this purpose as demonstrated in~\cite{Carvalho:2015ttv,Capozi:2019xsi,Aebischer:2019zoe}. Generally speaking, the aim of clustering is to group points in a data set based on their similarity. Analyzing the clusters (or corresponding representative benchmark points) can then reveal the main structures and features in the data without looking at all individual data points.

A useful analogy is to think of a global fit as a clustering problem that partitions the parameter space based on experimental observations into bins in $\chi^2$. The ``clusters'' in this case would be the selected confidence intervals and plotting them helps us understand what the data implies for the parameters. For example, the constraints on one of the parameters might depend on the value of a second one. In the simplest case this could reflect a linear correlation between the two variables.
This is a useful but incomplete summary of all available information. Supplementing the fit with other types of clustering can reveal additional features, as we demonstrate in the application sections. The mapping of clusters between parameter and observable space can help understand tensions between observables; clusters in parameter space can reveal correlations between parameters; the partitioning of parameter space anticipates  the impact of improved measurements, among others.

The work of~\cite{Carvalho:2015ttv,Capozi:2019xsi,Aebischer:2019zoe} focused for the most part on the identification of a set of benchmark points to be used in future studies. However, a main asset of clustering is that the interpretation of the outcome often provides new insights. In our case this means revealing information about the interplay between parameter and observable space, and a detailed examination of patterns in the predictions for the considered observables.
This is best achieved by combining an interactive selection of the method details with a comprehensive visualization of the outcome. Our work presents the tool {\tt pandemonium}, with a graphical user interface that allows a choice for the clustering settings and presents them with multiple displays of the outcome, providing a better understanding of how the clusters partition the parameter space and why. 
Thus, we can think of clustering as part of an exploratory workflow where one can try different clustering setups and study differences in resulting group assignment. If we also understand the implication of the selected settings, this comparison can reveal important aspects of the data.

When applied to global fits the information obtained with clustering is complementary to what we learn from the fit: whereas the fit reduces a large number of observables in the context of a model to a best fit point and its associated confidence intervals in the model parameter space; our clustering approach provides a map to understand how different observables induce different constraints on the parameter space. This is most instructive if the study is restricted to a subset of observables with the most impact. These can be chosen to be the ones driving the fit away/towards a preferred model; those which introduce tensions in the global fit; and so on. Such a subset can be preselected, if it exists, for example with the metrics developed in \cite{Capdevila:2018jhy}, as we do in Section~\ref{s:existingfit}. Alternatively, one can start with the complete set of observables and use the tools provided in the interface to select a subset. This reduction in the number of observables (or parameters) is necessary in order to concentrate on the most relevant features and is akin to dimension reduction via feature selection in machine learning.

Our paper is organized as follows. In Section~\ref{s:basics} we define coordinates in observable space that will serve as the objects to be clustered, and associate with them distance functions.  We then provide an introduction to clustering and different linkage methods. In Section~\ref{s:inter} we describe the different features of the graphical interface and illustrate how they can guide the interpretation of the complex interplay between parameter and observable spaces. 

In Section~\ref{s:existingfit} we discuss the case of the so-called neutral $B$ anomalies in the decay modes $B\to K^{(\star)} \ell^+\ell^-$. In this system, multiple observables in the angular distributions and branching ratios have been compared to the predictions of the standard model (SM) and global fits have consistently suggested disagreement at the $5\sigma$ level. Beyond the SM, the quark-level transition is described with an effective low energy Hamiltonian in which the Wilson coefficients are treated as free parameters. We use a suitable set of fourteen measurements (a 14-dimensional observable space) and predictions that depend on two parameters (for the most part) to study the problem with clustering methods. We then select an additional set of six possible future observables and compare their predictions to Belle~II projections in order to study their expected impact.  This comparison can thus be used to relate the resolving power of future measurements to the neutral B anomalies.

Ref.~\cite{Aebischer:2019zoe} previously introduced a python framework, {\tt ClusterKing}, that implements clustering parameter points based on the predictions for a single binned observable and we compare our results to that work in Section~\ref{s:compareCK}.
To that end we consider the decay modes $B \to D^{(\star)}\tau \nu$ which currently exhibit the so-called charged $B$ anomalies. We select two kinematic distributions that were also studied in \cite{Aebischer:2019zoe} in order to explicitly compare our framework to theirs.
Finally, in appendices we provide details relevant to the examples used in the applications.

\section{Clustering setup}\label{s:basics}

We are interested in the connection between the parameters of physics models and the resulting predictions for experimental observables. To this effect, each `data' point is defined by fixing the set of model parameters, and is represented in both parameter and observable  space.  Our goal is to look for patterns in observable space which are then investigated in parameter space. We begin by defining our notation.

\subsection{Notation}
\label{sec:notation}

When studying a particle physics model in the context of experimental observations we typically start from parameter space, for example defined in terms of Wilson coefficients in effective field theory, or via masses and couplings in a supersymmetric model. For a fixed point in parameter space we can then compute the model predictions, these are the theory predictions for experimental observables, for example branching ratios or angular observables. We can then score how well a parameter point fits the observed values in this observable space via the $\chi^2$ function. In the following we will formalize this notation to efficiently define the input to our clustering approach. Where possible the notation will follow that introduced in \cite{Capdevila:2018jhy}.

\textbf{Observable space:} consists of a set of observables $O_i$ that can be measured experimentally with a precision given by $\sigma_i^{exp}$, and we denote the experimentally measured value of $O_i$ by $E_i$. In many cases the experimental errors between observables will be correlated, and we describe uncertainties with the variance-covariance matrix $\Sigma_{ij}^{exp}$.

\textbf{Model parameter space:} the underlying theoretical description of the observables has a set of free parameters. Fixing the values of the parameters (we can denote them as the parameter vector of model $k$, $P_k$) then gives a fully specified model. For each model $k$ we can calculate the predictions for the considered observables $O_i$. We denote $X_{ki}$ the prediction of model $k$ for observable $O_i$. The theoretical calculation also introduces an uncertainty, $\sigma_i^{th}$, where for simplicity we neglect its dependence on the model parameters (i.e. we consider that $\sigma_i^{th}$ is the same for all possible $k$).\footnote{This assumption is often used to speed up calculations but can be tested and relaxed if found to fail as we will discuss later.}
Correlations between theoretical uncertainties of the different observables are again captured by the variance-covariance matrix, $\Sigma_{ij}^{th}$.

\textbf{$\chi^2$ function:} The agreement between model predictions and experimental observations is typically summarized with the $\chi^2$ function. When no correlations are present, the $\chi^2$ for model $k$ is given by
\begin{equation}
\chi^2_k = \sum_i \frac{(E_i - X_{ki})^2}{(\sigma_i^{exp})^2 + (\sigma_i^{th})^2},
\end{equation}
and including correlations by
\begin{equation}
\chi^2_k = \sum_{i,j} [E_i - X_{ki}] (\Sigma^{exp} + \Sigma^{th})^{-1}_{ij} [E_j - X_{kj}].
\end{equation}
Global fits are interested in finding the model $k$ that minimizes the $\chi^2$, resulting in the set of parameters that best describes the combination of observables $O_i$. We will denote this point by $BF$ with associated predictions $X_{BFi}$.

A simple way of ``clustering'' the parameter space is to bin the model points according to the value of $\Delta\chi^2_k=\chi^2_k-\chi^2_{BF}$. This is typically how the results of global fits are reported in parameter space, by drawing contour lines for fixed levels of $\Delta\chi^2$ corresponding to confidence intervals. 

\subsection{Measuring distance between models}
\label{sec:coord}

Before we can cluster models, we need to define how to measure the similarity (or rather dissimilarity) of two models. This dissimilarity is usually measured as a distance in a coordinate representation of the data points.
To characterize points based on their predictions we work with observable space information, i.e. the $X_{ki}$.

One possible distance metric could be defined analogous to the $\chi^2$ function:
\begin{equation}
d_{\chi^2}(X_k, X_l) = \sum_{i,j} [X_{ki} - X_{li}] (\Sigma^{exp} + \Sigma^{th})^{-1}_{ij} [X_{kj} - X_{lj}],
\label{eq:dchi2}
\end{equation}
or its square root, the Mahalanobis distance
\begin{equation}
d_M(X_k, X_l) = \sqrt{d_{\chi^2}(X_k, X_l)}.
\end{equation}
This definition is similar, but not identical, to what was used in~\cite{Aebischer:2019zoe}, because in our definition the covariance matrix does not depend on the model parameters, and the distance can thus be computed from a set of coordinates.

However, Eq.~\ref{eq:dchi2} is by no means the only distance function that can be used. Here we will be interested mostly in functions that can be implemented from a coordinate representation of the information allowing easy  comparison of different types of distance measures. A set of distance functions can be obtained from the general notion of a Minkowski distance based on the $p$-norm,
\begin{equation}
d_p(x, y) = \left(\sum_{i=1}^n |x_i - y_i|^p\right)^{1/p}.
\end{equation}
Where $p=2$ corresponds to the standard Euclidean norm, $p=1$ is the Manhattan distance, and $p\to\infty$ gives the infinity norm distance (also known as maximum or Chebyshev distance).

These different distance measures emphasize different aspects when assessing similarity, and that is why it is useful to be able to compare them. A larger $p$ generally implies higher sensitivity to outlying variables, with $p\to\infty$ representing the most extreme case, when the distance measure reflects only the largest difference across the $n$-dimensional space. The maximum distance would thus be used to emphasize dominant observables and the Manhattan distance would be preferred if we aim to reduce sensitivity to such points. Alternatively, we may wish to average the effects of all observables by using the Euclidean norm. In high dimensional spaces fractional distance metrics may be preferred~\cite{10.5555/645504.656414}. 

To use the different distance measures we define a coordinate representation that can capture the information used in the definition of $d_{\chi^2}(X_k, X_l)$. 
When neglecting covariance we define coordinates as
\begin{equation}
Y_{ki} = \frac{X_{ki} - R_i}{\sigma_i},
\label{eq:y1}
\end{equation}
with $R_i$ any fixed reference point (e.g. the origin, $R_i=0$, the experimental value $R_i=E_i$ or predictions for special points in parameter space, such as the SM), and $\sigma_i = \sqrt{(\sigma_i^{exp})^2 + (\sigma_i^{th})^2}$ the combined uncertainty. For this simple scenario, the $\chi^2$ distance is reproduced for $p=2$, i.e. using the Euclidean distance together with the coordinate representation of Eq.~\ref{eq:y1}.

In general correlation between observables can be large, and should not be neglected. Following \cite{Capdevila:2018jhy} we can define a coordinate system in observable space that is aware of these correlations. Here we introduce the coordinates
\begin{equation}
Y_{ki} = \sum_j \frac{1}{\sqrt{(\Sigma^{-1})_{ii}}} (\Sigma^{-1})_{ij} (X_{kj} - R_j),
\label{eq:y2}
\end{equation} 
to facilitate numerical calculations, and we will refer to Eqs.~\ref{eq:y1}~and~\ref{eq:y2} as pulls and pulls with correlations respectively.  Alternatively we could define the distance working with the square root of the inverse variance-covariance matrix directly to reproduce the Mahalanobis distance (see discussion in \cite{Capdevila:2018jhy}).

In both definitions we have allowed for a generic reference point $R_j$ which fixes the origin of our coordinate system. The value of $R_j$ is irrelevant when computing distances between points, and is therefore of no consequence to the clustering result as long as we assume fixed uncertainties (no model dependence of the $\Sigma^{-1}_{ij}$). This is often a good approximation which is used in global fits (see e.g.~\cite{Capdevila:2016ivx}), but will not always hold, e.g.\ when the dominant uncertainty is a Poisson error and the shape of the distribution is considerably different between model points as in the problems studied in~\cite{Aebischer:2019zoe}. In that case a distance can be directly computed between model predictions without introduction of a coordinate system, and thus without the need for a reference point.

When working with Eq.~\ref{eq:y2} the reference point should reflect the main point of interest, i.e.  the experimental value to characterize predictions based on existing observations or the best fit point when studying a global fit. For the analysis of future observables we might wish to compare to a special model, such as the SM or one suggested by a different fit.

The described coordinate representation and distance measures work well with the applications considered here. We note however that different (particle physics) problems might require adapting these definitions. One might also explore additional distances that have been introduced in the literature, typically designed to address specific problems and use cases, for example ones available in the \texttt{philentropy} R package~\cite{philentropy}.

\subsection{Hierarchical clustering}\label{s:clusters}

The distance function and the coordinates are used to compute the distance or dissimilarity matrix. For $K$ models that are compared based on their predictions $X_{ki},~k=1,\cdots K$ with distance metric $m$, this is the $K\times K$ matrix $d_m(k,l),~k,l=1,\cdots K$. With this information we can now use hierarchical clustering to group the models.

Hierarchical clustering starts by considering all data points as separate clusters and each step of the algorithm combines the two clusters that are most similar to each other. The process continues until all points are combined into a single cluster. In an ideal situation, the solution to the clustering problem consists of finding a number of clusters that ensures all points within a given cluster are similar to each other whereas points belonging to different clusters are not.

Different metrics have been developed to identify the best solution in terms of the number of clusters. In practice (and in the applications studied here) there may not be a clearcut grouping of points. This is especially true for data points that follow a continuous distribution with no discernible gaps between different clusters. However, even in such cases clustering can be useful and distill information as seen in our examples below. Section~\ref{s:stop} describes how we can pick a meaningful number of clusters in our setting.

 \subsection{Linkage}

While the dissimilarity between points is defined via the input to the algorithm, we also need to decide how to characterize dissimilarities between clusters with more than a single data point, the concept of linkage. This choice is crucial to the clustering outcome, which can vary considerably depending on the selection.

We start from $p$ $n$-dimensional points in the data set. A cluster $C_j$ defines a set of data points that have been grouped by the algorithm, with $j = 1,..., \kappa$ and $\kappa$ is the number of clusters at the current iteration. At the start $\kappa=p$ and at the final iteration $\kappa=1$. The preferred value of $\kappa$ is typically evaluated with a stopping criterion.

We denote $d(a,b)$ the distance  between two points $a,b$, and $d_{AB}$ as the distance between the clusters $A$ and $B$. Some of the commonly used linkage methods are defined as:

 \begin{itemize}
 \item Single linkage: the distance between clusters is computed as the shortest distance between any two points, $d_{AB} = min\{d(a,b): a\in A, b\in B\}$
 \item Complete linkage: the distance between clusters is computed as the longest distance between any two points, $d_{AB} = max\{d(a,b): a\in A, b\in B\}$
 \item Average linkage: in this method $d_{AB}$ is the average of all distances $d(a,b), a\in A, b\in B$. We will refer to this definition as ``unweighted'' average linkage. This is also called unweighted pair group method with arithmetic mean in the literature, and results in a ``weighting'' by size when combining clusters (a larger group will dominate the distance computation).

Alternatively there is the ``weighted'' average linkage (also called the McQuitty or weighted pair group method with arithmetic mean) which directly averages the distances, and thus drops the sensitivity to size when combining clusters. This means when $A,B$ have been combined to a cluster $C$ and we compute the distance of $D$ from $C$ it will be the average distance in the sense that $d_{DC} = \frac{d_{DA} + d_{DB}}{2}$.

 \item Ward linkage: Ward~\cite{ward} proposed to define clusters by minimizing an objective function, typically taken to be the within cluster dissimilarity. This means at each step we combine the two clusters which result in the smallest increase of this within cluster dissimilarity. In R two versions are available, \texttt{ward.D2} linkage squares the dissimilarities before updating, whereas \texttt{ward} linkage does not.
 
 \end{itemize}

Without a specific objective in mind, Ward linkage is typically preferred, because it leads to balanced clustering results. On the other hand, single or complete linkage are most useful in specific settings, with the first one typically producing large, spread out clustering results, and the latter producing tight clusters that can be close together. In the case of continuously distributed data, single linkage will sequentially merge all points into one large cluster and not produce any new insights, while complete linkage can give interesting results that emphasize large differences between clusters. Within this context we find that average linkage will often result in similar clustering outcomes as Ward linkage, but with a simpler interpretation of cluster distance.

\section{Interpretation of clustering outcome}\label{s:inter}

With our focus on clustering as an exploratory tool, the main outcome of the analysis is the interpretation of the results and the insights this provides into the interplay between the parameter and observable space. Additionally, we select benchmark points that are representative of the clusters and can be used in future analyses.

The interpretation of the clustering outcome is primarily done visually, in both parameter and observable space. This information is linked via a consistent color mapping to show the cluster assignment across views. Note that some visualizations discussed here would also be useful without any clustering information, but the grouping provided by the outcome simplifies comparison across the two spaces.

By using a set of interpretation diagnostics in an interactive interface we can explore the outcome with different cluster settings in detail. The interactive setting provides important benefits: the results can change drastically depending on the choices made, and we learn about the result by exploring different combinations and by directly comparing different settings.

\subsection{Parameter space}

Our primary interest is often to obtain information from the observables about the model parameters. Clustering  complements the $\chi^2$ analysis of a global fit, providing alternative partitioning of the parameter space based on grouping in the observable space. As discussed above, different settings  emphasize different aspects of the information.

For comparison to the clustering results, we also present a binning in $\Delta\chi^2$ divided into the requested number of clusters. As it is customary to report these intervals in units of Gaussian standard deviations, we first convert $\Delta\chi^2_k$ to their corresponding values in $\sigma$ under the assumption that  $\Delta\chi^2$ follows a $\chi^2$ distribution with a number of degrees of freedom given by the number of free parameters in the model. We then split the full range in $\sigma$ into $\kappa$ (the number of clusters) equidistant bins. To simplify the handling of  low interest regions corresponding to more than $5~\sigma$, we include everything above $5~\sigma$ into the last bin.

Problems with only two model parameters allow us to show the clustering outcome directly in the parameter plane. This can be augmented by highlighting points of special interest, for example the benchmark points selected for each cluster, the overall best fit point or selected models (such as the SM).
This graph can then be used to understand how each region in parameter space (or the corresponding benchmark point) maps onto the different patterns of predictions in observable space.

Visualizing parameter space becomes substantially more complicated when  the model contains more than two parameters.  Existing methods, like profiling or marginalizing, project all data points into a selected 2D plane, summarizing the information across the orthogonal space. To understand how the clusters split up the full parameter space conditional information is more appropriate (i.e.\ fixing the value of parameters in the orthogonal space). The simple solution adopted here is to 
display the clustering information in a plane for selected values of parameters not shown. With a regular grid this can be accomplished by showing a sequence of conditional plots (as was done in~\cite{Aebischer:2019zoe}). 
More generally we can use the new slice tour method~\cite{Laa:2019bap} to resolve the cluster assignment across higher dimensional spaces, but this option is beyond the scope of this work. 

\subsection{Observable space}

The clustering is performed in observable space, which thus defines the ``data'' space. This space is typically  high-dimensional, with large numbers of observables usually entering a global fit for example. To understand the outcome we need approaches to inspect the high-dimensional information patterns.
In our applications we use parallel coordinate plots, tour displays and non-linear dimension reduction, which is augmented with quantitative information (numerical summaries). Clustering specific visualizations, which were used during the interactive exploration, are described in the Appendix.

\subsubsection{High-dimensional visualization}

With any type of high-dimensional visualization we have to consider the different scales across the coordinate space.\footnote{To understand this consider that when drawing one or two variables we typically adjust the axis range to the range of values observed in the data, which could also be interpreted as a scaling of the data with the overall range. Similarly we could scale each variable in a multivariate dataset with the range or standard deviation to ensure a common scale across the different directions.} While typically scaling of each variable separately is recommended, this is not the case here. As seen in the description of the coordinate systems, all variables are already scaled based on physical information (the total uncertainties, potentially taking correlations into account). The spread along each variable is thus determined by the variance in predictions for that observable, obtained across the grid, normalized to the total uncertainty. Any remaining differences in scale are thus physically meaningful. They will enter the calculation of dissimilarity and thus should be represented in the plots.

We do however center each variable independently to avoid distraction: a constant offset will not influence the clustering and can be distracting in the visualizations. However, this means that we cannot directly read-off the agreement with the experimentally observed value when using Pull coordinates.

\textbf{Parallel coordinate plot:} a static visualization of the data points in all $n$ coordinates is achieved by mapping each variable to a parallel line. For each data point we draw a line connecting its values along those parallel coordinates. This display gives a good overview of patterns in the data, for example to detect correlations between the coordinates, or to understand differences between groups. We use the R package \texttt{GGally}~\cite{GGally} to generate this graph, and we map the cluster assignment to color and highlight the cluster benchmark points. This gives a good overview, but is limited to bivariate relations and simple patterns.

\textbf{Grand tour display:} using an animated sequence of two-dimensional projections we can show the distribution of the data across the full $n$-dimensional space~\cite{As85,tourr}. The tour shows randomly selected projections that are smoothly interpolated to provide the viewer with a continuous rotation of the high-dimensional distribution \cite{buja2005computational}. Watching the animation shows the data from all possible viewing angles, and the viewer can extrapolate from the observed low-dimensional shapes and patterns to the high-dimensional distribution. For example, we can use tours to understand grouping, identify multivariate outliers or to understand (non-linear) correlations between the variables. We have previously used the grand tour to gain insights into the grouping of particle physics observables and to identify outliers in high-dimensional space in~\cite{Cook:2018mvr}.
Tour visualizations could also be augmented with information from the dendrogram, as illustrated in~\cite{model-vis}, but this is beyond the scope of this work.

\textbf{Dimension reduction:} instead of showing the full $n$-dimensional space, we can map the high-dimensional distribution onto a 2-dimensional plane for plotting using non-linear dimension reduction techniques. These are typically machine learning methods that aim to find the mapping for which the inter-point distance in the low-dimensional space is as similar as possible to the distance in the $n$-dimensional space. This is often useful for the visualization of clustering, but the results from these methods are not easily interpretable in terms of the original variables, and it can be helpful to compare the results to those in a tour \cite{lee2020casting}. Here we consider three methods: t-SNE \cite{tsne}, UMAP \cite{mcinnes2018umap} and local linear embeddings \cite{lle}.

\subsection{Summary information}

Ideally clusters should be tight (points within a cluster are very similar to each other) and well separated (the individual clusters are very different from each other). This is often best evaluated visually, as described above.
In addition, we also use numeric summaries to characterize the clustering outcome. There are two types of summaries that we consider.

First, we can define statistics that characterize the cluster size and separation. This includes simple measures like the radius and diameter of a cluster, and also comparisons of within to between cluster dissimilarities that are often used to decide the preferred number of clusters. For our applications measures with a physical interpretation are most relevant, measures used in the statistics literature are described in the Appendix.

A different type of summary is to characterize each cluster by one representative benchmark point. This is the parameter combination that best captures the observable space predictions of all points in the cluster. The benchmark can be used for cluster comparisons, and as a representative in further analyses.

\subsubsection{Cluster benchmark points}\label{ss:benchpts}

The benchmark point $c_j$ for a cluster $C_j$ is selected as the point which minimizes
\begin{equation}
f(c, C_j) = \sum_{x_i \in C_j} d(c, x_i)^2.
\label{eq:bp}
\end{equation}
We can then define the radius of the cluster $C_j$ as
\begin{equation}
r_j = \max_{x_i \in C_j} d(c_j, x_i).
\end{equation}

We can think of the radius as a measure of accuracy with which the benchmark point represents the full cluster. The interpretation will depend on the coordinate and distance definition.
For example, when working with coordinates that mimic a $\chi^2$ function (in combination with Euclidean distance), we can relate the cluster radius to a confidence interval with the appropriate number of degrees of freedom. 

We may also wish to characterize the similarity of the full cluster, irrespective of the benchmark point. For this case we can define the cluster diameter, which is given as the largest distance between any two points in $C_j$. 

\subsubsection{Stopping criterion}\label{s:stop}

In our applications, the coordinates are continuous functions of parameters that can be measured with a certain uncertainty. This suggests a different take on the number of clusters than using the typical cluster statistics described in the Appendix. Instead, we want to choose clusters such that points that are grouped together are experimentally indistinguishable at some level. This can be quantified with the maximum cluster radius statistic. 
For example, we may use a stopping criterion based on the radius such that the benchmark point is representative of the cluster up to a selected level in $\sigma$, or based on the diameter to ensure a selected level of similarity between all points that are grouped together.
This can be complemented by also checking the distances between benchmark points: a good solution should have distinguishable benchmark points that are representative of their cluster. 

Recall that the overall range in distances between data points (determining the radius) depend on both the anticipated measurement uncertainty and the parameter range that we choose to explore. The latter, as well as the sampling grid spacing, must therefore be chosen judiciously, incorporating any existing information, to obtain a meaningful clustering result.

\section{Application: effective Hamiltonian for $B\to K^{(\star)}\ell^+\ell^-$}\label{s:existingfit}

The effective low energy Hamiltonian responsible for the quark-level transition $b\to s\ell^+\ell^-$ is written in terms of Wilson coefficients and four fermion operators in the general form (see for example \cite{Buchalla:1995vs}):
\begin{eqnarray}
{\cal H}_{\rm eff} = -\frac{4G_F}{\sqrt{2}}V_{tb}V^\star_{ts}\sum_{i,\ell=\mu,e}C_{i\ell}(\mu){\cal O}_{i\ell}(\mu),
\label{effH}
\end{eqnarray}
where $C_{i\ell}$ denote Wilson coefficients, ${\cal O}_{i\ell}$ four-fermion operators and $\ell$, can be electrons or muons. 

Multiple studies over the last decade have suggested that the SM predictions for the Wilson coefficients $C_{i\ell}$ might not be in agreement with experiment and that the discrepancies can be accommodated by allowing some form of new physics parameterized into these coefficients (see for example \cite{Capdevila:2017bsm} and references therein). 
We consider four operators in this paper (only the two unprimed ones for most of the study):
\begin{eqnarray}
{\cal O}_{9\ell} = \frac{e^2}{16\pi^2}(\bar{s}\gamma_{\mu}P_Lb)(\bar\ell \gamma^\mu\ell),&&{\cal O}_{9^\prime\ell} = \frac{e^2}{16\pi^2}(\bar{s}\gamma_{\mu}P_Rb)(\bar\ell \gamma^\mu\ell), \nonumber \\
{\cal O}_{10\ell} = \frac{e^2}{16\pi^2}(\bar{s}\gamma_{\mu}P_Lb)(\bar\ell \gamma^\mu\gamma_5\ell),&&{\cal O}_{10^\prime\ell} = \frac{e^2}{16\pi^2}(\bar{s}\gamma_{\mu}P_Rb)(\bar\ell \gamma^\mu\gamma_5\ell).
\end{eqnarray}
Within the standard model the values of the corresponding Wilson coefficients are $C^\text{SM}_{9,10}=4.07,-4.31$ and $C^\text{SM}_{9^\prime,10^\prime}=0$ for both muons and electrons. New physics is parametrized in a model independent way as deviations in these coefficients from their SM values, $C_{i\ell} \equiv C_i^\text{SM}+C^\text{NP}_{i\ell}$ ($i=9^{(')},10^{(')})$. For simplicity we ignore other possible coefficients, assume that any new physics affects only the muons, and take the parameters to be real ignoring the possibility of CP violation. Notationally, the parameters used in the clustering exercise are the deviations from the SM, the $C^\text{NP}_{i}$ at a scale $\mu =4.8$~GeV.

Equation~\ref{effH} describes a multitude of experimental observables in several decay modes of the $B$ meson in terms of  the Wilson coefficients. These coefficients are thus the parameter set of our clustering study. Global fits have been used to extract their preferred values from all existing measurements. For example, the study in  \cite{Capdevila:2017bsm} considers 175 observables to fit six of the Wilson coefficients. These fits have resulted in values that differ from the SM expectation by around five standard deviations and have therefore attracted much attention in the literature.  In this section we study this system as a clustering exercise, and will need to reduce its dimensionality. Recall that within the clustering analysis we are interested in the detailed relations between parameter and observable space. We therefore begin by deliberately reducing the dimensionality of these spaces based on prior knowledge, selecting the parameters and observables of primary interest and later on explore the impact of additional parameters in subsection \ref{s:fourp} and the impact of additional observables in subsection \ref{s:moreobs}. To this end we consider only two parameters, namely $C_9$ and $C_{10}$ (except for subsection~\ref{s:fourp} where we add two more) because these have been singled out as the most important ones by the global fit studies.  Similarly we will reduce the number of observables to a set of fourteen, listed in Table~\ref{t:obs} to work with a two-dimensional parameter space and a fourteen-dimensional observable space.

Twelve of the selected observables relate to aspects of the angular distribution in $\bar B\to \bar K^\star (\to \bar K \pi) \ell^+\ell^-$: binned values of $P_2$ and $P_5^\prime$ defined as in Eqs.~259,~260 of  \cite{Kou:2018nap}. The other two will be the ratios $R_K$ and $R_{K^\star}$ defined as
\begin{eqnarray}
R_{K^{(\star)}}= \frac{BR(B\to K^{(\star)} \mu^+\mu^-)}{BR(B\to K^{(\star)}e^+e^-)}.
\end{eqnarray}
The selection relies on ranking the observables by their importance in constraining the different directions in parameter space as measured by the metrics we introduced in \cite{Capdevila:2018jhy}.\footnote{That same analysis studied the impact of correlations, thus providing guidance when selecting the appropriate coordinate representation for this study. } Of this set:

\begin{itemize}

\item $P_5^\prime$  has been signaled by multiple studies as a major contributor to the discrepancy between the SM and the global fits, in particular the bins labeled by ID $4,5$ in Table~\ref{t:obs}. They are mostly important for determining $C_9$ and their ranking showed little sensitivity to correlation effects.

\item $P_2$ is singled out by the pull and residual analysis, with the bins with ID $10,11$ in Table~\ref{t:obs} being important for determining $C_9$ (especially when correlations are included).

\item For both $P_5^\prime$ and $P_2$ we use all the existing $q^2$ bins that have been measured, instead of just those that deviate from the SM, to get a more complete picture of the distribution.

\item $R_K$ (ID $13$) and $R_{K^\star}$ (ID $14$) directly test lepton universality and are most important in determining $C_{10}$. The rankings with or without correlations single out $R_K$ as being important,  whereas $R_{K^\star}$ ranks high only when correlations are included. The importance of $R_K$ according to these metrics increased after the 2019 LHCb update \cite{Aaij:2019wad}. As discussed in \cite{Capdevila:2019tsi}, the new measurement means that $R_K$ is now completely dominant for the determination of $C_{10}$ and that becomes apparent in this study. The same arguments have shown their importance in other studies \cite{Alda:2020okk}.

\end{itemize}
This set will permit us to study the tension observed in the global fits between $R_K$, certain bins of $P_5^\prime$, and to a certain extent certain bins of $P_2$.

To generate theoretical predictions for the observables, as well as to obtain the averages of existing experimental measurements, we rely on the package {\tt flavio} \cite{Straub:2018kue}. We first select the region of parameter space to be studied assuming that our interest is to explore the space extending between the SM, $(C_9,C_{10})=(0,0)$, (which is the expected result) and the two dimensional best fit of \cite{Descotes-Genon:2015uva},  $(C_9,C_{10})=(-1.08,0.33)$, (which is in reasonable agreement with other global fits).

We sample this parameter space on a $28\times 12$ grid corresponding to equally spaced (by 0.05) values of $C_{9}$ between $[-1.2,0.15]$ and $C_{10}$ between $[-0.1,0.45]$ to produce a set of 336 model points. The coarseness of the grid reflects a compromise between desired resolution and response time in the interactive tool.\footnote{Knowledge about the problem, external to the tool, is required to choose a range of interest. If one were to choose for example $[-1,1]$ for both $C_{9,10}$, a large sector of that space would lie outside $5\sigma$ from the measurements, as would be visible when inspecting the $\Delta\chi^2$ bins.}

For each model point we obtain a prediction, along with a corresponding theoretical uncertainty, $X_{ki}\pm \sigma_{ki}$ ($k=1,2,\cdots, 336,~i=1,2,\cdots ,14$) in the notation of the previous section.
We then combine this information with the experimental values $E_i\pm \sigma_i^{exp}$. To include correlations we use the inverse covariance matrix $\Sigma^{-1}$, which is also calculated by {\tt flavio} and which includes the known experimental correlations as well as the theoretical correlations at the SM point. With this information we obtain coordinates $Y_{ki}$ as defined in Eq.~\ref{eq:y1} or Eq.~\ref{eq:y2} with $R_i=E_i$.

\subsection{Comparison with $\chi^2$ and global fit results}\label{s:compchi2gf}

For a first overview we start with a comparison of the results of clustering to those from the global fit. For this set of observables one finds a BF point $(C_9,C_{10})=(-0.81,0.12)$ that lies $\Delta\chi^2\approx 16.6$, or about $3.7 \sigma$ from the SM. An evaluation of the $\chi^2$ function on the grid finds the minimum at the point $(C_9,C_{10})=(-0.8,0.1)$ and this point, along with the SM,  $(C_9,C_{10})=(0,0)$ are highlighted in the figures below as an asterisk and an open circle respectively.\footnote{In some of our results we use only a subset of observables from the list in   Table~\ref{t:obs}, this changes the BF and the position of the asterisk on the plots.}

The $\chi^2$ function can be thought of as a one-dimensional response to the input in observable space  that can be used to cluster the points and can be mapped onto parameter space. The maximum $\Delta\chi^2$ for  points on our grid is 18.66 which, for two degrees of freedom, corresponds to $3.92\sigma$. We use this to partition the space into $\kappa$ clusters by placing boundaries at $3.92\sigma/\kappa$ intervals.

\begin{figure}[h]
\centering{
\includegraphics[scale=0.75]{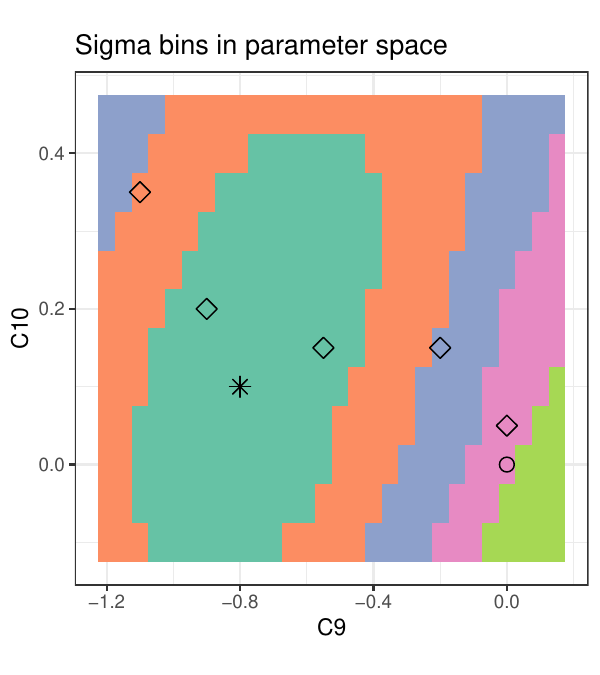}}
\caption{The results of classifying the model points into 5 regions with boundaries at $0.78\sigma,~1.57\sigma,~2.35\sigma,~3.13\sigma,$ and $3.92\sigma$  from the BF.}
\label{f:compchi}
\end{figure}

\begin{figure}[h]
\centering{
\includegraphics[scale=0.4]{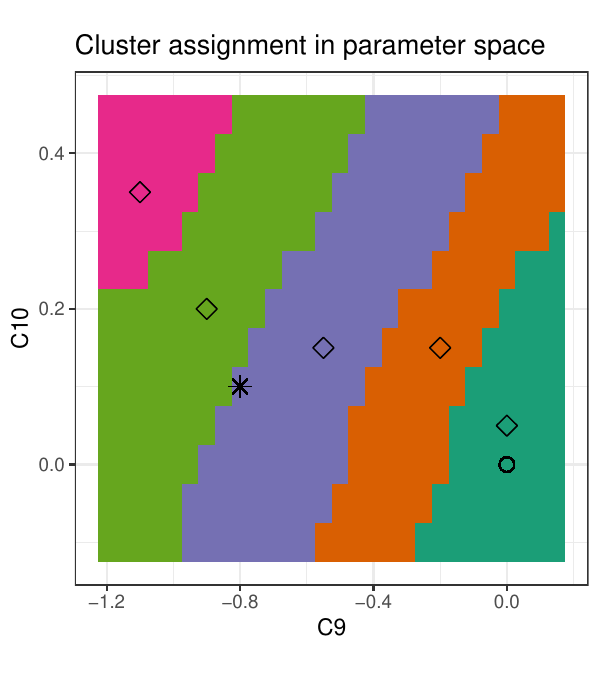}
\includegraphics[scale=0.57]{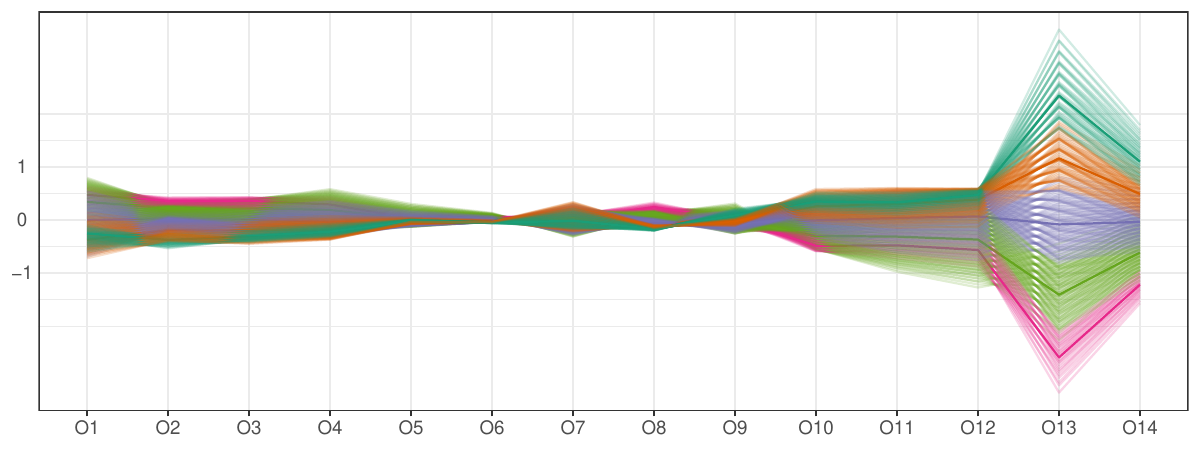}}
\caption{Clustering with Ward.D2 linkage (which minimizes the variance within clusters) and Euclidean distance on pulls from experiment (left panel), and the corresponding parallel coordinates for all 14 observables (right panel) with color code matching. The darker line for each color in the parallel coordinate plot marks the cluster benchmark (also indicated on the left, with an open diamond symbol).}
\label{f:compchi-pc}
\end{figure}

We begin with a comparison between the $\chi^2$ based classification and one possible outcome of hierarchical clustering in Fig.~\ref{f:compchi} and Fig.~\ref{f:compchi-pc}. For this first comparison we choose as coordinates the pulls from experiment with correlations included, Eq.~\ref{eq:y2} with $R_i=E_i$, and Ward.D2 linkage with Euclidean distance for $\kappa=5$. 
The regions classified by $\chi^2$ and shown on the left panel, are then lying at up to  $0.78\sigma,~1.57\sigma,~2.35\sigma,~3.13\sigma,$ and $3.92\sigma$ from the BF.
Recall that here we are interested in equidistant bins in $\sigma$, where the number of bins is chosen to match the number of clusters investigated in the following. Therefore, the resulting ellipses do not match the classically used $\sigma$ intervals.

The number of clusters used, $\kappa=5$, gives a value of adjusted Rand index (ARI) near .3 when the clustering is compared to the $\chi^2$-based classification. The largest ARI for these clustering parameters, about 0.5, occurs for $\kappa=2$ (i.e. the clustering result is most similar to the binning in $\chi^2$ with the minimum number of clusters).  

The clustering result shown in Fig.~\ref{f:compchi-pc} (left) already illustrates that this procedure contains complementary information to the $\chi^2$ classification. Although the clusters roughly have a similar shape, the $\chi^2$ considers both sides of the BF to be equivalent. This is why the ARI index is maximized for only two clusters in this case. 
The parallel coordinate display, shown in the right panel of Fig.~\ref{f:compchi-pc}, can be used to understand the difference between the two classifications. The light green and purple clusters both have large overlap with the $0.78\sigma$ region in Fig.~\ref{f:compchi}, but the values of $R_K$ and $R_{K^\star}$ (coordinates $O_{13}$ and $O_{14}$ on the plot) clearly distinguish between them. Similarly the pink and brown clusters of the right panel share partial overlaps with the $1.57\sigma$ and $2.35\sigma$ regions, but the parallel coordinate plot shows which observables separate them. Comparisons such as this one provide insight into distinguishing parameter points with identical $\chi^2$.

Note here that the clustering was a necessary step to enable us to make the connection between the parameter space (as shown on the left in Fig.~\ref{f:compchi-pc}) and the observable space represented in the right plot of Fig.~\ref{f:compchi-pc}: we use it to partition the parameter space into regions that map in a meaningful way onto the observable space, and connect the information using color in the two displays.

Interpretation of the different scales appearing in the parallel coordinate display can also be of interest. Recall that here we are comparing centered coordinate values as defined in Eq.~\ref{eq:y2}, and the y-axis of Fig.~\ref{f:compchi-pc} (right) thus indicates the mean value (zero after centering) and moving one unit away from the mean value. Since these units include information about the uncertainties as well as their correlations, the variance along each direction can be interpreted directly as a measure of how relevant this observable might be in the clustering: a larger spread means considerable variation in predictions when taking into account all sources of uncertainty, and thus these observables can resolve different regions in parameter space. On the other hand, little variation indicates that the current uncertainties make this observable less useful in distinguishing the considered parameter points. This can be augmented with information from the tour display, in particular to better understand correlations between predicted values.

\begin{figure}[h]
\centering{
\includegraphics[scale=0.75]{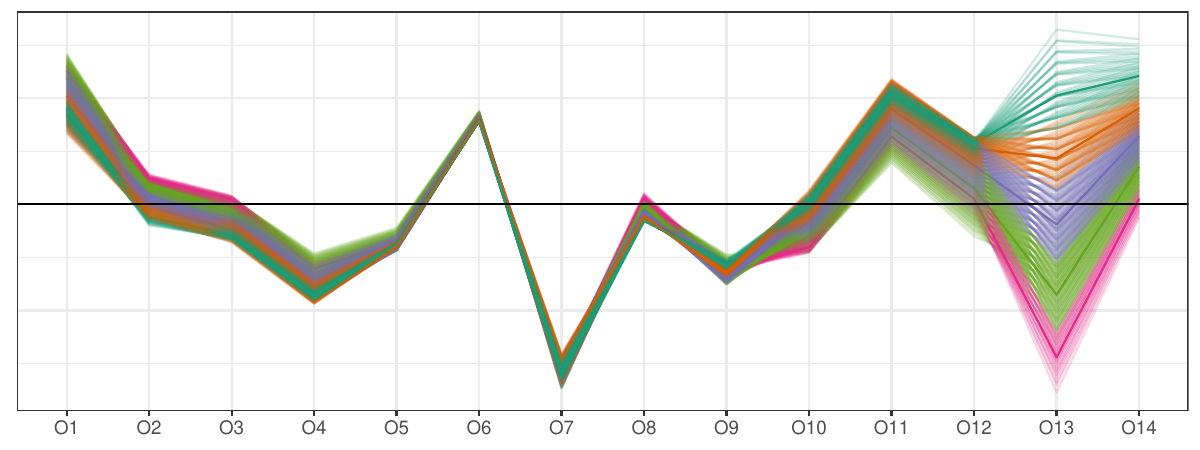}}
\caption{As Fig. \ref{f:compchi-pc} but without centering. The black horizontal line marks the position of the reference point, in this case the central value of the averaged experimental results.}
\label{f:compchi-pc2}
\end{figure}

Clustering can also shed light into internal tensions in a global fit. The parallel coordinate plot without centering shown in Fig.~\ref{f:compchi-pc2} serves to make this point. In this figure, information about the central value of the experimental result for each observable is retained as the origin of the vertical axis, and is shown as a black horizontal line (all pulls vanish when evaluated at the reference point). For example the light green cluster is closer to experiment than the purple one for many observables but not for $O_{1,2,13}$. In particular the bins of $P_5^\prime$ that contribute most to the discrepancy between SM and BF ($O_{4,5}$) pull in a different direction than $R_K$ ($O_{13}$).

For each cluster, it is possible to find a representative defined by its centroid (see Eq.~\ref{eq:bp}), which can then be used as a benchmark for comparative studies. In Fig.~\ref{f:compchi} the positions of these benchmarks in parameter space are marked by diamonds, and they are highlighted in the parallel coordinate plot. The parameter values for each of these benchmarks can also be read off from the ``benchmarks'' tab on the app, which is most useful in problems with more than two parameters. 

\subsection{Clustering choices}\label{s:choosingpar}

In arriving at our first clustering result, we had to make a number of choices: the coordinate representation and distance metric, the linkage method and finally the number of clusters to be shown. These choices could have a large impact on the resulting clustering, and they should be informed by the underlying problem and question. In practice we suggest to look at the distribution in observable space and to try out different settings to complement existing knowledge to make the most of the result. Here we illustrate this by discussing the number of clusters while keeping the other settings fixed.

We first use the tour display to study the distribution of points in the 14D observable space revealing that the points are distributed continuously on a 2D surface, see animation \href{https://uschilaa.github.io/animations/pandemonium1.html}{here}. This can be understood as a consequence of our setup: our model points sit on a uniform grid in parameter space and all the observables are smooth continuous functions of the parameters. This means that we cannot expect clearly separated clusters and most of the cluster statistics typically used to determine a preferred number of clusters will not be effective here. However, since our coordinates are normalized to the total uncertainty, the cluster size has a physical interpretation: we can select the number of clusters that results in grouping points that are considered indistinguishable at a selected level of confidence. 

\begin{figure}[h]
\includegraphics[scale=0.4]{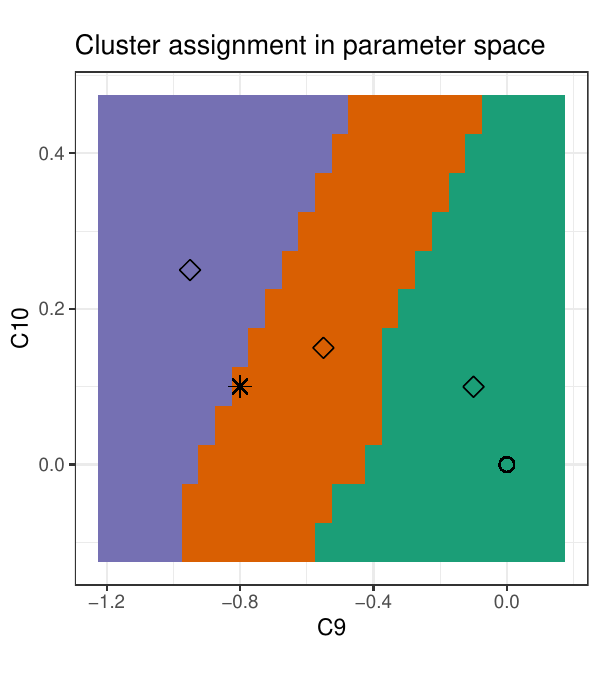}\includegraphics[scale=0.57]{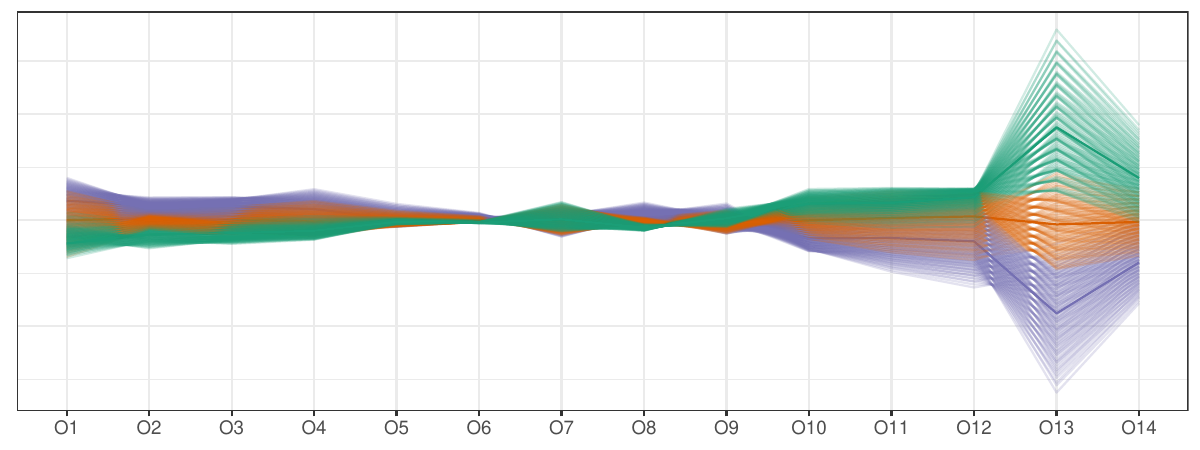}
\includegraphics[scale=0.4]{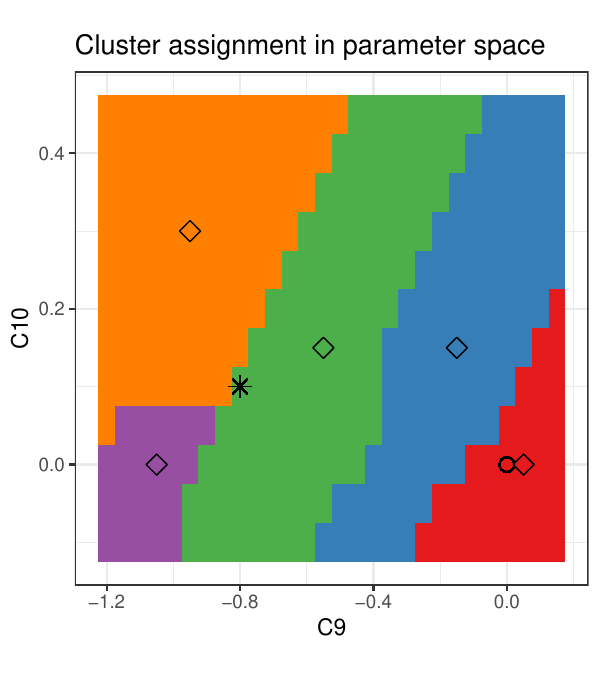}\includegraphics[scale=0.57]{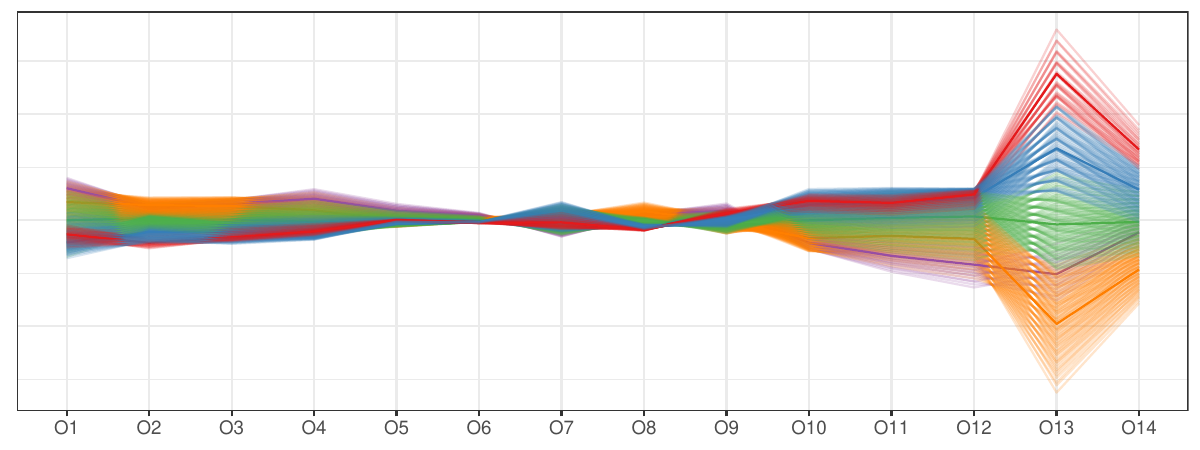}
\includegraphics[scale=0.4]{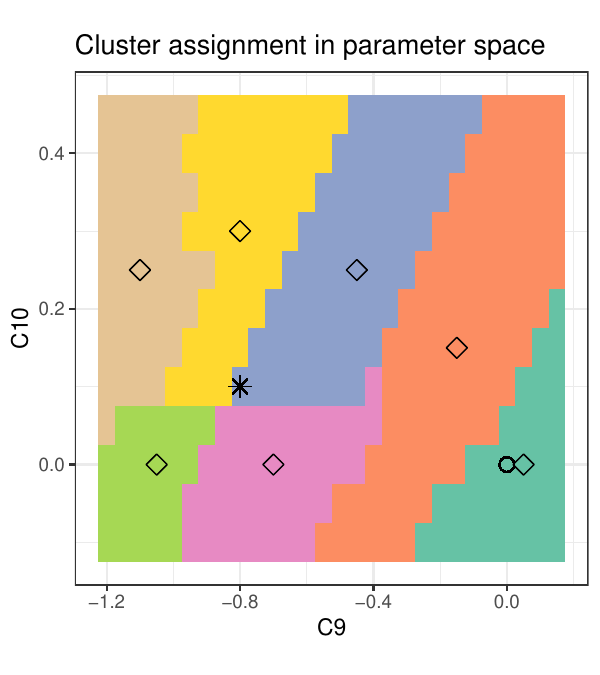}\includegraphics[scale=0.57]{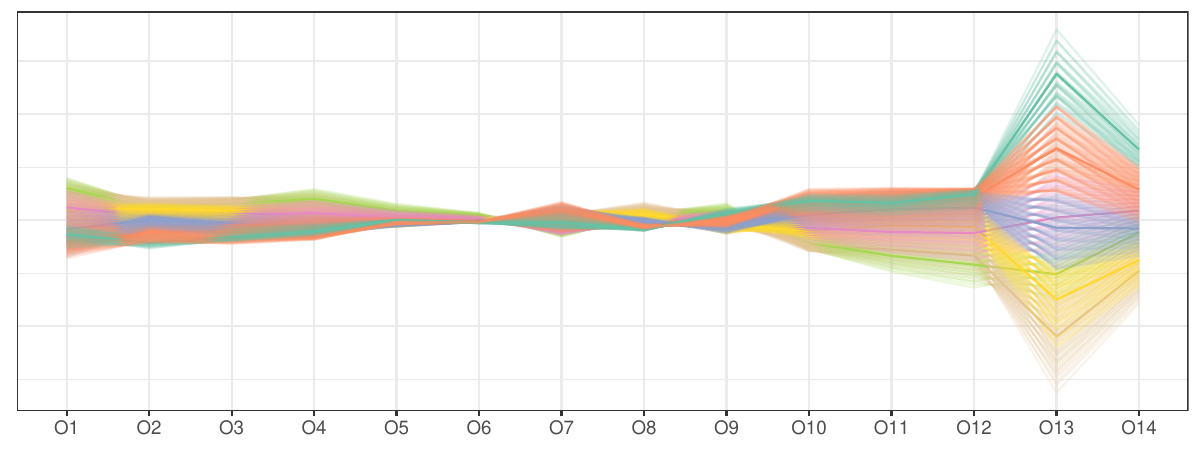}
\caption{The results of three, five and seven clusters using Euclidean distance on the pulls with covariance and average linkage in parameter space (left panel) with the corresponding parallel coordinate plots (right panel).}
\label{f:euc}
\end{figure}

\begin{figure}[h]
\centering{\includegraphics[scale=0.45]{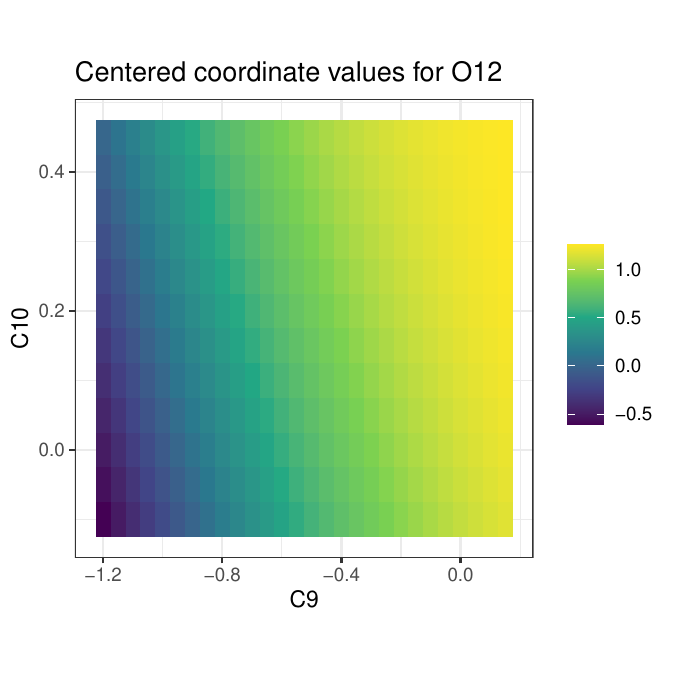}
\centering{\includegraphics[scale=0.45]{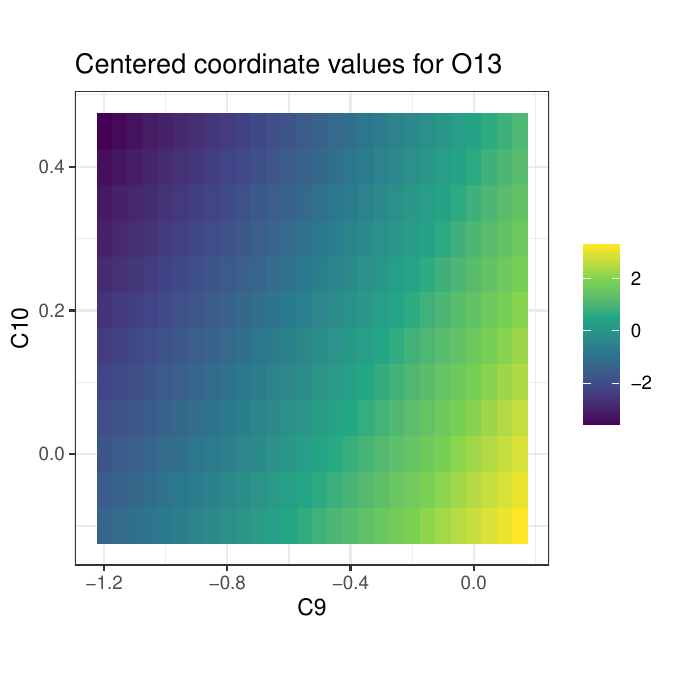}}
}
\caption{Variation of observables 12 (left, $P_2$ at high $q^2$) and 13 (right, $R_K$) with the two model parameters. Note that the color scale shows the variation in terms of the centered pulls calculated with correlations included (Eq.~\ref{eq:y2}), thus zero indicates the mean value, and any variation away from the mean is measured in those units.}
\label{f:rkvar}
\end{figure}

Partitioning this data set necessarily involves some arbitrariness, introducing splits along the 2D surface. To illustrate how this is not necessarily a shortcoming we now consider using the average linkage method on the pull coordinates with covariance and Euclidean distance. 
In  Fig.~\ref{f:euc} we show a sequence of three, five and seven clusters. This sequence shows the first two partitions  occurring roughly along the lines $C_{10}-C_9\approx 1~(0.6)$, corresponding to fixed values of $R_{K^{(\star)}}\sim 0.76~(0.86)$ respectively.\footnote{Approximate linearized equations for these two observables are $R_K \approx R_K^{\star}\approx 1 +0.24(C_9-C_{10})$ when other Wilson coefficients are set to their SM values \cite{Kou:2018nap}.} The variation in the predictions for $R_K$ is shown in Fig.~\ref{f:rkvar} (right), and  confirms the conclusion that this split happens along the direction of constant $R_K$ ($O_{13}$). 

In the next step two splits occur. First, the dark green group is partitioned in two (red and blue cluster in the second row) at some fixed value of $R_K$ as can be confirmed in the right panel, which shows the two new groups separated mostly in $O_{13}$. Second, the purple group splits in two (now bright orange and purple) roughly at $C_{10}\sim 0.07$. The parallel coordinate plot shows this being due mostly to certain bins of the $P_2$ observable ($O_{11,12}$). Note here that while the highlighted benchmark points of the two groups do differ in $O_{13}$, the two groups overlap in this observable.

Finally in the last step, the bright orange cluster splits into what is now shown in brown and yellow near $C_9\sim -.95$ and the bright green cluster splits into what are now the light purple and pink clusters near  $C_{10}\sim 0.07$. These splits are more subtle and due to the combination of multiple observables.

\begin{figure}[h]
\centering{
\includegraphics[scale=0.4]{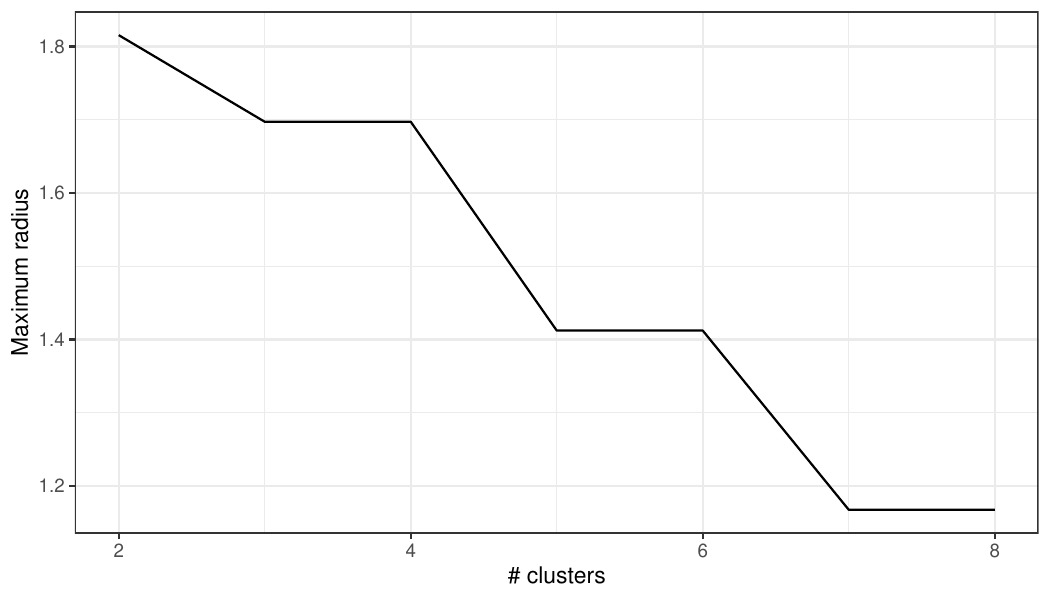}~\includegraphics[scale=0.4]{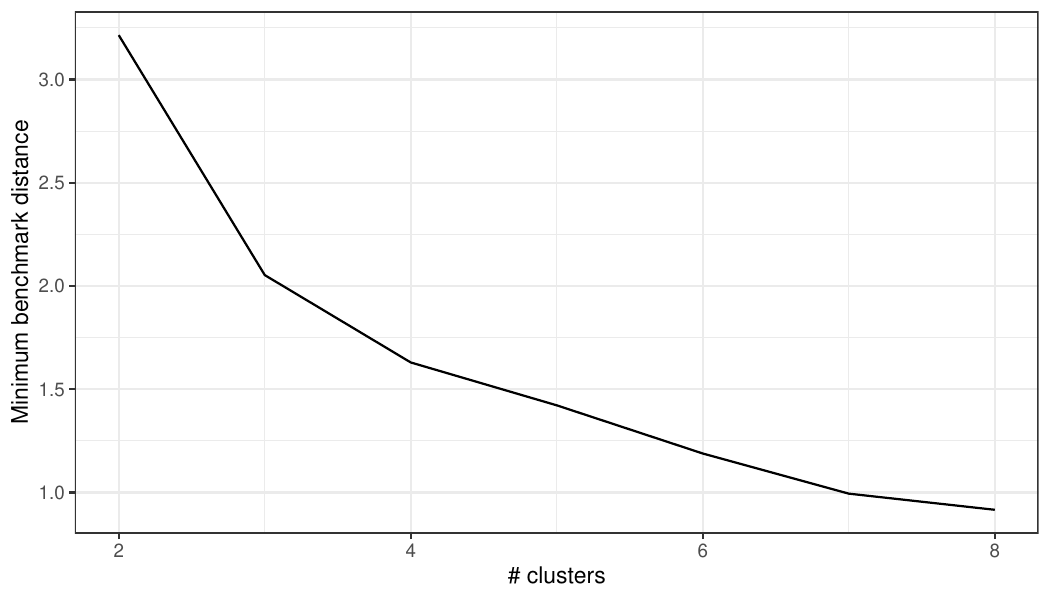}}
\centering{
\includegraphics[scale=0.4]{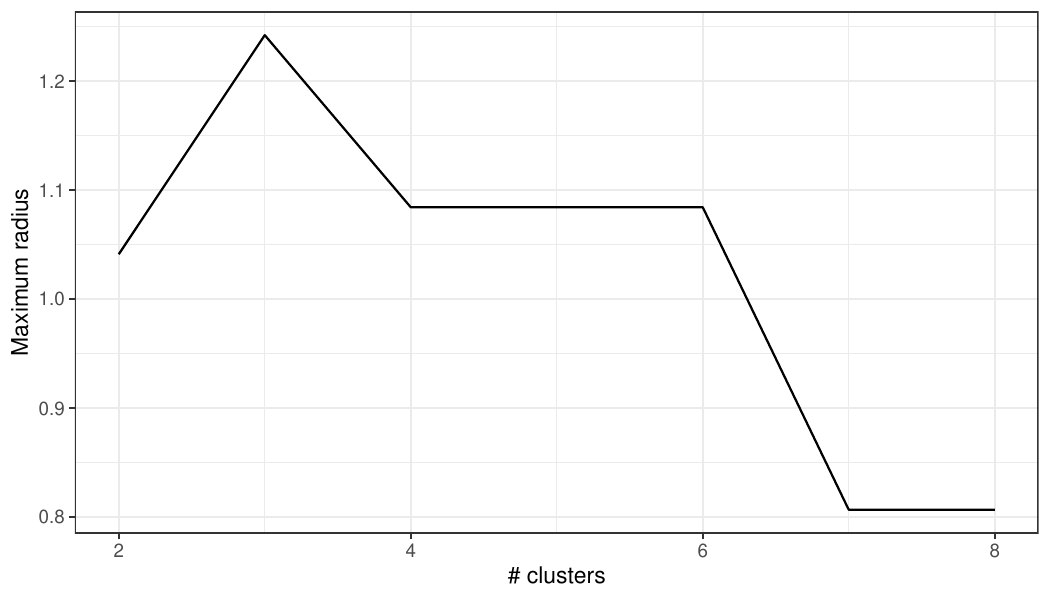}~\includegraphics[scale=0.4]{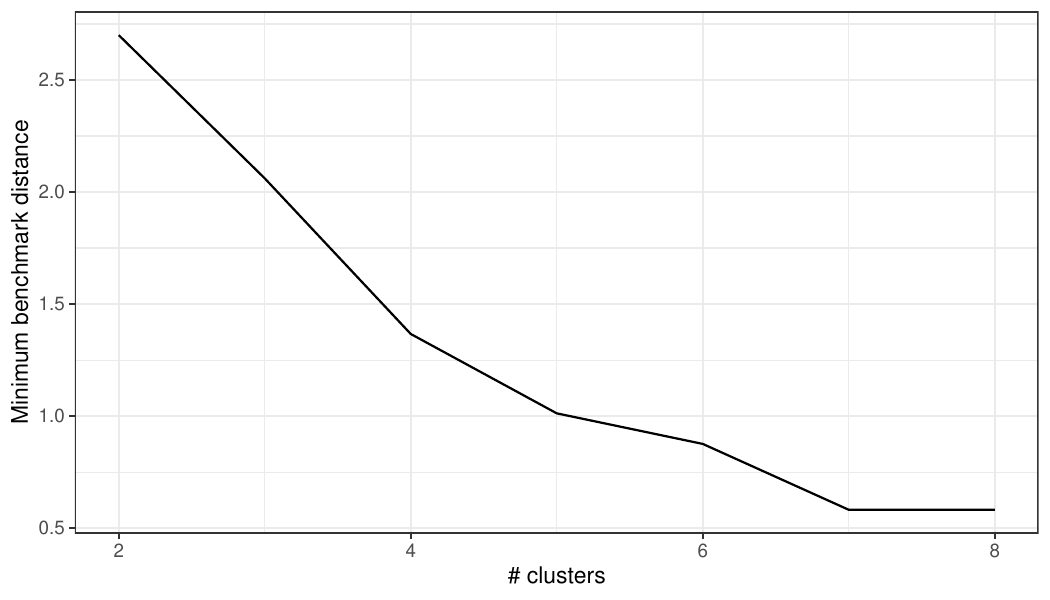}}
\caption{Maximum cluster radius (left) and minimum benchmark distance (right) when using pull coordinates with correlations and Euclidean distance with average linkage (top);  maximum distance with complete linkage (bottom).}
\label{f:rdmax}
\end{figure}

The optimal number of clusters in this case can be decided by considering their size and separation. To this end we show in the top panel of  Fig.~\ref{f:rdmax} the maximum cluster radius (i.e. the largest distance any point has from its representative benchmark) and the minimum benchmark distance (the smallest separation between two benchmarks) as functions of the number of clusters. These two metrics capture the similarity of points within a given cluster and the dissimilarity between different clusters respectively. Following our discussion in Section~\ref{ss:benchpts} we interpret the cluster radius as a confidence level contour.  The square of the Euclidean distance on the pulls\footnote{It would be exact if we used instead the coordinates $\delta_\sigma$ defined in \cite{Capdevila:2018jhy}.} is a good approximation to $\Delta\chi^2$, implying for this example that 
a cluster radius $r_j\lesssim 1.5$ corresponds to a cluster $C_j$ of points that are indistinguishable at the $1\sigma$ level from their benchmark point $c_j$\footnote{Because at 68\% c.l. $\Delta\chi^2=2.3\approx 1.5^2$ for two degrees of freedom.}. The left panel of Fig.~\ref{f:rdmax} shows that with 5 clusters all radii are below this value. The smallest number of clusters satisfying this condition corresponds to the desired partitioning: the simplest solution in which all the points within a cluster lie at most within $1\sigma$ of the benchmark. Similarly, the minimum benchmark distance statistic, shown in the right panel of Fig.~\ref{f:rdmax}, reveals that for 5 clusters any two benchmarks are separated by at least $d(c_i,c_j)\gtrsim 1.5$ so they are distinguishable at the $1\sigma$ level. These two statistics combined, indicate a preferred solution of five clusters for this problem.  
Alternative criteria to determine the number of clusters can also be constructed; for example using the maximum cluster diameter which gives  the maximum separation between any two points within a given cluster. For these five clusters, this value is about 3, which for two degrees of freedom corresponds to about $2.5\sigma$.

A somewhat different interpretation is possible using maximum distance. In this case choosing a maximum cluster radius of one indicates that none of the predictions differ from the benchmark point by more than one unit of combined theoretical and experimental uncertainty. In the bottom panel of  Fig.~\ref{f:rdmax} we show the corresponding metrics for maximum distance with complete linkage on the pulls. The maximum radius statistic suggests four to six clusters whereas the minimum benchmark distance suggests five clusters. The five clusters for this choice of distance and linkage differ from the center panel of Fig.~\ref{f:euc} mostly in that all boundaries now follow the correlation implied by $R_K$ and $R_{K^\star}$. 

\subsection{Correlation, collective effects and dominant observables}

Next we want to explore how other choices affect the results, and how we can connect this to physics insights. For simplicity here we restrict our study to results with three clusters, pull coordinates and average linkage. We  vary the distance metric used and whether correlations are included in the coordinate definition.

\begin{figure}[h]
\includegraphics[scale=0.4]{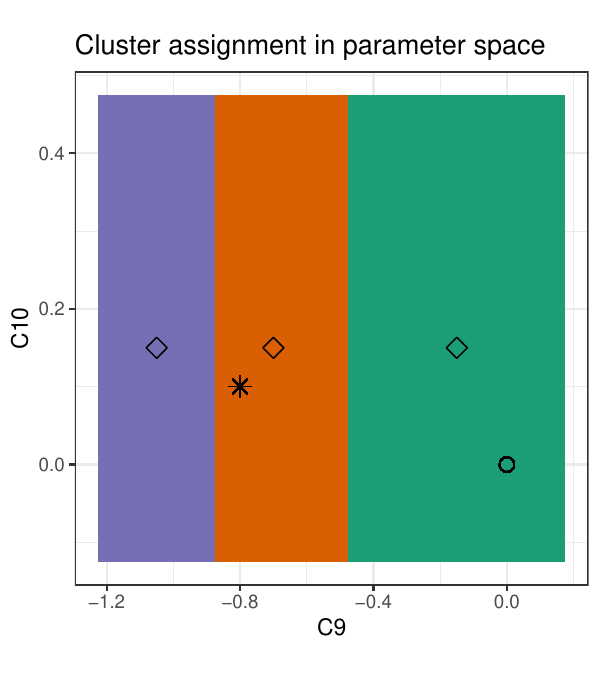}\includegraphics[scale=0.6]{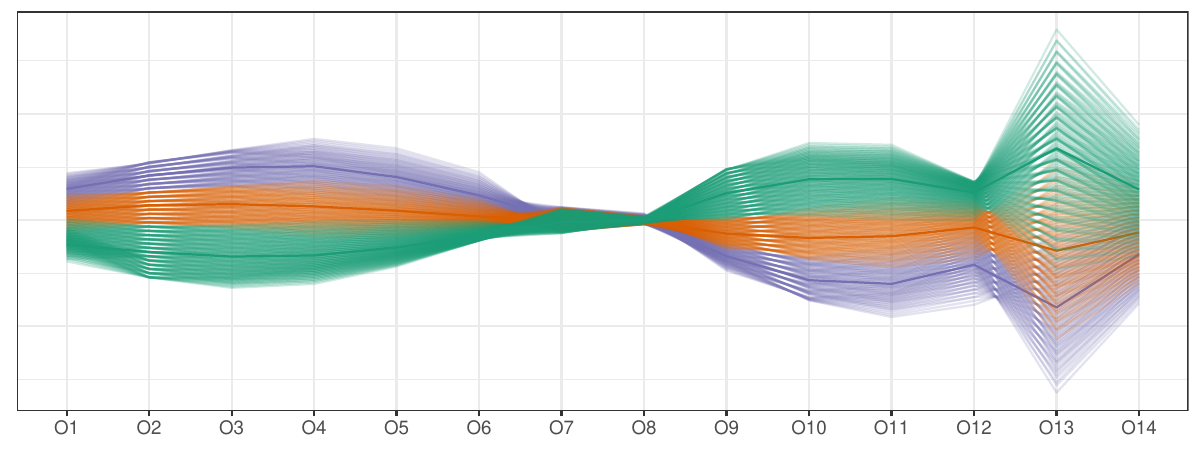}
\caption{The results of three clusters using Euclidean distance on the pulls without covariance and average linkage in parameter space (left panel) with the corresponding parallel coordinate plots (right panel).}
\label{f:euc2}
\end{figure}

First, we repeat the clustering shown in the upper panel of Fig.~\ref{f:euc}, but computing the coordinates without correlations. This is shown in Fig.~\ref{f:euc2} (left) and we see that the partitioning now occurs exclusively along $C_9$ (while there is of course still some dependence on $C_{10}$ is is much smaller than that on $C_9$, and thus is not visible at the selected grid coarseness and number of clusters, see Fig.~\ref{f:new_depc10} and the associated discussion in the appendix). Indeed, the effect of correlations is known to be important for the global fit of $b\to s\ell^+\ell^-$ observables. The parallel coordinate plot shown in Fig.~\ref{f:euc2} (right) shows that these clusters are neatly separated along the first twelve observables but mix for $R_K$ and $R_{K^\star}$ ($O_{13,14}$). Thus, the angular observables have gained importance when correlations are neglected (the reduced importance of these observables in the global fit when accounting for correlated errors was previously noted with the metrics developed in~\cite{Capdevila:2018jhy}).

These results clearly show that the angular observables suggest a different partitioning of the parameter space compared to $R_K$ and $R_{K^\star}$ ($O_{13,14}$), and the results depend critically on having one of them  dominate. These insights could also be obtained by experimenting with different linkage methods, or by  dropping the dominant observables $R_K$ and $R_{K^\star}$ from the input.

\begin{figure}[h]
\includegraphics[scale=0.5]{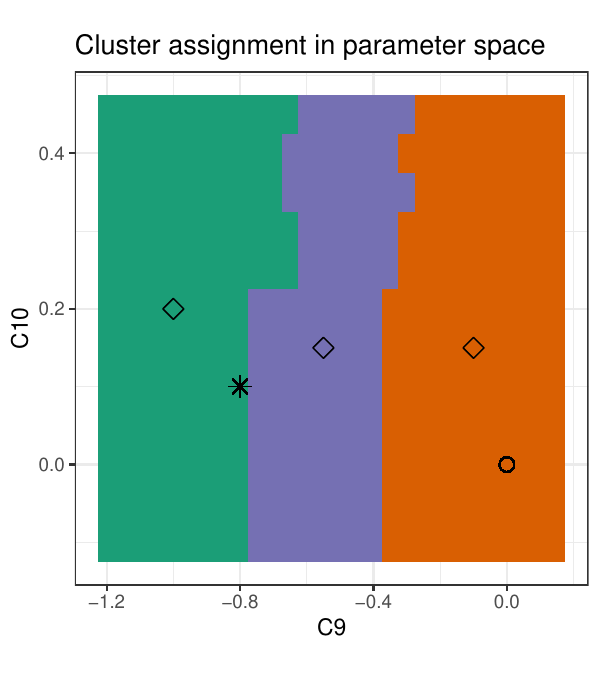}
\includegraphics[scale=0.5]{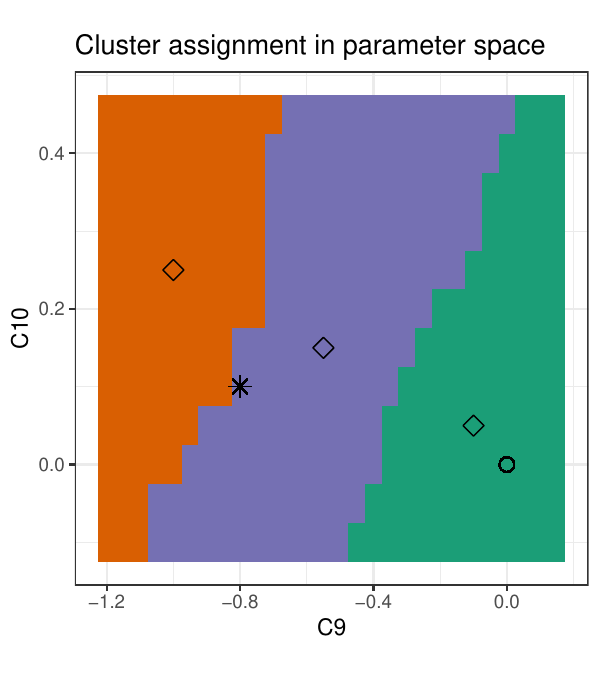}
\includegraphics[scale=0.5]{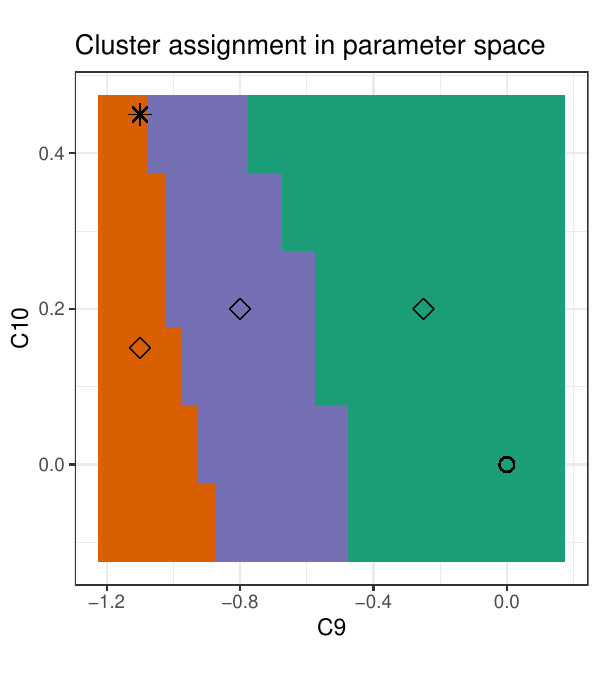}
\caption{The results of three clusters using average linkage and pull coordinates with different distance metrics: Manhattan distance with covariance (left panel), Maximum distance without covariance (middle panel) and Euclidean distance with correlations, but dropping  $R_K$ and $R_{K^\star}$ (right panel). For the associated parallel coordinate plots see Fig.~\ref{f:3clwpc} in the Appendix.}
\label{f:3cl}
\end{figure}

For example, we can explore the partitioning of the parameter space when including the correlations in the coordinate definition, but using Manhattan distance, or when dropping correlations but using Maximum distance. These results are shown in Fig.~\ref{f:3cl} (left and middle). We find that Manhattan distance, which de-emphasizes dominant observables shows results similar to those in Fig.~\ref{f:euc2} (i.e. when the importance of angular observables is exaggerated by neglecting correlations). Similarly, when placing the emphasis on dominant observables by using Maximum distance, we find that the patterns match those of Fig.~\ref{f:euc} even when correlations are not included.

Finally, we can compare these results with the clustering obtained after explicitly removing $R_K$ and $R_{K^\star}$ from the dataset. This is shown in Fig.~\ref{f:3cl} (right) and produces a different pattern.\footnote{Note that the BF point changes in the right panel as a result of having dropped two observables.} Because we have dropped the two observables that previously dominated the result and induced a positive correlation between the two parameters, we now observe the negative correlation induced by some of the angular observables instead. As an example we show the variation in predictions for observable 12 (highest $q^2$ bin of $P_2$) in Fig.~\ref{f:rkvar} (left).  

While these comparisons of course do not show exact matches, the overall patterns observed give useful hints that can aid the physics interpretation of the importance of different observables, and the types of patterns they induce in the clusters in parameter space. 

\subsection{Beyond fixed errors}

One of the assumptions we have made up to now is that the theoretical uncertainties for all observables are independent of the model parameters. This is clearly an approximation, but one that is often used in global fits to simplify the optimization. For the observables considered here this assumption was assessed to be valid in~\cite{Descotes-Genon:2015uva} by direct comparison of the covariance matrix, but the detailed impact on the results was not studied.\footnote{See footnote 12 on page 25 of~\cite{Descotes-Genon:2015uva}.}

Here we study how the computation of theory uncertainties and correlations affect the clustering
outcome. The first thing to notice is that when we use model independent errors in our coordinate definitions (Eqs.~\ref{eq:y1}~and~\ref{eq:y2}), the distance between two models does not depend on the reference point $R_i$. Instead, when the errors are model dependent the distance calculation does depend on the reference point chosen. Here we will work with pull coordinates with respect to the experimentally measured values.

Uncertainties are evaluated via sampling of the nuisance parameters and this introduces a further statistical error that differs between parameter points.   Finally, because our implementation for the computation of coordinates on the fly assumes fixed input for the covariance matrix, we have to compute the coordinate representation outside the app and load the resulting values as user defined coordinates. This section thus serves as an example of how to use the interface when the desired coordinate definition is not covered by those introduced in Section~\ref{sec:coord}.

\begin{figure}[h]
\centering{\includegraphics[scale=0.3]{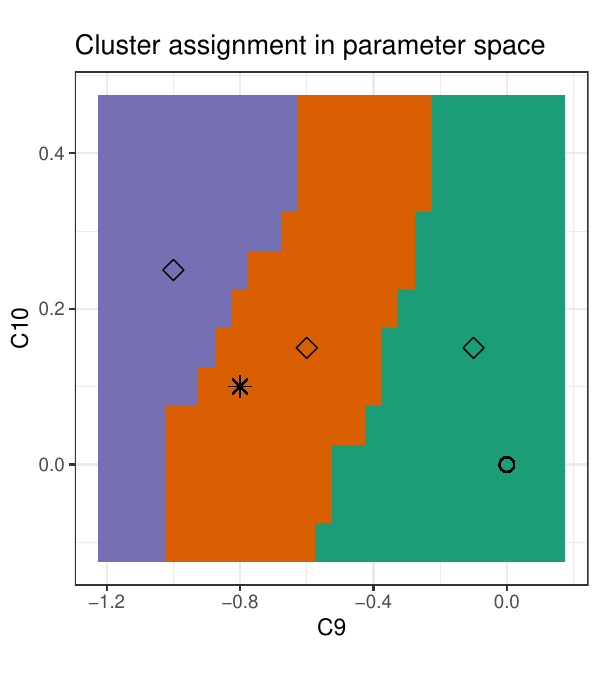}\includegraphics[scale=0.3]{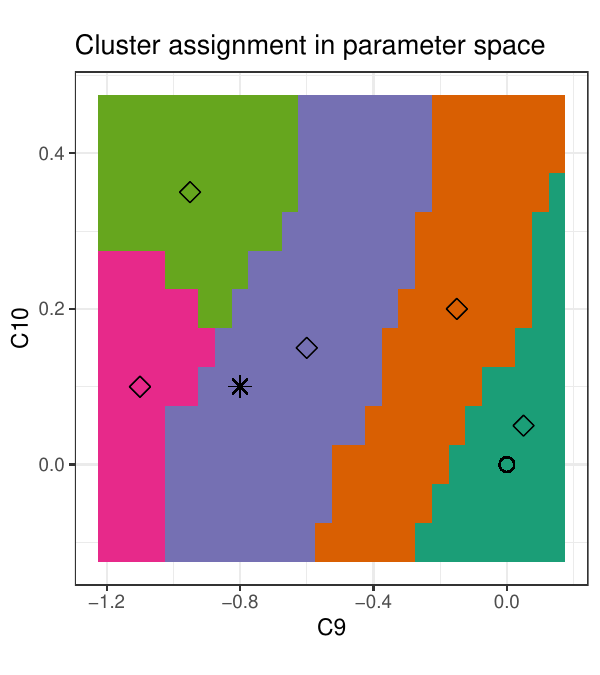}\includegraphics[scale=0.4]{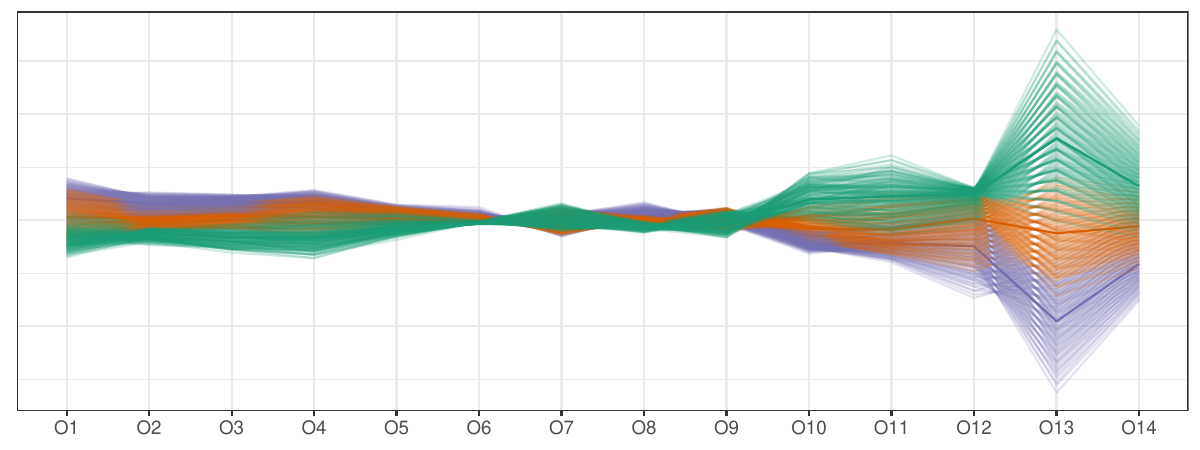}}
\caption{The results of three (left) and five (centre) clusters using Euclidean distance on the pulls with correlations and average linkage with theoretical errors evaluated at each model point shown in parameter space and the corresponding parallel coordinate plot for three clusters (right panel).}
\label{f:eucX}
\end{figure}

To illustrate the results we reproduce the simple clustering used before: three and five clusters using average linkage and Euclidean distance, but now computing the coordinates using the model specific covariance matrix for each parameter point. The resulting clustering in $C_9$ vs $C_{10}$ is shown in Fig.~\ref{f:eucX} (left and middle), together with the corresponding parallel coordinate plot for 3 clusters (right).\footnote{To reduce the statistical uncertainty introduced by sampling of the nuisance parameters we use $N=2000$ when evaluating the new physics uncertainty within {\tt flavio}.}

At small absolute values of the Wilson coefficients (i.e. near the SM point) the result looks similar to what was found with fixed errors: it indicates a correlation of $C_9$ and $C_{10}$ that results in approximately fixed values of $R_K$. This is expected since the fixed errors had been evaluated at the SM point. However, we note that the boundaries change slightly as we move away from the SM, most notably the partitioning between purple  and brown clusters which shifts towards larger (negative) values of $C_9$. As the number of clusters increases, the differences with Fig.~\ref{f:euc} become more evident. This is, of course, expected as finer details become more important and is illustrated with the five cluster partition in Fig.~\ref{f:eucX}. 

\subsection{More than two parameters}\label{s:fourp}

In general there will be more than two parameters affecting the predictions. To generalize these tools to incorporate that possibility we need to visualize more than two dimensions in both parameter and observable space. One possibility to study the resulting clusters would use two separate tour displays with the capability to slice \cite{Laa:2019bap,Laa:2020wkm}. This is beyond the scope of the present work but will be considered in a future publication.

However, we can use the present tool to visualize clusters with two dimensional slices of parameter space provided the parameter scan is on a regular grid, because this allows us to select points in a slice using an exact condition on the additional parameters. To see how this works we continue with our example of  neutral B-anomalies, where the global fit studies suggest that two additional parameters, $C_{9^\prime}$ and $C_{10^\prime}$, may also play a non-negligible role. 

We reproduce the settings from Section~\ref{s:choosingpar}, i.e. we use average linkage with Euclidean distance on the pulls and split the data into three clusters.

\begin{figure}[h]
\centering{\includegraphics[scale=0.4]{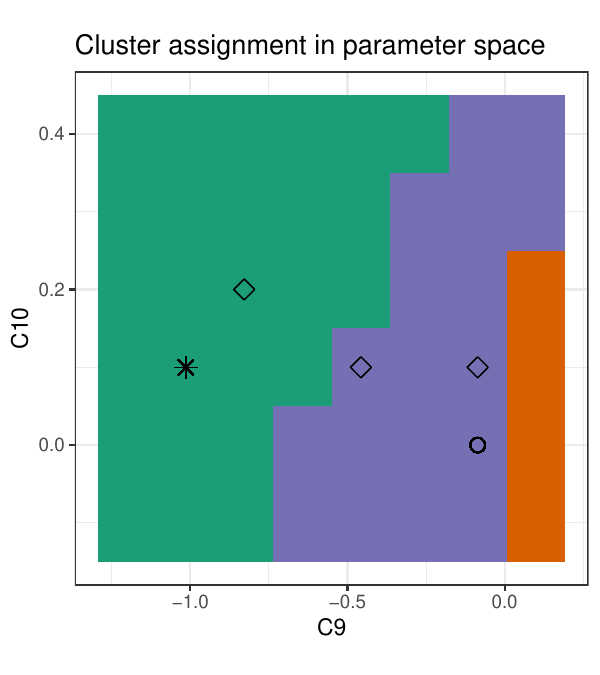}\includegraphics[scale=0.4]{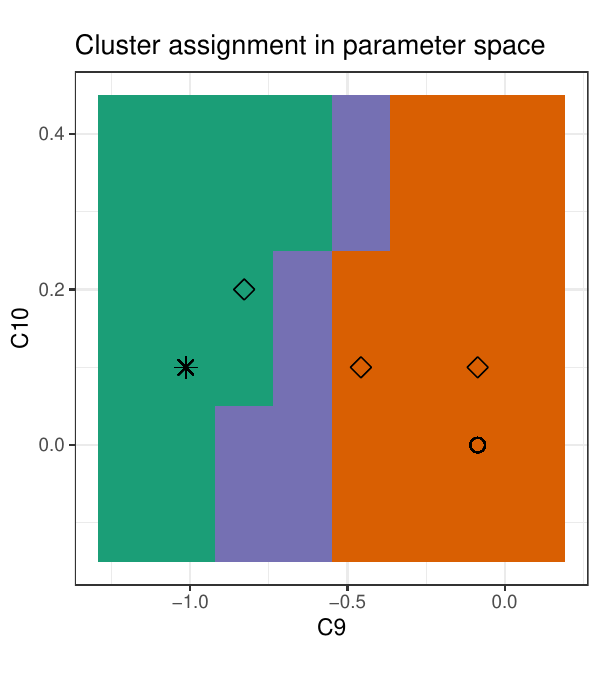}\includegraphics[scale=0.4]{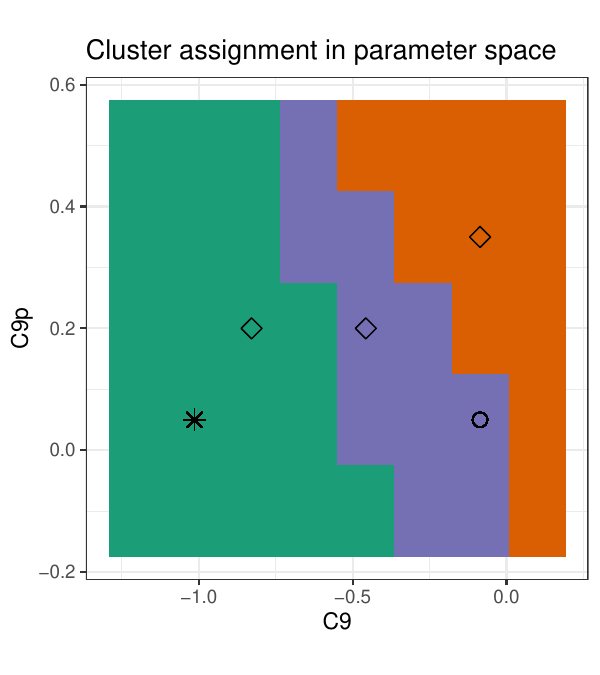}\includegraphics[scale=0.4]{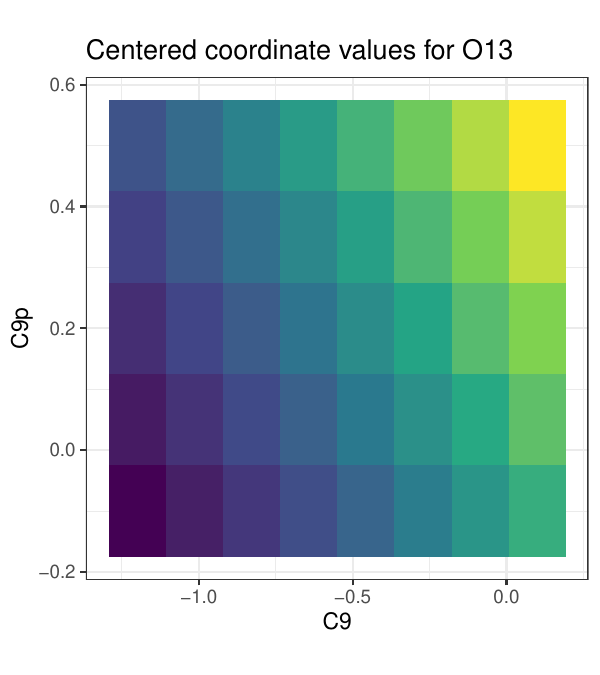}}
\caption{Three cluster separation of four parameter models with average linkage and Euclidean distance of pulls with correlations. Slices on $C_{9}-C_{10}$ plane at  $C_9^\prime = -0.1$ and $C_{10}^\prime =0$ (left-most) and $C_9^\prime = 0.5$ and $C_{10}^\prime =0$ (second left). Slice on the  $C_{9}-C_{9}^\prime$ plane at  $C_{10} = 0.2$  and $C_{10}^\prime =0$ (second right). Variation of predictions for $O_{13}$ across the $C_{9}-C_{9}^\prime$ plane (right).}
\label{f:euc4d}
\end{figure}

In Fig.~\ref{f:euc4d} we illustrate the clustering in the $C_{9}-C_{10}$ plane with two slices with $C_{10}^\prime =0$ and $C_9^\prime = -0.1$ (left panel) and then  $C_9^\prime = 0.5$ (center-left panel). The by now familiar correlation along fixed values of $R_K$ appears, but this time the size of the clusters changes as we vary  $C_9^\prime$ revealing the influence of the additional parameters. Next we show the $C_{9}-C_{9}^\prime$ plane using a slice with $C_{10} = 0.2$ and $C_{10}^\prime =0$ (center-right panel). The correlation seen in this slice can be traced primarily to $R_K$ as well, as can be seen in figure showing the variation in predictions for $O_{13}$ across the $C_{9}-C_{9}^\prime$ plane (right panel).  

\begin{figure}[h]
\centering{\includegraphics[scale=0.4]{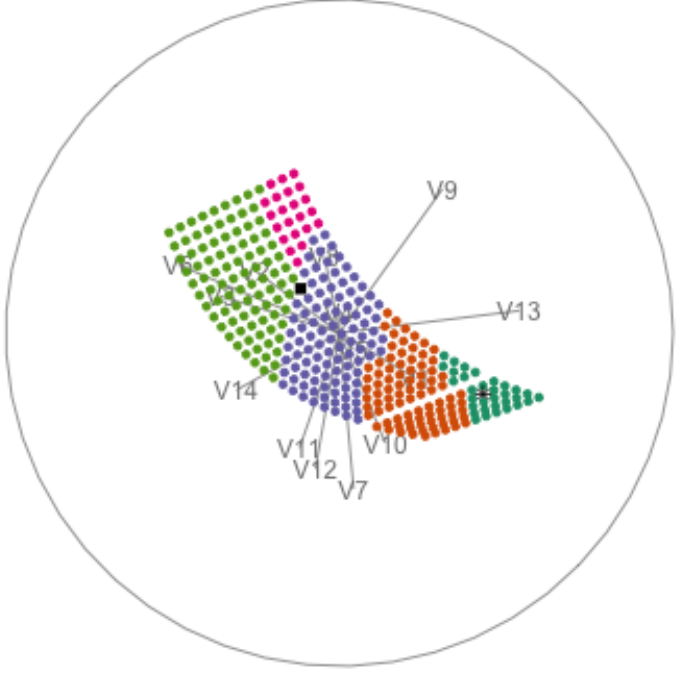}\includegraphics[scale=0.4]{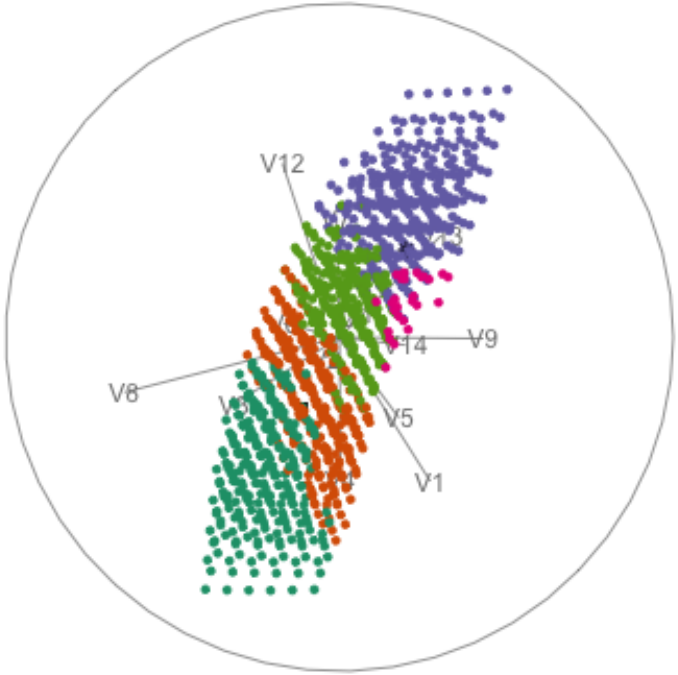}\includegraphics[scale=0.4]{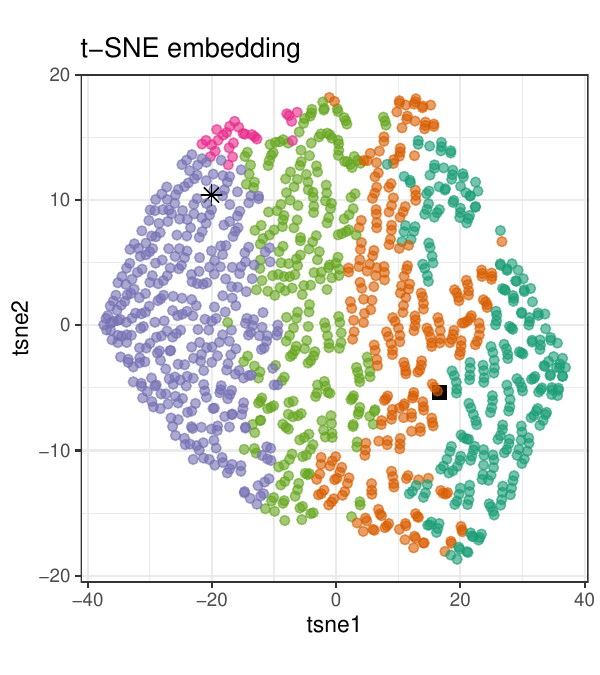}}
\caption{Still image from the tour showing five clusters in 14D observable space for models with two parameters (left, see animation \href{https://uschilaa.github.io/animations/pandemonium1.html}{here}) and four parameters (middle, see animation \href{https://uschilaa.github.io/animations/pandemonium2.html}{here}). For comparison we also show the 2D view obtained with t-SNE which is not resolving the clusters for this example.}
\label{f:tour24}
\end{figure}

It is suggestive to compare the clustering in observable space for the cases where the models depend on two and four parameters. With two parameters the predictions sit on a  2D manifold in 14 dimensions, whereas with four parameters they span a 4D manifold. This is easy to understand as fourteen functions of $n$ parameters  correspond to a parametric representation of an $n$D manifold in 14 dimensions. Interestingly, noise from the marginalization of nuisance parameters shows up in these structures as ``thickness'' of the expected dimensionality.\footnote{In our model evaluation using {\tt flavio}, this can be controlled setting the number of random evaluations of each observable.}  In Fig.~\ref{f:tour24} we illustrate this with two still plots from tours of five clusters in observable space for two and four parameters containing links to animated gifs. The plot in the right panel shows a 2D display of the four parameter data in observable space. Since our example does not contain clearly separated clusters, the points also appear continuously spread out, with some artificial structure introduced from the non-linear mapping which is not reflecting the clustering outcome. For example, clusters are spread across small gaps that appear in the display. In this case the information provided by the tour is more useful.

\begin{figure}[h]
\centering{
\includegraphics[scale=0.3]{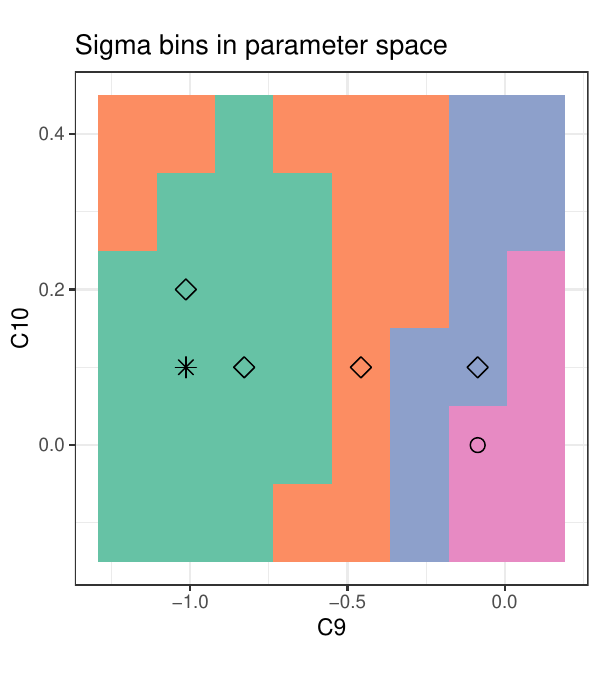}\includegraphics[scale=0.3]{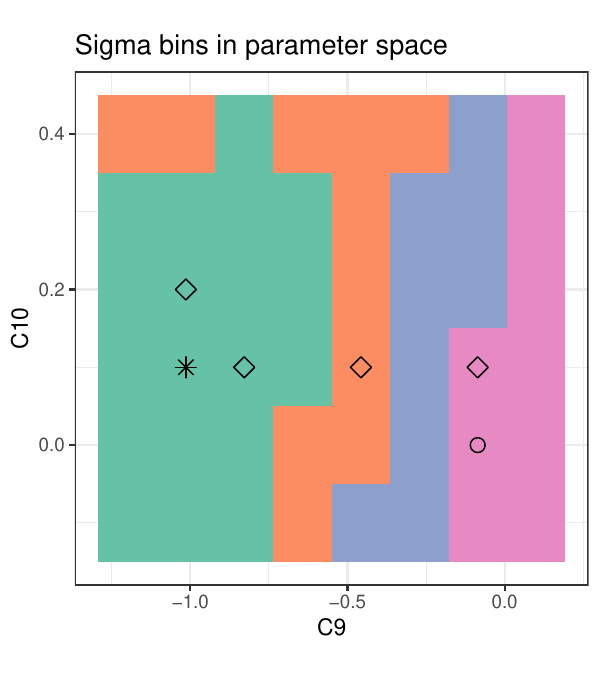}\includegraphics[scale=0.3]{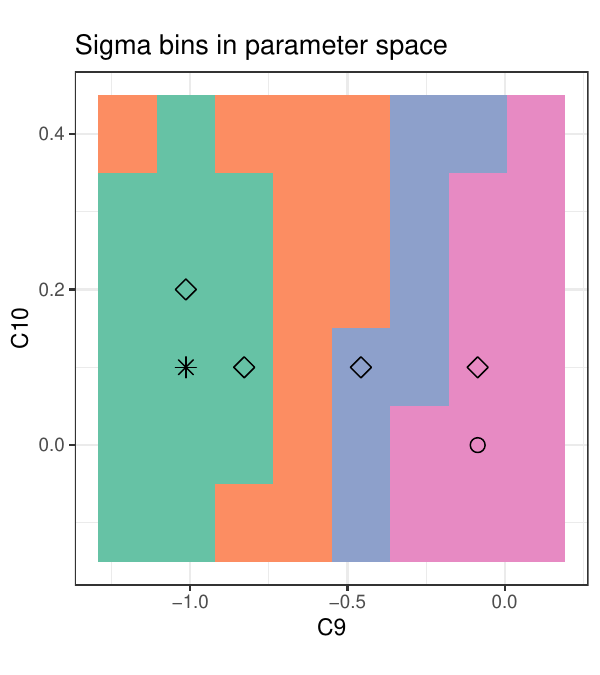}\includegraphics[scale=0.3]{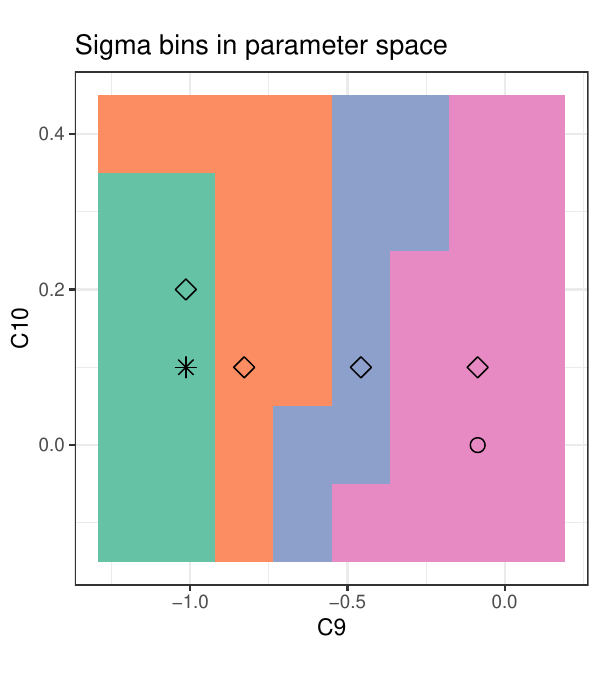}}
\centering{\includegraphics[scale=0.3]{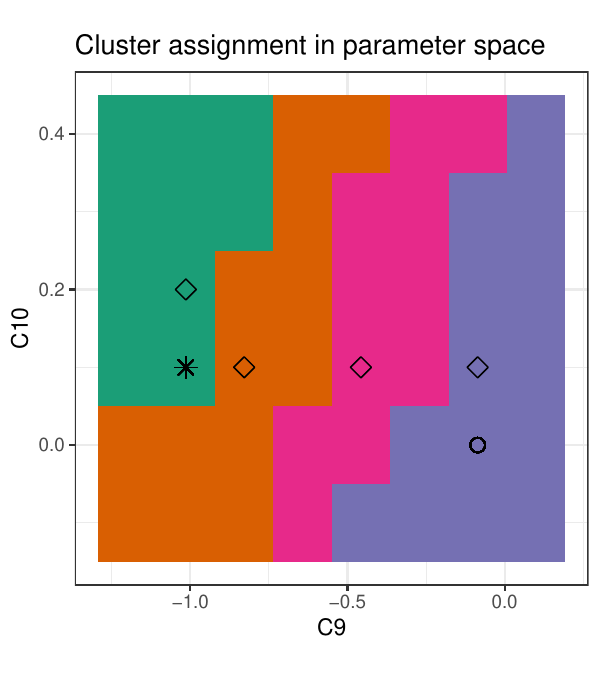}\includegraphics[scale=0.3]{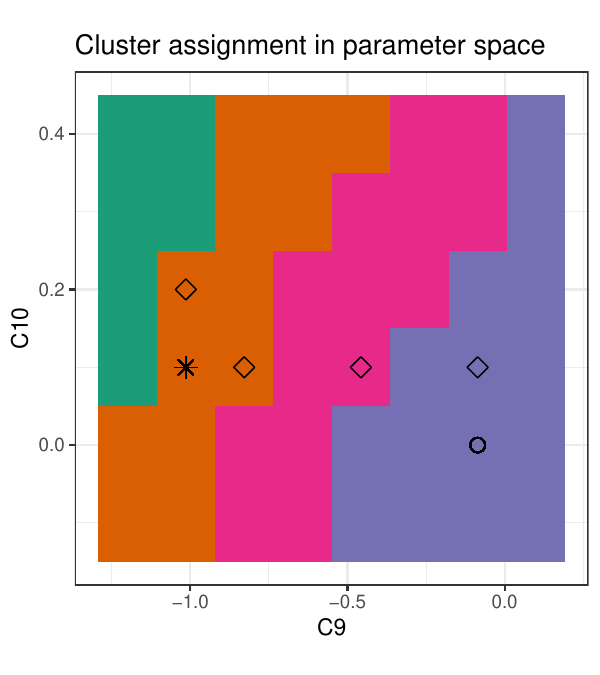}\includegraphics[scale=0.3]{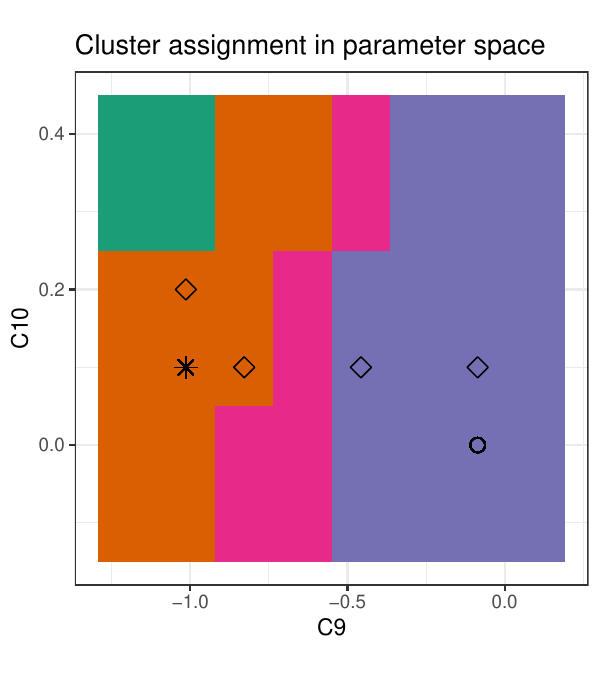}\includegraphics[scale=0.3]{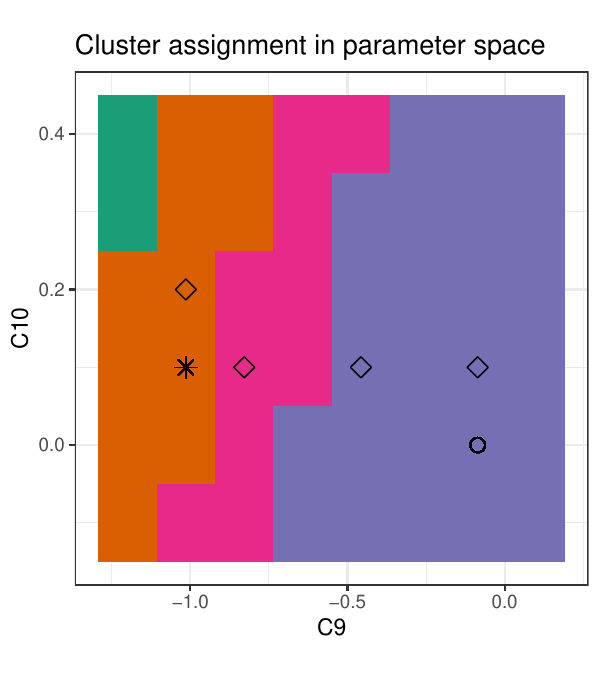}}
\caption{Partitioning into four clusters of data with four Wilson coefficients: above confidence level contours; below: clusters with average linkage and Euclidean distance.}
\label{f:rh}
\end{figure}

In this setting the $\chi^2$ function varies very slowly along certain directions in the $C_{9^\prime}=C_{10^\prime}$ plane and this affects its minimization. The clustering is not affected by this, but can result in hyper-cylindrical clusters. In Figure~\ref{f:rh} we show four slices in the $C_9-C_{10}$ plane corresponding to $C_{10^\prime}=-0.2$ and $C_{9^\prime}=-0.1,~0.05,~0.2,~0.35$ respectively. The top row illustrates how the $\chi^2$ function varies slowly along this direction while the bottom row illustrates a similar cluster behavior. But this particular functional dependence does not affect the clustering algorithm, which is grouping together points with similar predictions. The clustering outcome can however illustrate the limited sensitivity along certain directions.

\subsection{Additional observables}\label{s:moreobs}

Up to now we have only considered a small subset of the 175 observables included in the global fit of \cite{Capdevila:2016ivx}. The selection was guided by the fit results and its detailed analysis in \cite{Capdevila:2018jhy}. We now explore how additional observables could provide complementary information if included in the cluster analysis. To this end we consider a much larger subset that is easily accessible via an implementation in  {\tt flavio}, 89 observables, and we perform a preliminary analysis selecting observables for further study. From the observables entering the fit of \cite{Capdevila:2016ivx} we have included in our set of 89 the LHCb measurements except some of the observables describing the angular distribution of $B_s\to\Phi \mu\mu$ which are not yet defined in {\tt flavio}. None of the excluded observables  were found to have a major impact in our study of \cite{Capdevila:2018jhy}.

\begin{figure}[h]
\includegraphics[scale=0.4]{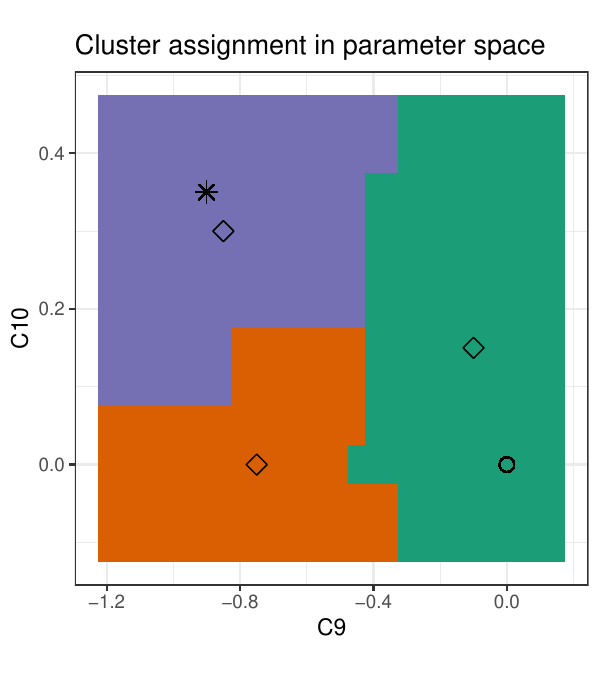}\includegraphics[scale=0.6]{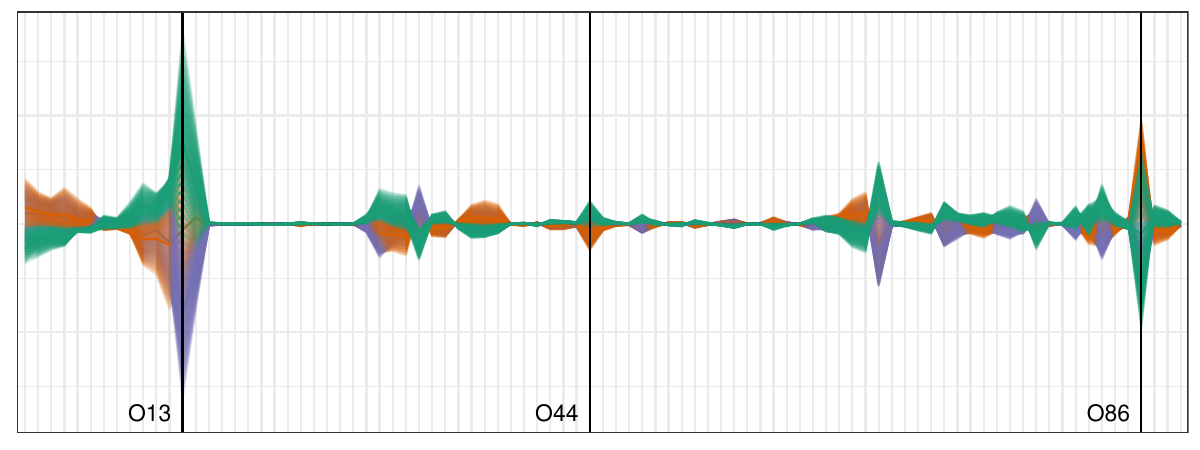}
\caption{The results of three clusters using Euclidean distance on the pulls with covariance and average linkage in parameter space (left panel) with the corresponding parallel coordinate plots (right panel) for all 89 observables.}
\label{f:euc89}
\end{figure}

\begin{figure}[h]
\includegraphics[scale=0.5]{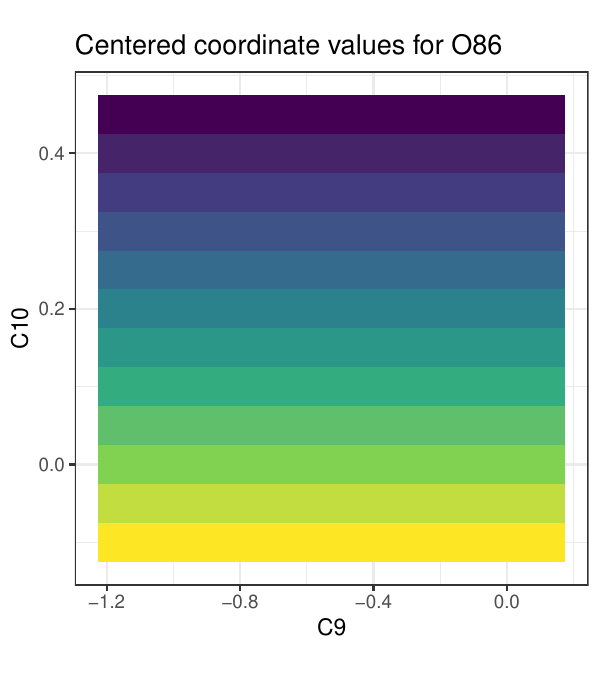}\includegraphics[scale=0.5]{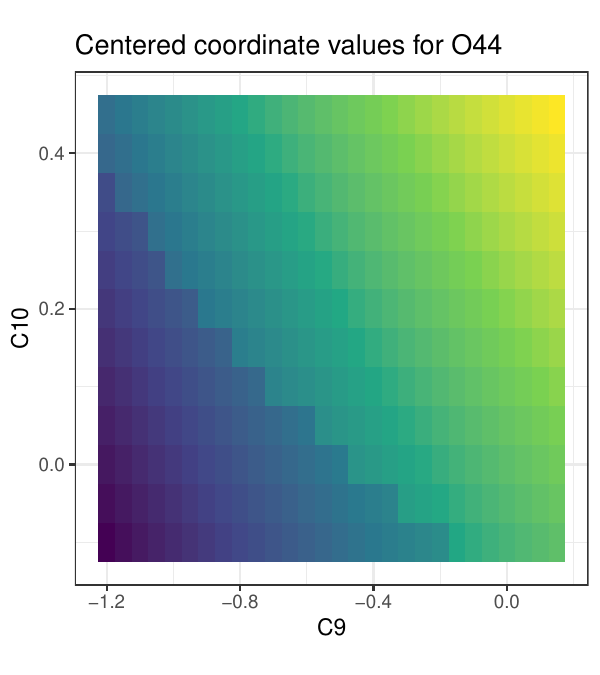}\includegraphics[scale=0.5]{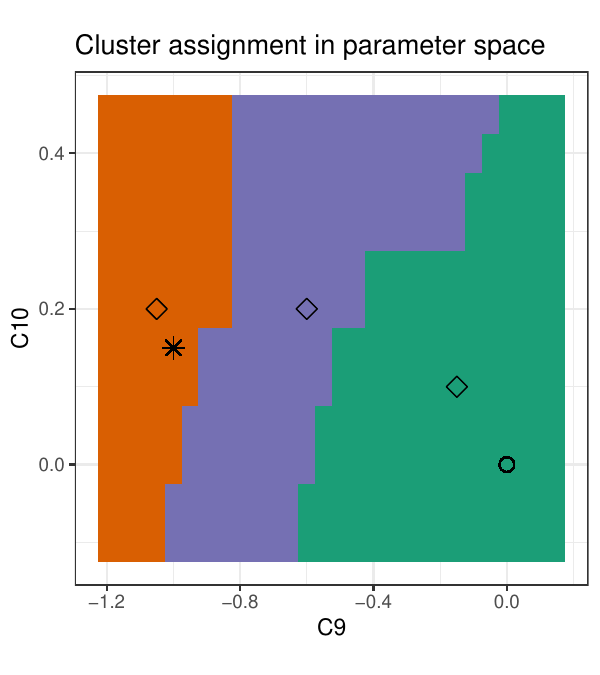}
\caption{Variation of the prediction for the two selected observables $Br(B_s\to\mu^+\mu^-)$ (left panel) and $P_4^\prime(B_0\to K^*\mu^+\mu^-)[0.1-0.98]$ (middle panel). The variation is shown in terms of the pulls calculated without correlation effects. The right panel shows the partition of parameter space when only the first 85 observables are kept.}
\label{f:euc85}
\end{figure}

In Fig.~\ref{f:euc89} we show the results of three clusters with the same settings as in Fig.~\ref{f:euc} allowing for direct comparison: average linkage with Euclidean distance on the pulls including all known correlations. The left panel displays the partition of parameter space and the right panel the associated parallel coordinates including all 89 observables with the original 14 of Table~\ref{t:obs} labeled $[1-14]$. The cumulative effect changes the partitioning seen in Fig.~\ref{f:euc}. The parallel coordinate plot shows that although $R_K$ is still dominant, there are a few other important observables which we have placed near the end of the list. The largest effect would come from number 86, $Br(B_s\to\mu^+\mu^-)$, which is also known to play an important role in the global fit. For example, if we keep only the first 85 observables, the partition of parameter space shown in the right panel of Fig.~\ref{f:euc85} is very similar to that in Fig.~\ref{f:euc} with only our original subset of 14 observables validating our choice. On the other hand, the large set of observables provides increased resolution in the partitioning of parameter space. With the same criteria used so far, the resolution increases from four clusters with the dominant 14 observables to nine clusters with all 89. For completeness we show this clustering in the parameter space in Fig.~\ref{f:new_9clu} in the appendix.

\begin{figure}[h]
\includegraphics[scale=0.4]{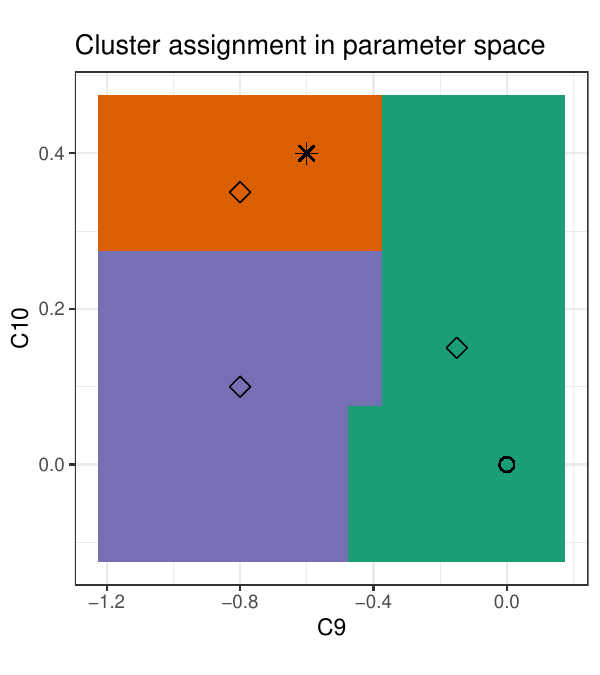}\includegraphics[scale=0.6]{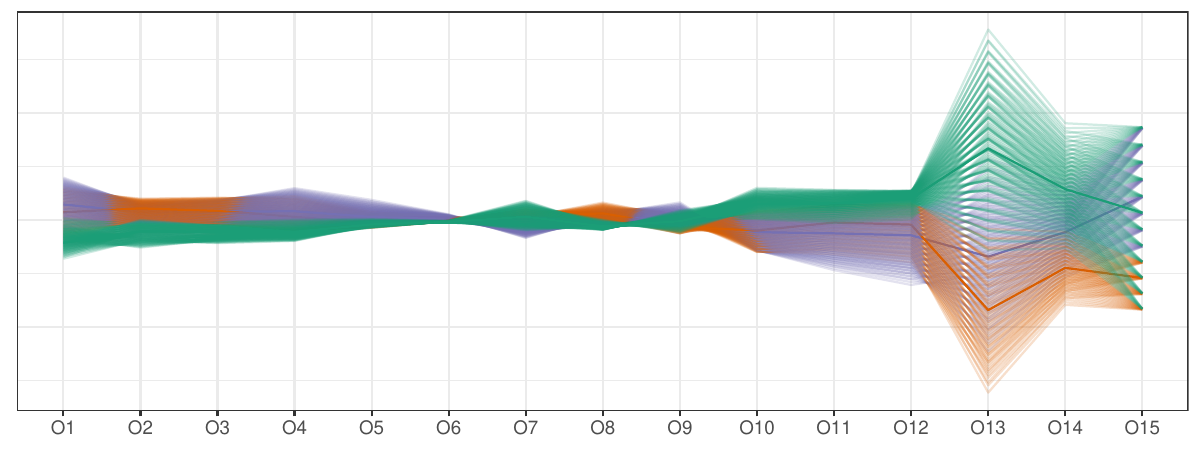}
\includegraphics[scale=0.4]{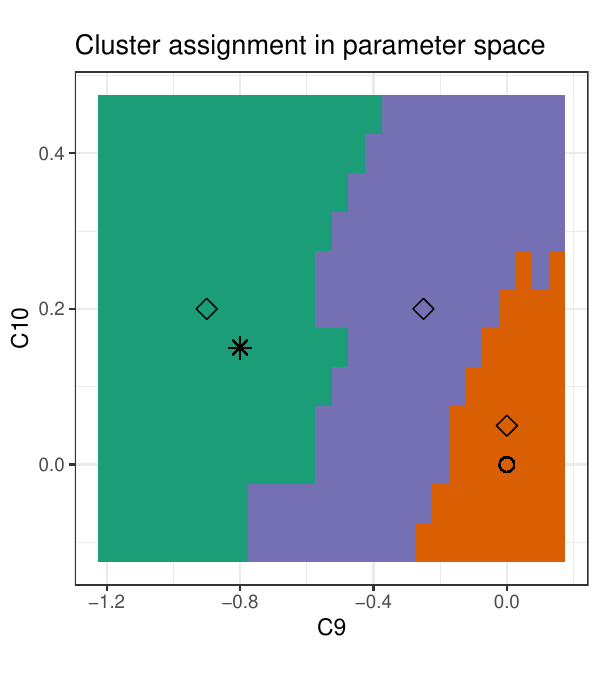}\includegraphics[scale=0.6]{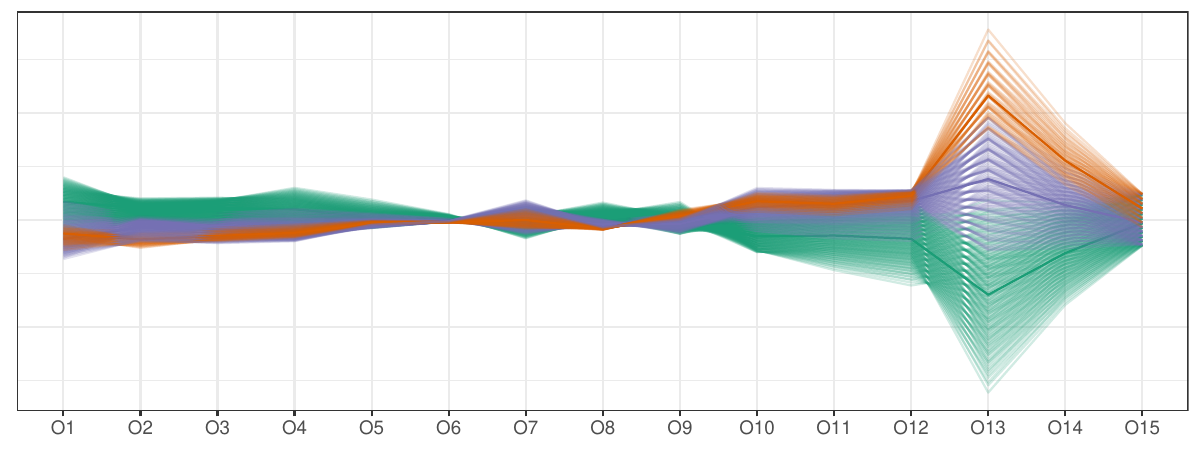}
\caption{The results of three clusters using Euclidean distance on the pulls with covariance and average linkage in parameter space (left panel) with the corresponding parallel coordinate plots (right panel) for the 14 observables in Table~\ref{t:obs} plus $Br(B_s\to\mu^+\mu^-)$. The top panel corresponds to the combination of measurements from {\tt flavio} and the bottom panel to the combination from   \cite{Alguero:2019ptt}.}
\label{f:euc86}
\end{figure}

In Fig.~\ref{f:euc86} we illustrate the effect of adding $Br(B_s\to\mu^+\mu^-)$ ($O_{15}$ in this plot) to the set of observables in  Table~\ref{t:obs}. The left panel shows the partitioning of parameter space into three clusters using Euclidean distance on the pulls with covariance and average linkage and the right panel shows the corresponding parallel coordinates. The latter indicates that the importance of $Br(B_s\to\mu^+\mu^-)$ is almost as large as that of $R_K$ and that it produces a very different clustering. The left panel of Fig.~\ref{f:euc85} shows the variation of $Br(B_s\to\mu^+\mu^-)$ across the parameter space indicating that it mostly partitions the space along $C_{10}$. It has been pointed out before that there are different treatments for the combination of measurements for this observable in the literature and that they lead to very different conclusions. With {\tt flavio} we are using an average of experimental results $Br(B_s\to\mu^+\mu^-) = (2.81^{+ 0.24}_{-0.22})\times 10^{-9}$ whereas  \cite{Alguero:2019ptt}, for example, finds  $Br(B_s\to\mu^+\mu^-) = (2.94 \pm 0.43)\times 10^{-9}$. The difference in these errors leads to different conclusions at present as can be seen by comparing the top and bottom row of Fig.~\ref{f:euc86}.
Note here that the apparent difference in crossing between predictions for the last two observables is a result of the different scales, but the relation between the two is fixed by the theoretical expression. When looking at the parallel coordinate plot with rescaled axes (rescaling each coordinate to have variance of one before plotting, this version always presented in the app alongside the unscaled plot), we can see that the behavior is the same in both cases.

\begin{figure}[h]
\includegraphics[scale=0.4]{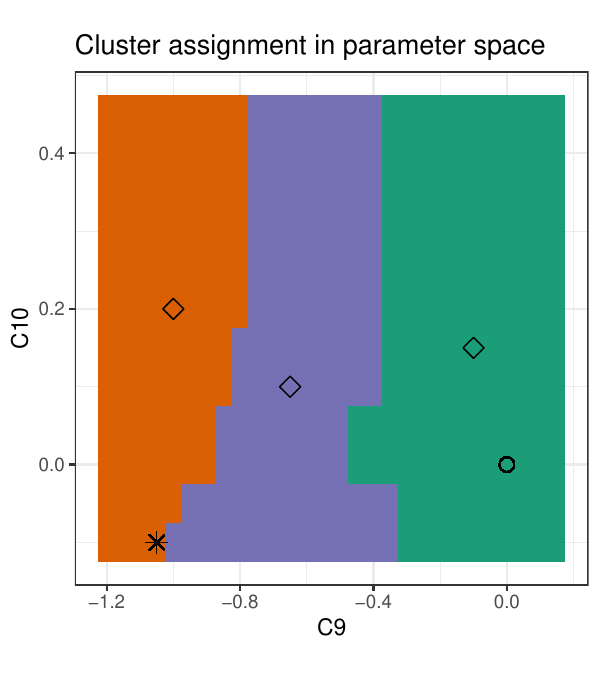}\includegraphics[scale=0.6]{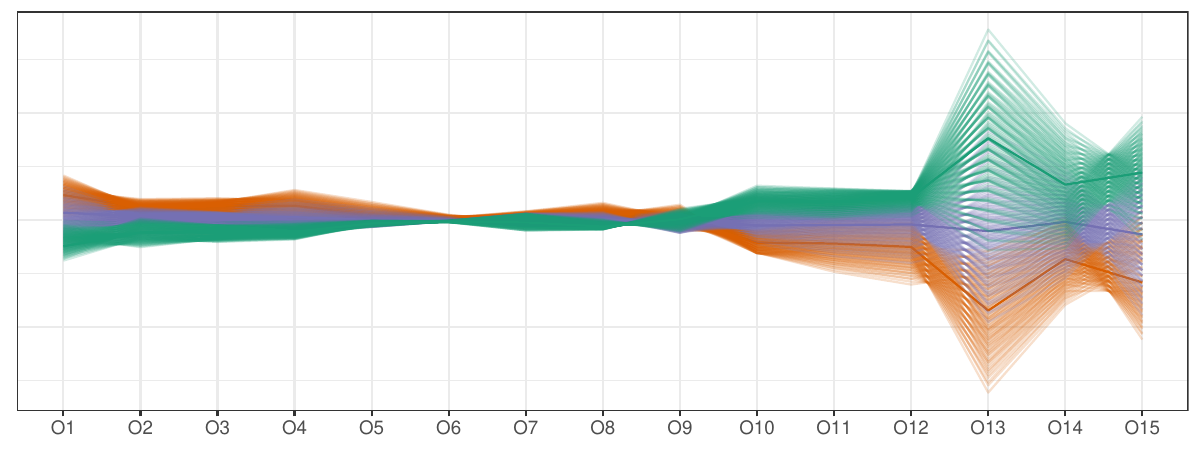}
\caption{The results of three clusters using Euclidean distance on the pulls with covariance and average linkage in parameter space (left panel) with the corresponding parallel coordinate plots (right panel) for the 14 observables in Table~\ref{t:obs} plus a {\it projection for a possible future measurement} of $P_4^\prime(B_0\to K^*\mu^+\mu^-)[0.1-0.98]$ as described in the text.}
\label{f:euc44}
\end{figure}

In Fig.~\ref{f:euc44} we illustrate the effect of adding $P_4^\prime(B_0\to K^*\mu^+\mu^-)[0.1-0.98]$  ($O_{15}$ in this plot) to the set of observables in  Table~\ref{t:obs}. With the current LHCb measurement, $P_4^\prime(B_0\to K^*\mu^+\mu^-)[0.1-0.98] = 0.135\pm 0.118$, \cite{Aaij:2020nrf}, the error in this observable is too large to have a significant impact on the fit. However, the central panel of Fig.~\ref{f:euc85}, showing the variation of this observable across the parameter space, indicates a very interesting pattern with a parameter correlation orthogonal to that seen in $R_K$. This indicates that if the error can be reduced by a factor of four by future measurements, this observable can provide very useful information as we illustrate {\it under this assumption of a reduced error} in the left panel (partition of parameter space) and right panel (parallel coordinates) of Fig.~\ref{f:euc44}, for three clusters using Euclidean distance on the pulls with covariance and average linkage. 

\subsection{Assessing the impact of future measurements}\label{s:futurefit}

Another application of clustering is to assess the impact of future measurements, both on clustering problems and on global fits. In this case clustering can also provide information to plan or prioritize future measurements. Different models can be compared before measurements are made to understand how different predictions map onto different regions of parameter space. This can be used, for example, to select benchmark points for detailed study.

When measurements do not exist, additional choices must be made when selecting coordinates. The reference point $R_i$ should be set based on a preferred model, for example the SM or a best fit point obtained in a global fit to existing measurements.\footnote{However, we recall that the reference point does not enter the distances when assuming fixed (i.e. model-independent) uncertainties and using pull coordinates.} The uncertainties used to normalize the coordinates can be estimated by combining theoretical uncertainties with anticipated statistical uncertainties in future measurements. Moreover, since the latter correspond to an expected luminosity, the clustering outcome can show how the partitioning of parameter space varies with luminosity.

We now continue the study of processes which can be described with Eq.~\ref{effH} by considering modes  that have been proposed in the literature but not yet measured. Many examples of this kind are listed in \cite{Capdevila:2016ivx}; and the Belle collaboration has already reported some preliminary results for a few of them \cite{Wehle:2016yoi}. 

We use here lepton flavor violating observables that directly compare angular distributions in the decay $B\to K^\star \mu^+\mu^-$ to the corresponding ones in the decay $B\to K^\star e^+e^-$,
\begin{eqnarray}
Q_2\equiv P_2^\mu-P_2^e, && Q_5\equiv P_5^{\prime \mu}-P_5^{\prime e}.
\end{eqnarray}
These observables directly test lepton universality like $R_{K^{(*)}}$ do, providing a direct test for physics beyond the SM. In addition, they directly  compare the $P_{2}$ and $P_5^{\prime}$ observables that have shown deviations from the SM for final state muons and electrons.
We use the set of six observables listed in Table~\ref{t:futobs}, which satisfy the following criteria:
\begin{itemize}
\item There exist projections for the sensitivity that can be achieved by Belle II \cite{Kou:2018nap}.
\item They were singled out as important by the pull and residual analysis of \cite{Capdevila:2018jhy}. In addition, it has been suggested that $Q_5$ can play an important role in separating different NP scenarios  \cite{Alguer__2019}. 
\end{itemize}
The theoretical predictions for these observables, along with their corresponding uncertainty, is computed for the same model points of the previous section with the aid of {\tt flavio}, and we limit ourselves to the two parameter case ($C_{9}$, $C_{10}$) and with all theoretical uncertainties evaluated at the SM point.

\begin{figure}[h]
\centering{\includegraphics[scale=0.4]{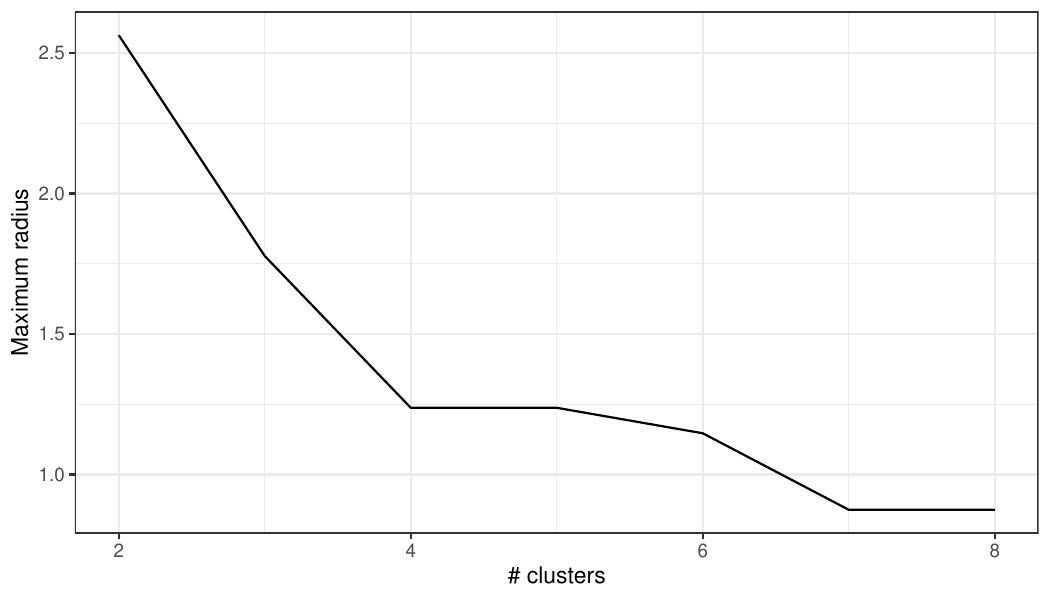}\includegraphics[scale=0.4]{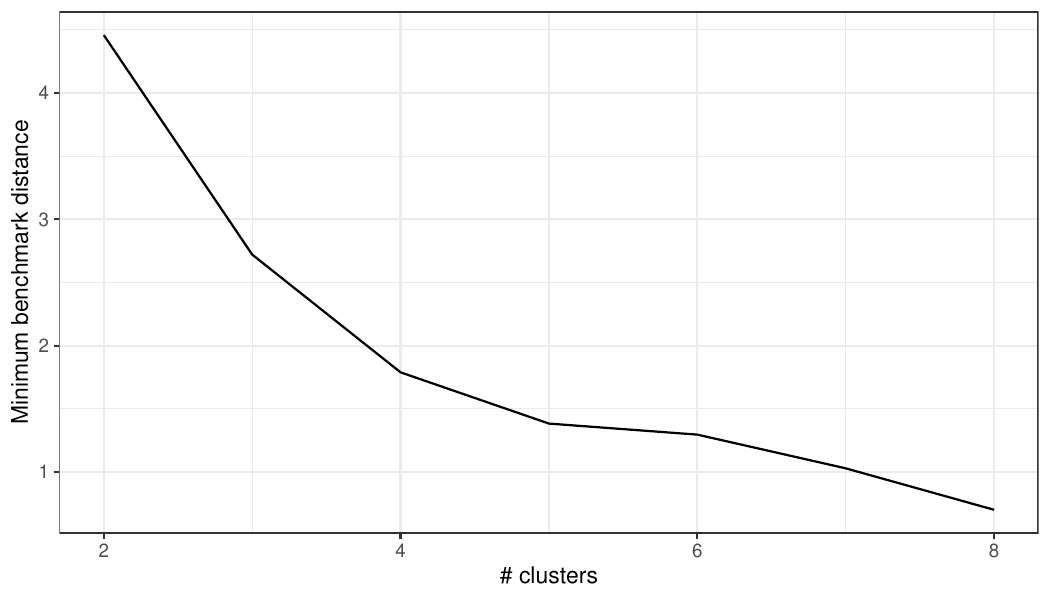}}
\caption{Maximum cluster radius (left) and minimum benchmark separation (right) for average linkage and Euclidean distance on pulls with correlations for 50~ab$^{-1}$ as a function of number of clusters. }
\label{f:stats}
\end{figure}

We begin by clustering the pulls with the SM as a reference point, an estimated experimental covariance matrix for 50~ab$^{-1}$ from Eq.~\ref{guesscov}, and using average linkage with Euclidean distance.\footnote{The goal of the Belle II experiment is to accumulate an integrated luminosity of 50~ab$^{-1}$ by the middle of the next decade, so we will use this number as a benchmark.} For these choices, we show in Fig.~\ref{f:stats} the maximum cluster radius and minimum benchmark separation as a function of the number of clusters. The same arguments of Section~\ref{s:choosingpar} tell us that the optimal number of clusters for this observable set is four.

\begin{figure}[h]
\centering{\includegraphics[scale=0.3]{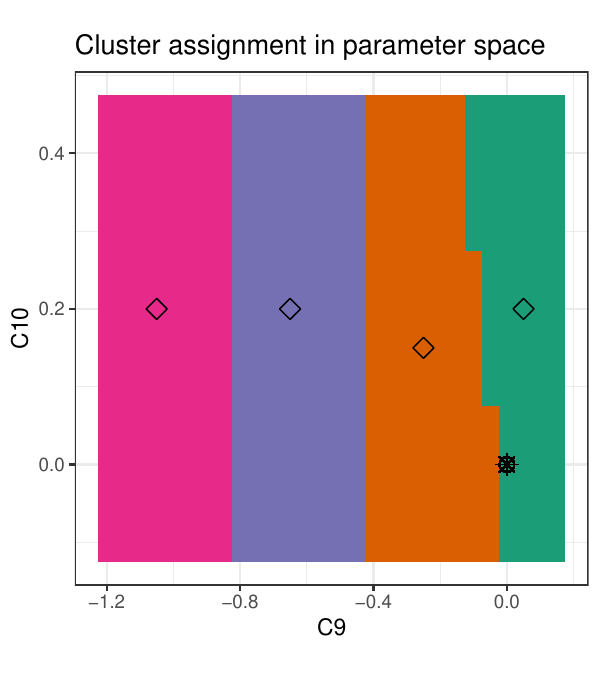}\includegraphics[scale=0.45]{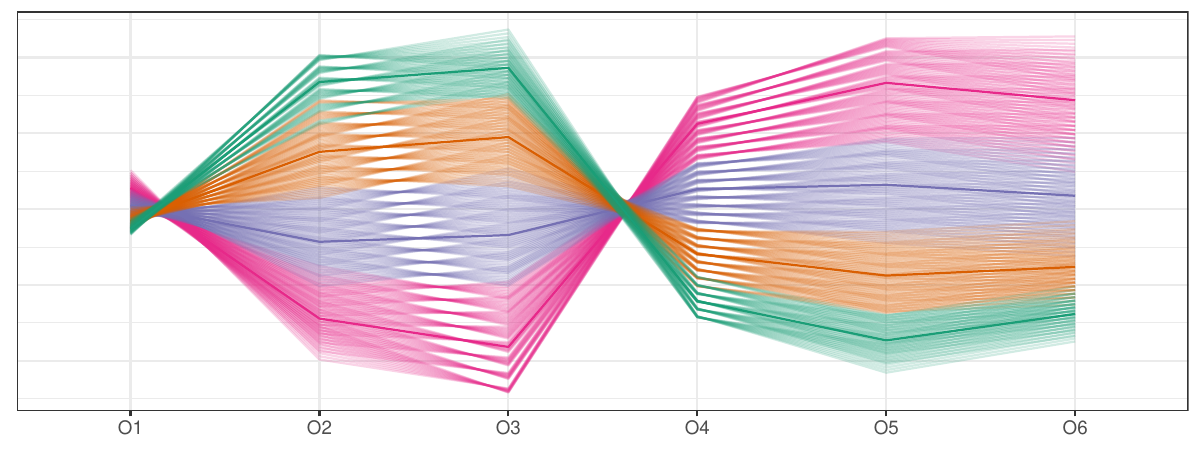}\includegraphics[scale=0.3]{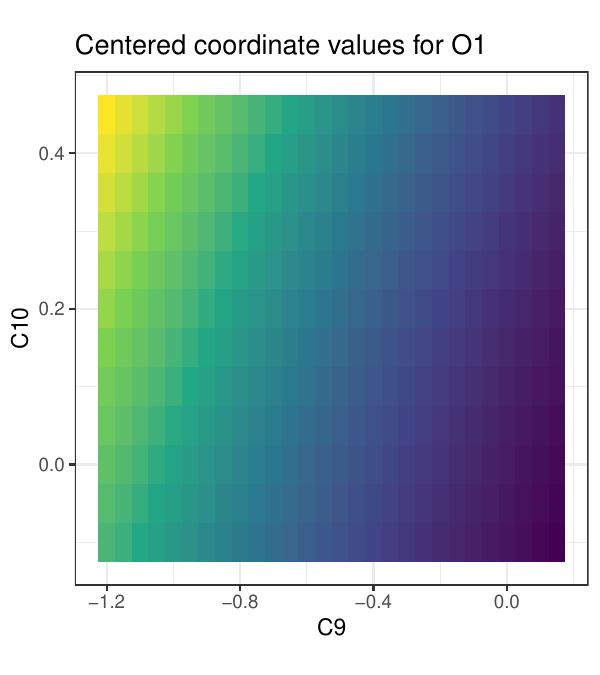}}
\caption{Four clusters using average linkage and Euclidean distance on pulls from the SM with estimated experimental covariance matrix as in Eq.~\ref{guesscov} for the observables in Table~\ref{t:futobs} (left) with corresponding parallel coordinates (center). Variation in predictions across the parameter space for $O_1$ (right).}
\label{f:fut2}
\end{figure}

The corresponding clusters in parameter space are shown in Fig.~\ref{f:fut2} along with the corresponding parallel coordinate plot. The position of the cluster benchmarks (indicated by the diamonds) and the shape of the cluster boundaries  already suggest that these observables will be mostly sensitive to $C_9$. The parallel coordinate plot shows that all the observables considered in Table~\ref{t:futobs}, with the possible exception of the lowest $q^2$ bin for $Q_2$ ($O_1$), provide a clean separation of the clusters. The variations in predictions across parameter space indicate that $O_1$ (also shown in Fig.~\ref{f:fut2}) has the best chance of providing resolving power for $C_{10}$, but to increase its importance, the relative precision of this measurement with respect to the others must increase. $O_6$ also shows a slight correlation between $C_9$ and $C_{10}$ but has a different pattern than $O_1$.

Apart from comparing clusters, we can also use the parallel coordinate plot to understand correlations between the observables more generally. In this example we see clear positive correlation between most of the observables, and negative correlation between observables 3 and 4.

Clustering with complete linkage and maximum distance yields similar results, reinforcing the conclusion that there are no standout coordinates among this set. Variations are also minor if the correlations are not included, suggesting that they are not important at the level implied by our estimate in Eq.~\ref{guesscov}. 

The projected errors for a luminosity of 5~ab$^{-1}$ are larger by about a factor of 3. Consequently we expect the resolution to be about three times worse, indicating that the same region of parameter space shown in Fig.~\ref{f:fut2} cannot be resolved with this data. 

\subsection{Future measurements in the context of an existing set}

Future measurements of new observables will not be considered in isolation, instead they will be added to the existing set. We now illustrate the impact of measuring the set of observables in Table~\ref{t:futobs}  on the clustering problem of Section~\ref{s:existingfit}. Of course, this is only a partial study, as a complete analysis for physics projections would also require us to estimate future improvements in the measurements of the original set of 14 observables.

In light of our previous discussion we expect several things to happen when the two sets are combined: the resolution measured by the stopping criterion should increase; the original clusters of Section~\ref{s:existingfit} will change to reflect the dominant patterns in the complete set of observables. These changes may be significant if the new observables provide the dominant constraints. It is instructive to compare the combination using two values for the Belle II luminosity, 5~ab$^{-1}$ and 50~ab$^{-1}$.

\begin{figure}[h]
\centering{\includegraphics[scale=0.4]{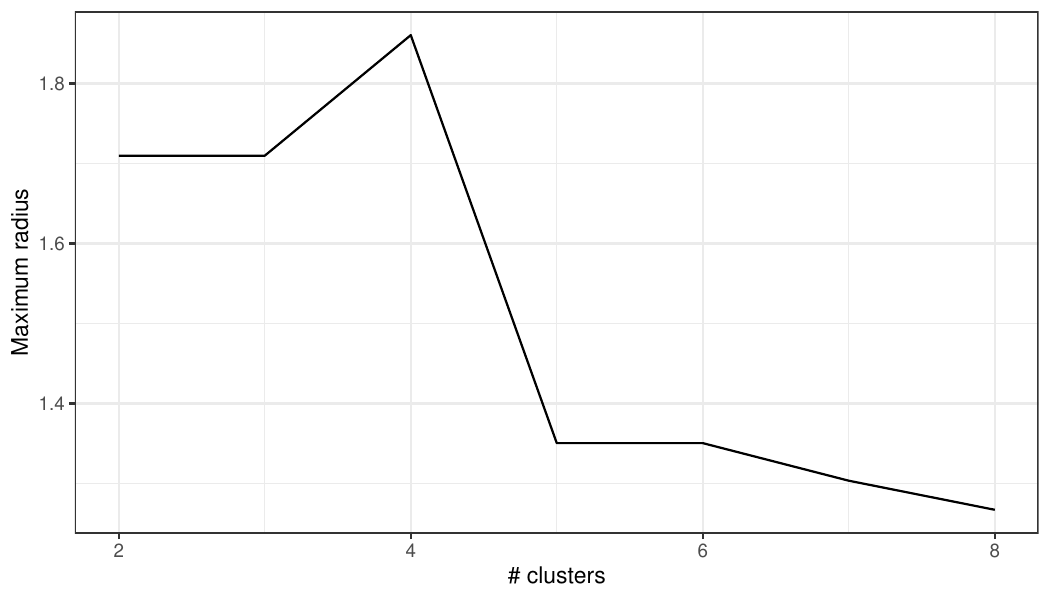}\includegraphics[scale=0.4]{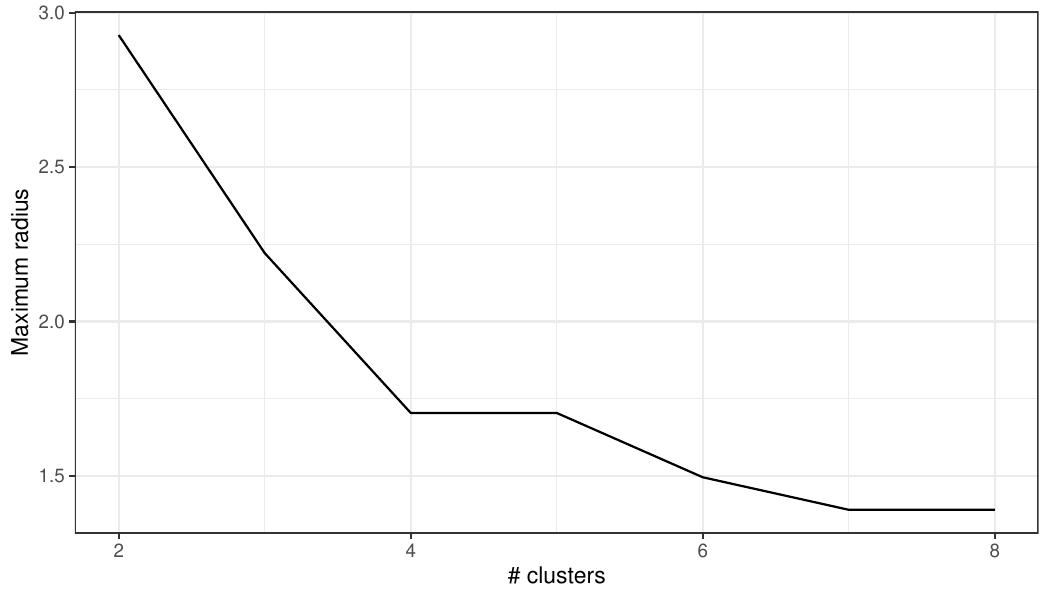}}
\caption{Maximum cluster radius for average linkage and Euclidean distance on pulls with correlations as a function of number of clusters for the combined observables of Tables~\ref{t:obs}~and~\ref{t:futobs}. The left  (right) panel shows the results for 5~ab$^{-1}$ (50~ab$^{-1}$) for the future measurements.}
\label{f:combined}
\end{figure}

In Fig.~\ref{f:combined} we show the maximum cluster radius for average linkage and Euclidean distance on pulls with correlations as a function of the number of clusters. The covariance matrix is a diagonal combination of the two matrices previously used (the $14\times 14$ matrix of Section~\ref{s:existingfit} and the $6\times 6$ matrix of Section~\ref{s:futurefit}), it ignores correlations between the original and new sets of observables. Requiring the maximum radius to be below $\sim 1.5$ (since we still have only two parameters), the suggested number of clusters grows from five in Section~\ref{s:existingfit} to six for the 50~ab$^{-1}$ projections. The minimum benchmark separation in this case also increases to about 1.7.

\begin{figure}[h]
\centering{\includegraphics[scale=0.5]{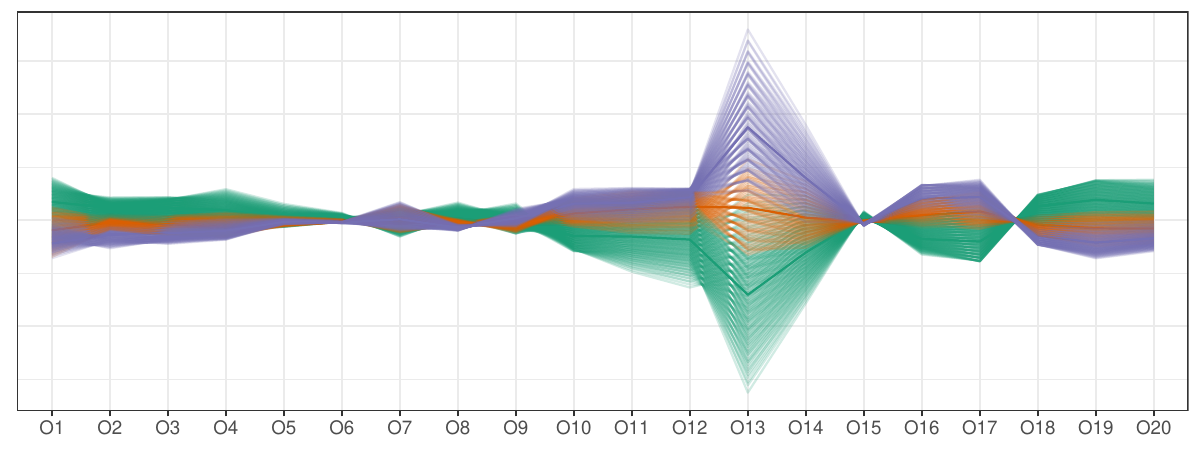}}
\centering{\includegraphics[scale=0.5]{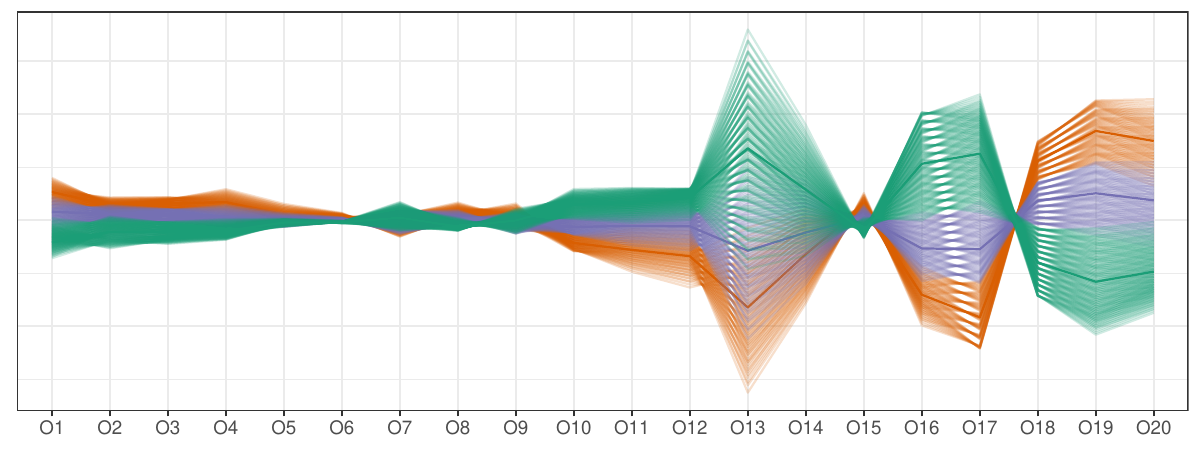}}
\caption{Parallel coordinates for three clusters. The top (bottom) panel shows the results with 5~ab$^{-1}$ (50~ab$^{-1}$) for the future measurements.}
\label{f:combinedPC}
\end{figure}

The parallel coordinates  reveal several of the new observables may reach similar importance to $R_K$ at 50~ab$^{-1}$ but not at 5~ab$^{-1}$.\footnote{Here we have adjusted the observable ID, the new observables are now numbered 15 to 20 in the same order as indicated in Table~\ref{t:futobs}.} Their combined effect is likely to dominate the clustering in this situation at the higher luminosity (but recall that this plot does {\it not} include future projections for the first fourteen coordinates which include $R_K$). In Fig.~\ref{f:combinedPC} we illustrate this using only three clusters for simplicity.

\begin{figure}[h]
\centering{\includegraphics[scale=0.4]{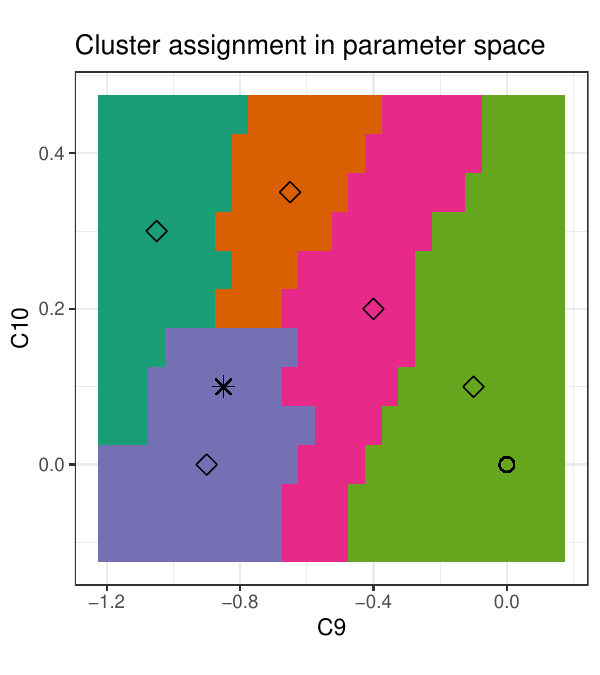}\includegraphics[scale=0.4]{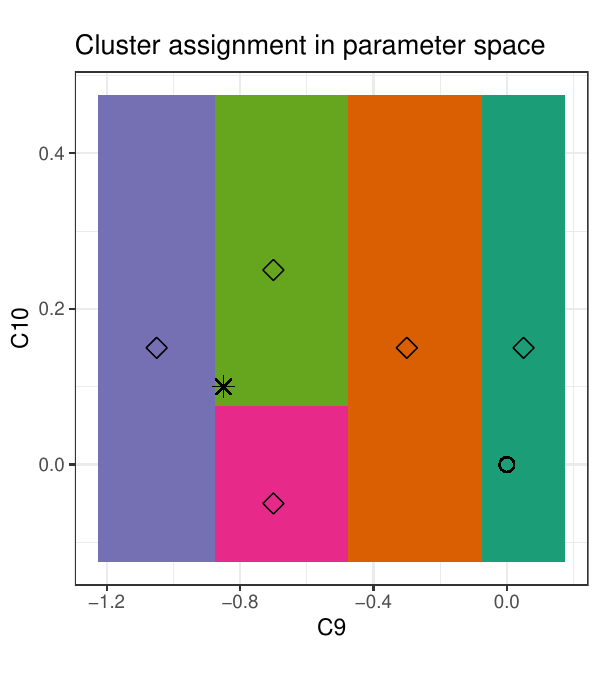}\includegraphics[scale=0.4]{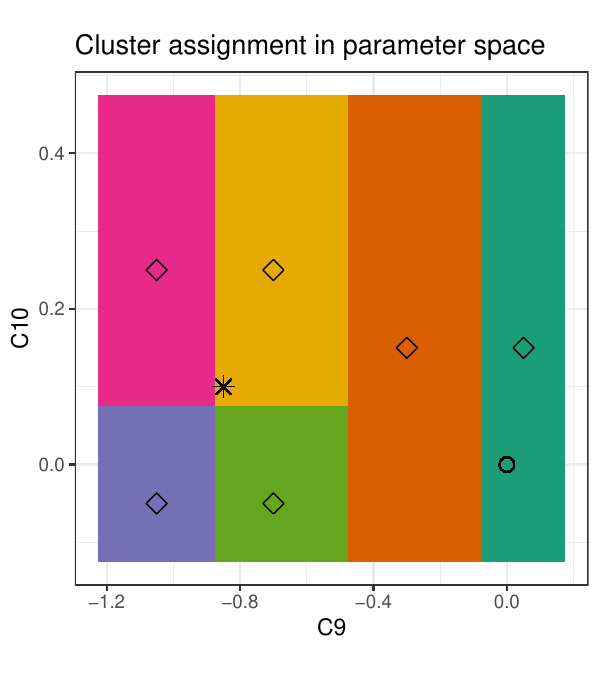}}
\caption{The results of five and seven clusters using Euclidean distance on the pulls with covariance and average linkage in parameter space with the combined observables of Tables~\ref{t:obs}~and~\ref{t:futobs}. The left (center) panel shows 5 clusters assuming 5~ab$^{-1}$ (50~ab$^{-1}$) for the future observables, and the right panel six clusters assuming 50~ab$^{-1}$.}
\label{f:combinedWC}
\end{figure}

The effect of the new observables at the two values of luminosity is shown  in Fig.~\ref{f:combinedWC}.  The left panel shows five clusters assuming 5~ab$^{-1}$ for Belle II. The results look similar to Fig.~\ref{f:euc}, particularly with respect to the shape of the boundaries, indicating that $R_K$ is still dominant, as can also be seen in the top panel of Fig.~\ref{f:combinedPC}. In contrast, with a projected luminosity of 50~ab$^{-1}$ for Belle II, $R_K$ is no longer dominant and we illustrate the partition into five (center) and six (right) clusters in Fig.~\ref{f:combinedWC}. The best fit marker in that figure assumes that the future measurements will fall on the current best fit. 

\subsection{Summary of results}
\label{s:psum}

We have used the new interactive tool {\tt pandemonium} to cluster the predictions for fourteen observables obtained from models with two (and four) parameters.  Salient points of this study include:

\begin{itemize}

\item Clustering the observables partitions the parameter space into distinct regions with a mapping between each region and its corresponding observable predictions. {\tt Pandemonium} offers a range of tools to study that mapping interactively.

\item The discriminating power of the clustering outcome is determined by the number of clusters. We propose that this number is given by the minimum number of clusters for which all points within a cluster $C_j$ lie at most within $1\sigma$ of its corresponding benchmark $c_j$  while benchmarks are separated by at least $1\sigma$ . Clustering the pulls with Euclidean distance and average linkage for this problem,  partitions the parameter region $(-1.2\leq C_9\leq 0.14) \wedge (-0.1 \leq C_{10}\leq 0.45)$ into five clusters. We have shown that this can be increased to six clusters including six new observables with the expected sensitivity at Belle II with 50~ab$^{-1}$.

The tool allows the user to choose a different stopping criterion. We have proposed an alternative that uses maximum distance with complete linkage which also results in five clusters for this example. 

\item Observables that dominate the clustering are those that have the largest variation in prediction relative to their uncertainty across the parameter range. In this set, $R_K$ (ID 13) (and to a lesser extent $R_{K^\star}$, ID 14) plays an outsized role in partitioning parameter space. The parallel coordinates (including correlations) show that this observable has the largest variance in pulls, i.e. the range in predictions is large compared to the total uncertainty, and this is what makes it a ``dominant'' observable.  The dominance of $R_K$  is evident as it determines the shape of the boundaries between main clusters.

\item We have illustrated several ways to showcase the effect of a dominant observable, such as combining complete linkage with maximum distance. Alternatively, it is also possible to suppress dominant observables, for example using the Manhattan distance. These methods provide insights into the effect of removing a dominant observable and in this way study internal tensions in a global fit. For example, the global fits show that  removing $R_K$ significantly shifts the BF value of $C_{10}$ upwards. In the clustering, this shift is accompanied by a change in the shape of the inter cluster boundaries indicating a very different correlation between $C_9$ and $C_{10}$ with and without $R_K$.

\item Finer partitioning occurs along an approximately constant value of $C_{10}$ (going from 4 to 5 clusters) or $C_9$ (beyond 5 clusters). The parallel coordinates suggest that this is a collective effect from multiple observables. In this way the cluster boundaries reveal correlations between parameters that are not obviously connected to a single observable. The clustering is in effect visualizing the combination of all the equations that connect the parameters to the observables.

\item  The tool offers the possibility to compare clustering with or without correlations between the observables. In this example, this shows that if correlations are ignored, the relative importance of the angular observables with respect to $R_{K^{(\star)}}$ increases and the main clustering now depends mostly on $C_9$. This observation is consistent with the pull and residual analysis of \cite{Capdevila:2018jhy}.

\item Clustering pulls with fixed errors does not capture the reference point. However, the position of the reference point in observable space is the origin and this can be visualized in parallel coordinate plots without centering. 

\item Clustering can help understand tensions between observables in a global fit. For example, the best fit point shown as a star lies near the intersection of two clusters in the right panel of Fig.~\ref{f:compchi}. The two bins of $P_5^\prime$ that contribute most to the discrepancy between SM and experiment prefer the light green cluster,  whereas $R_{K}$ prefers  the purple cluster. In fact, Fig.~\ref{f:compchi-pc2} indicates that the clusters where $P_5^\prime[4-6]$ and P$_5^\prime[6-8]$ move closer to their experimental values, take $R_{K}$ further away from its experimental value, clearly exposing the tension between these two observables.

\item Clustering with more than two parameters, in combination with slice displays, helps us understand multivariate relations, as demonstrated for the parameters $C_9$, $C_{10}$ and $C_9^\prime$. Inspecting the slices provides additional information compared to the profiled results usually presented when fitting more than two observables by accurately capturing the shape of the boundaries as a function of the parameters in the  space orthogonal to the viewing plane. A more obvious example of this can be seen in the example presented in the next section. In particular the first two panels of Fig.~\ref{f:coordCK} illustrate a change in the correlation between two parameters ($C_{VL}$ and $C_{SL}$) as a third one ($C_T$) changes. 

\item Without prior knowledge to preselect the most important observables it is also possible to cluster based on a much larger set of observables. This can be used as a preliminary step for selecting smaller subsets for further study. Although the large set can also be studied with the tools in {\tt pandemonium}, plots showing variations of individual observables across parameter space have to be done separately. The clustering results with large  numbers of observables give an overall picture and improve the resolution; but subsets must be considered to study details such as internal tensions. 

\item A study with a large set of observables can also be used to preselect a smaller subset for further study. For example, looking at 89 observables we can see that $Br(B_s\to\mu^+\mu^-)$ and $P_4^\prime(B_0\to K^*\mu^+\mu^-)[0.1-0.98]$ could play an important role under certain conditions and understand why. The importance of the former was already known and depends on how the different existing measurements are combined. The importance of the latter is inferred from its variation across parameter space and we estimate that a reduction in the current experimental error by about a factor of 4 is needed for this observable to have full impact on the global fit.

\end{itemize}

While the main focus here is on new insights that can be obtained through the investigation of the clustering outcome, we can also perform cross-checks to see if the results are aligned with what is expected based on previous studies in the literature. Consider as an example the case when $C_9$ and $C_{10}$ are the only non-zero WC for new physics. In this scenario it is known that $Br(B_s\to\mu^+\mu^-)$ depends only on $C_{10}$, whereas $R_{K^{(*)}}$ depends on $C_9-C_{10} $. The addition of $Br(B_s\to\mu^+\mu^-)$ thus provides sensitivity to $C_{10}$ in the clustering study, as can be seen directly in Figure~\ref{f:euc85}. When its error is sufficiently small for its pull to compete with that of $R_K$ in importance, it produces main partitions along this direction as illustrated in Figure~\ref{f:euc86}.

\section{Kinematic distributions in $B\to D^{(\star)} \tau \nu$ decay and comparison with {\tt ClusterKing}}\label{s:compareCK}

In this section we compare our work to the Python package {\tt ClusterKinG}, which was recently introduced to cluster kinematic distributions \cite{Aebischer:2019zoe}. The interactive environment that we have described here can also be used to cluster kinematic distributions as we show below. Our work, as implemented in the {\tt pandemonium} package, offers the following advantages
\begin{itemize}

\item Flexibility to choose linkage methods, distance functions and numbers of clusters according to a variety of validation indices.

\item Visualization of the resulting clusters in both parameter and observable space, making use of different tools from the visualization literature that also allow for direct inspection of high dimensional distributions.

\item Easily include more than one distribution, which allows model differentiation in multiple directions in parameter space.

\end{itemize}

We use two of the examples in \cite{Aebischer:2019zoe} for this comparison. They both relate to new physics  that may affect the processes $B\to D^{(\star)} \tau^- \bar \nu_\tau$ (listed in Table~\ref{t:compCK}). The models contain three parameters, the Wilson coefficients $C_{VL}$, $C_{SL}$ and $C_T$ in the low energy effective Hamiltonian responsible for the $b\to c\tau^-\nu_\tau$ quark level transition.
We generate predictions for the same grid of Wilson coefficients as  \cite{Aebischer:2019zoe}: ten equally spaced points for each parameter within the bounds $ -0.5 \leq C_{VL},C_{SL} \leq 0.5$, $-0.1 \leq C_T \leq 0.1$ using {\tt flavio}.  The ranges chosen for the scan are motivated by the fits in \cite{Murgui:2019czp}. The dominant uncertainties in future experiments are assumed to be statistical and their estimate is outlined in Appendix~\ref{aCK}. The {\tt ClusterKinG} package is designed to compare the {\it shapes} of two distributions (histograms) and its clustering is based on the distance function\footnote{Apparently there is a typo in Eq. 2.1 of \cite{Aebischer:2019zoe}.}
\begin{eqnarray}
\chi^2(H_1,H_2)=\sum_{i=1}^N\frac{(N_1n_{2i}-N_2n_{1i})^2}{N_2^2\sigma_{1i}^2+N_1^2\sigma_{2i}^2}.
\label{chi2dis}
\end{eqnarray}
Each histogram corresponds to a model point for one of the kinematic distributions mentioned above and $N_{1,2}$ are the respective normalization. 

The distance function in Eq.~\ref{chi2dis} differs from the ones we have considered so far in two important ways:
\begin{itemize}
\item it compares the shapes of two distributions but ignores their overall normalization;
\item it cannot be implemented solely from coordinates because it depends on two independent quantities at each point, $n_i$ and $\sigma_i$.
\end{itemize}
The first point reflects a choice: an experiment may be better suited to  measure precisely the shape of a  distribution rather than its normalization. However, the difference between predictions of different models may be more prominent in integrated rates and this can be captured by the options in our work as we illustrate below.
The second point can be circumvented with a user defined distance matrix which is calculated outside the app and imported as a csv file as we do in this section. 

We begin with a direct comparison to \cite{Aebischer:2019zoe} using the nine bins in the $\cos\theta_\tau$ distribution and matching their clustering parameters: complete linkage for six clusters and an externally calculated distance matrix corresponding to Eq.~\ref{chi2dis}.  To estimate the errors we assume a yield of 1000 events to assign statistical uncertainties to each bin and add a 10\% systematic error in quadrature. With these choices we find benchmarks that agree within errors with those reported in \cite{Aebischer:2019zoe}.\footnote{Specifically, although the WC values of these benchmarks are not identical to those in \cite{Aebischer:2019zoe}, the predictions at the two sets of benchmarks are practically the same. There are several factors that can account for small numerical differences, such as the version of {\tt flavio} used to compute the predictions, and the specific Poisson and statistical errors assigned to each bin. Note that the errors discussed in the text in \cite{Aebischer:2019zoe} do not match the code in their appendix.\label{footdata}} 

\begin{figure}[h]
\centering{\includegraphics[scale=0.4]{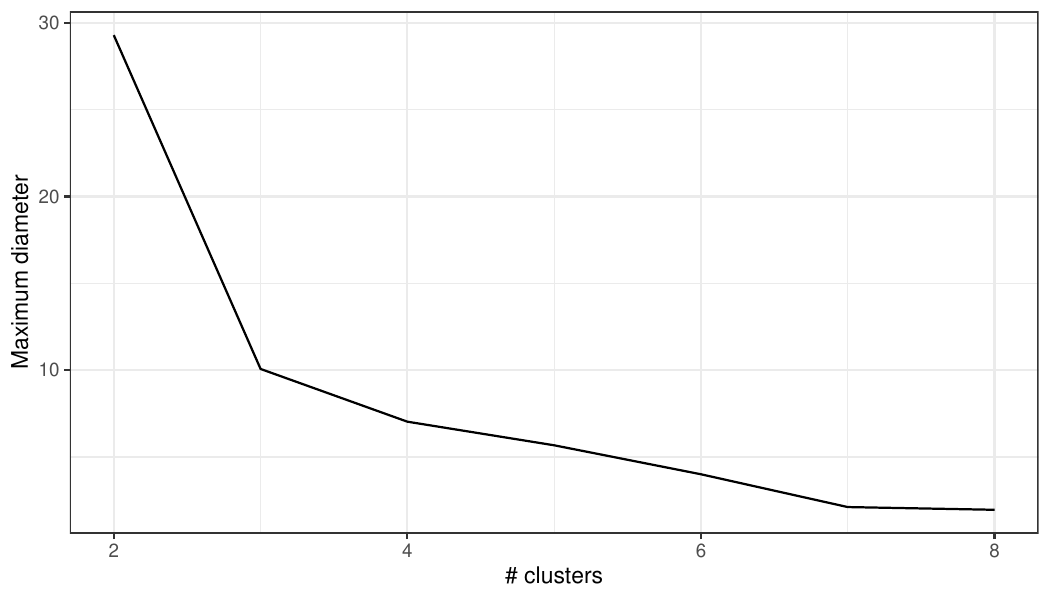}\includegraphics[scale=0.4]{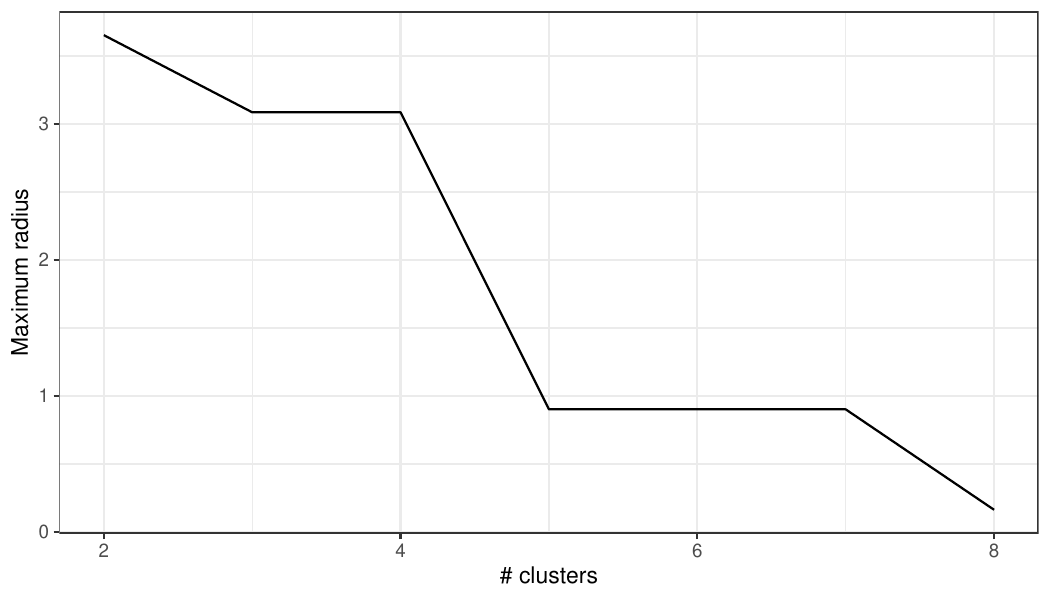}}
\caption{{\tt ClusterKinG} stopping criterion to determine the number of clusters (left panel) and maximum cluster radius as a function of the number of clusters for complete linkage and distance function Eq.~\ref{chi2dis} for the $\cos\theta_\tau$  distribution in $B\to D^{\star} \tau^- \bar \nu_\tau$.}
\label{f:statsvsCK}
\end{figure}

{\tt ClusterKinG}  uses an automatic stopping criteria to determine the number of  clusters to be six whereas  this is a choice in our case. That stopping criterion can be reproduced within our framework in terms of the maximum cluster diameter.  Requiring that all points inside the cluster be indistinguishable, the criterion stops clustering at the smallest number of clusters for which the ${\rm maximum~diameter} \leq N$ ($N$ is the number of bins) for the distance function $d(X_k,X_l)=\chi^2(H_1,H_2)$. In Fig.~\ref{f:statsvsCK}  we show the maximum cluster diameter and radius as a function of the number of clusters for complete linkage and distance function  Eq.~\ref{chi2dis}.  Applying the condition ${\rm maximum~diameter} < 9$ to {\it our data} (see footnote~\ref{footdata}) suggests four clusters instead of six, but a careful look at the figure shows that the criterion is approximately satisfied already at three clusters and does not change much between four and six clusters. On the other hand we see a large drop in cluster radius when going from 4 to 5 clusters, suggesting that 5 clusters may be the preferred solution. 
This illustrates an advantage in determining the number of clusters after visually inspecting multiple criteria. 

It is important to emphasize at this point that this stopping criterion has a very different interpretation from what we are doing, and is mentioned here solely for the purpose of comparing to our interface. {\tt ClusterKinG} is comparing distributions corresponding to different parameter choices by asking how likely it is that both histograms are drawn from the same underlying normal distribution, and thus how likely it is that any difference is attributable to random fluctuations. 

In previous sections we have proposed a different criterion based on maximum cluster radius (see the discussion in Section~\ref{s:choosingpar}) with a different interpretation. In this example there are three parameters in the models, and if we use the coordinates of \cite{Aebischer:2019zoe} there is one constraint (the normalization of the distribution). We thus use a $\chi^2$ distribution with two degrees of freedom to evaluate the confidence interval and require a maximum cluster radius of $2.3$ which suggests that five, six or seven clusters are equally valid. When a stopping criterion exhibits flat behavior as in this example, one might decide to trade complexity for interpretability and pick the simplest scenario, the smallest number of clusters in the range where the index is approximately constant.

\begin{figure}[h]
\centering{\includegraphics[scale=0.4]{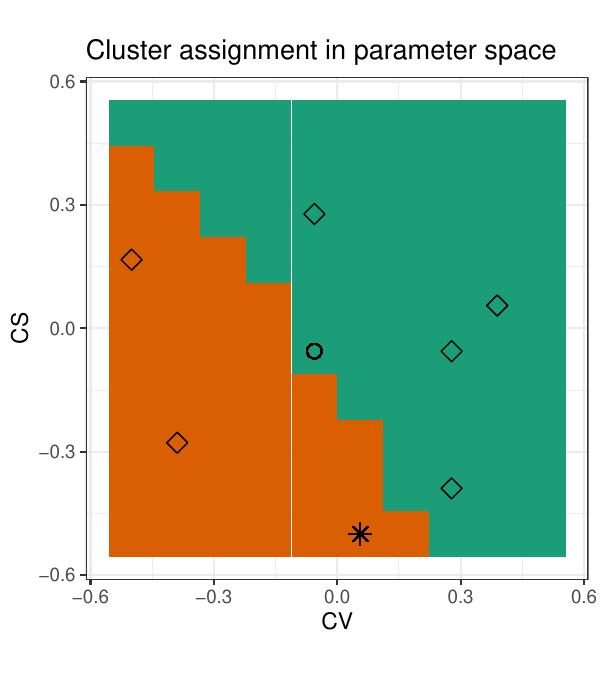}\includegraphics[scale=0.4]{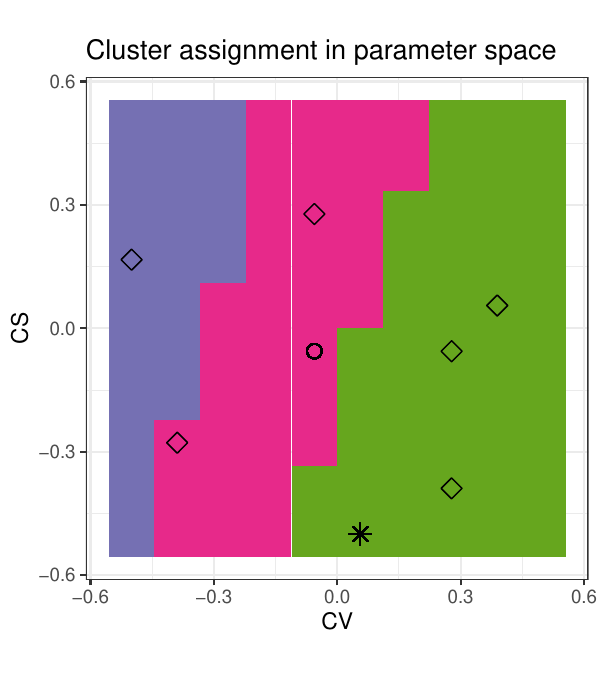}\includegraphics[scale=0.4]{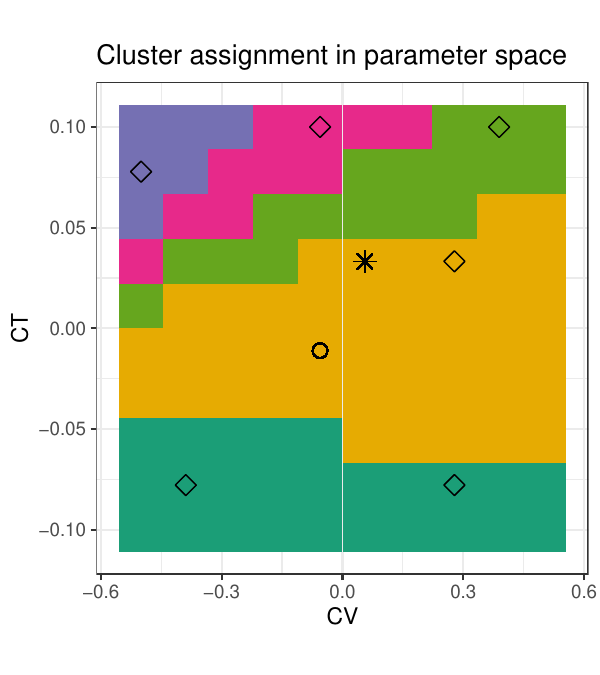}}
\centering{\includegraphics[scale=0.6]{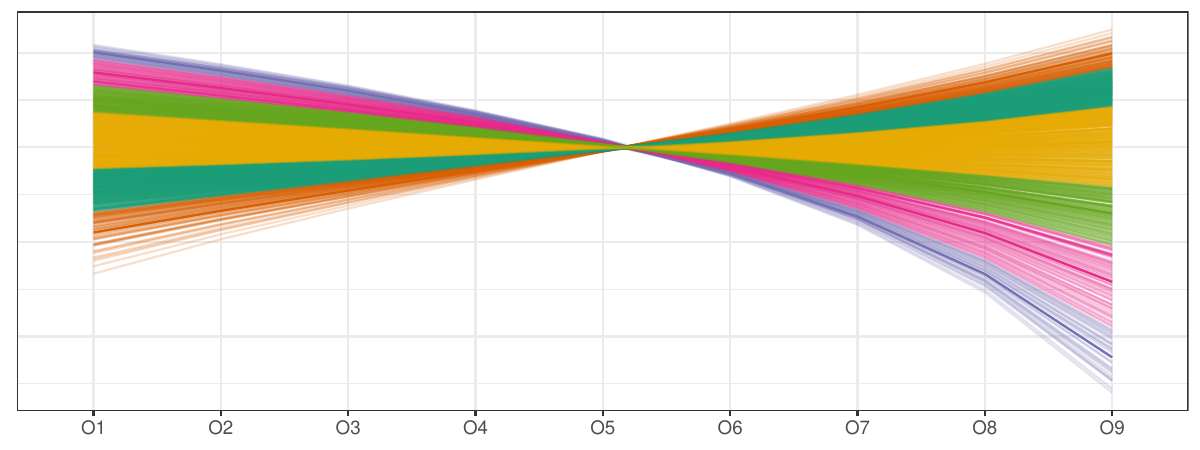}}
\caption{Six clusters with complete linkage and user defined distance function Eq.~\ref{chi2dis} for the $\cos\theta_\tau$ distribution: slices in parameter space are shown in the $C_{VL}-C_{SL}$ plane with $C_T=-0.1$ (left) and  $C_T=0.1$ (center);  and in the $C_{VL}-C_{T}$ plane with $C_{SL}=0.5$ (right). The parallel plot for user defined coordinates is shown in the bottom panel.}
\label{f:coordCK}
\end{figure}

In Fig.~\ref{f:coordCK} we show three slices in parameter space along with the parallel coordinates for six clusters using complete linkage and the user defined distance in Eq.~\ref{chi2dis}. The slices show good agreement with Figure~1 of \cite{Aebischer:2019zoe}.  The parallel coordinate plot shown in the same figure for {\it user defined} coordinates (see appendix~\ref{aCK}) is roughly equivalent to the benchmark distribution plots produced by {\tt ClusterKinG} (e.g. Figure~2 of \cite{Aebischer:2019zoe}). 

Going beyond a direct comparison with {\tt ClusterKinG} we have access to all the options discussed previously. For example we can combine more than one kinematic distribution by labeling the bins with sequential coordinates as listed in the Appendix. To avoid unnecessary complications associated with correlated errors, we illustrate this by combining the $\cos\theta_\tau$ distribution of $B\to D^{\star} \tau^- \bar \nu_\tau$ with the $q^2$ distribution of the branching ratio $B\to D \tau \nu$. In this case we are interested in the full information about the model carried by the observables, including the normalization, and we revert to using the pull coordinates. 

\begin{figure}[h]
\centering{\includegraphics[scale=0.6]{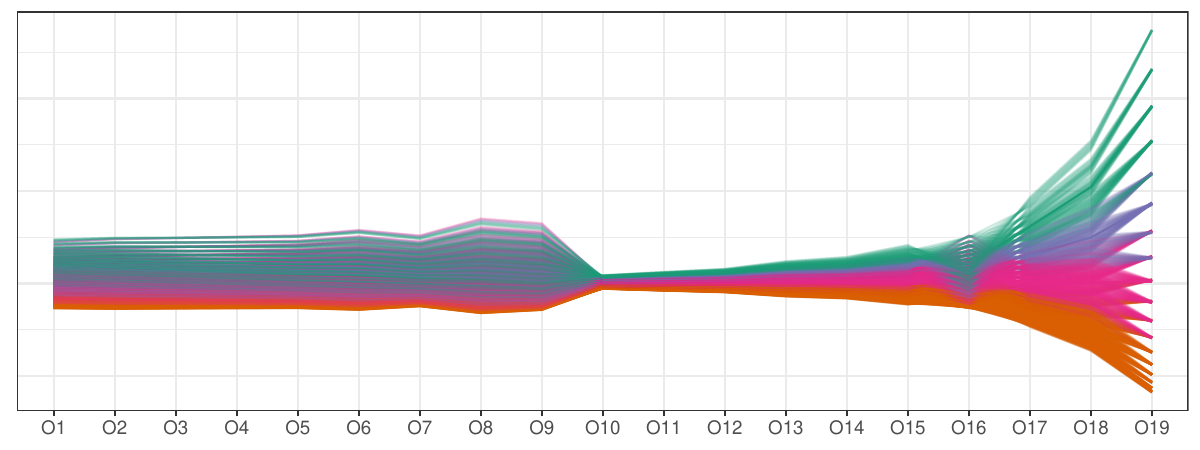}}
\caption{Parallel coordinates of the combined distributions  $\cos\theta_\tau$ in $B\to D^{\star} \tau^- \bar \nu_\tau$ and $q^2$ in $B\to D \tau \nu$ for four clusters using average linkage and Euclidean distance on the pulls.}
\label{f:semilq}
\end{figure}

The first thing to notice is that the resolving power of the combined distributions, as measured for example by the maximum cluster radius, increases to more than eight clusters. For simplicity then, we choose four clusters to illustrate a few possibilities. In Fig.~\ref{f:semilq} we show the parallel coordinates  using average linkage and Euclidean distance on the pulls. 

The first nine coordinates can be directly compared to Fig.~\ref{f:coordCK}. Note that while the lines showing the different models were crossing in a parallel coordinate plot of the user defined coordinates (i.e. for normalized histograms) this does not happen here, because the normalization of the distributions is different for each model. The same effect enhances the importance of the first few and last few bins in Fig.~\ref{f:coordCK} and diminishes that of $O_{4-6}$ for example. Using the pulls, all the nine bins of the $\cos\theta_\tau$ distribution in $B\to D^{\star} \tau^- \bar \nu_\tau$ have comparable importance in the clustering. The $q^2$ distribution of the branching ratio for $B\to D \tau \nu$ has a different behavior. The last two bins appear as the dominant coordinates and would have a large effect on the clustering. As before, we can choose to enhance the importance of these coordinates by using complete linkage and maximum distance, or suppress it by using Manhattan distance.

\begin{figure}[h]
\centering{\includegraphics[scale=0.5]{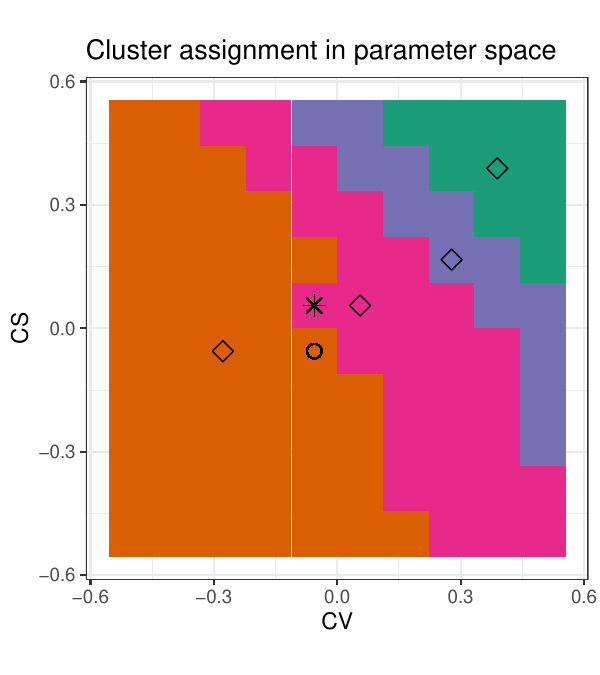}
\includegraphics[scale=0.5]{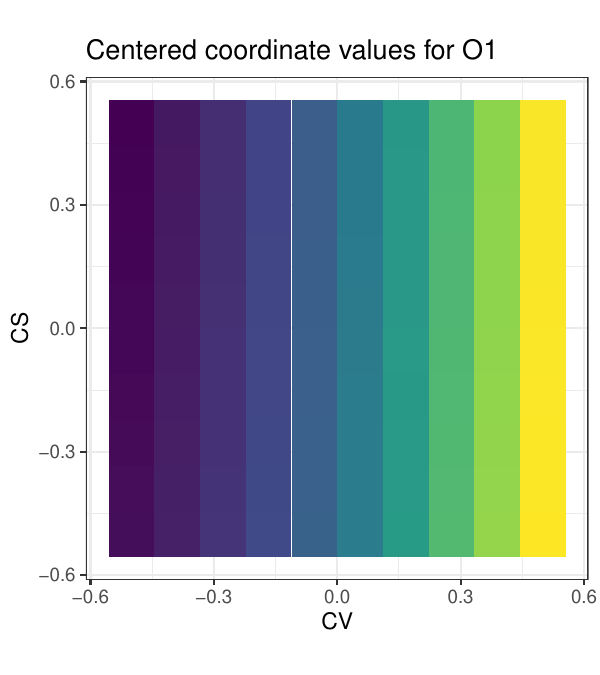}
\includegraphics[scale=0.5]{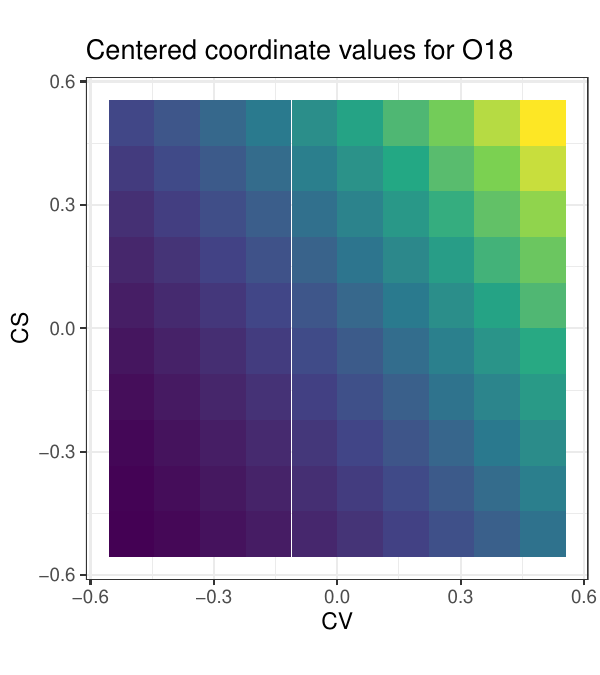}}
\centering{\includegraphics[scale=0.5]{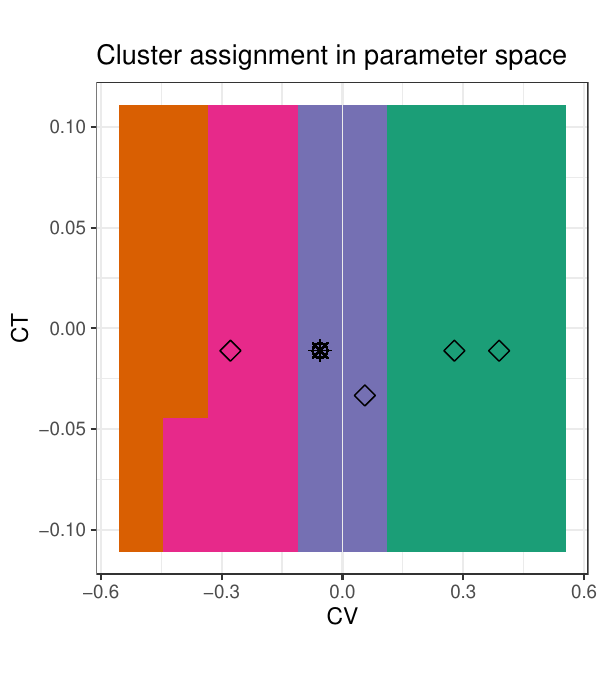}
\includegraphics[scale=0.5]{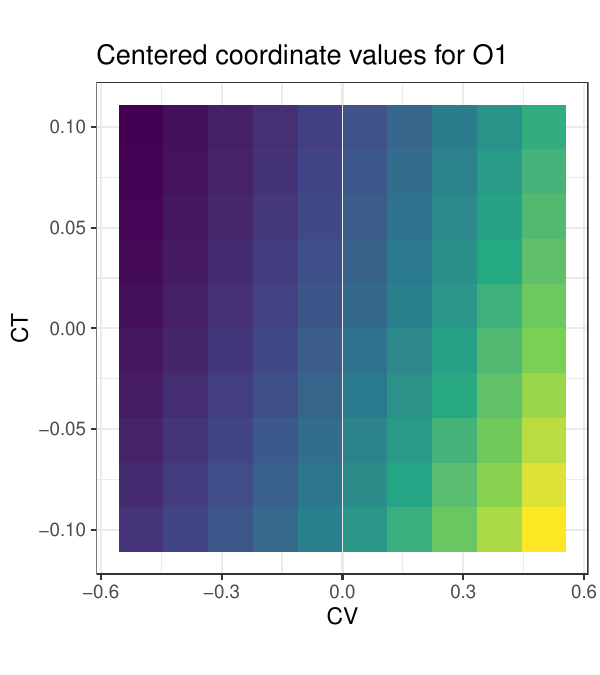}
\includegraphics[scale=0.5]{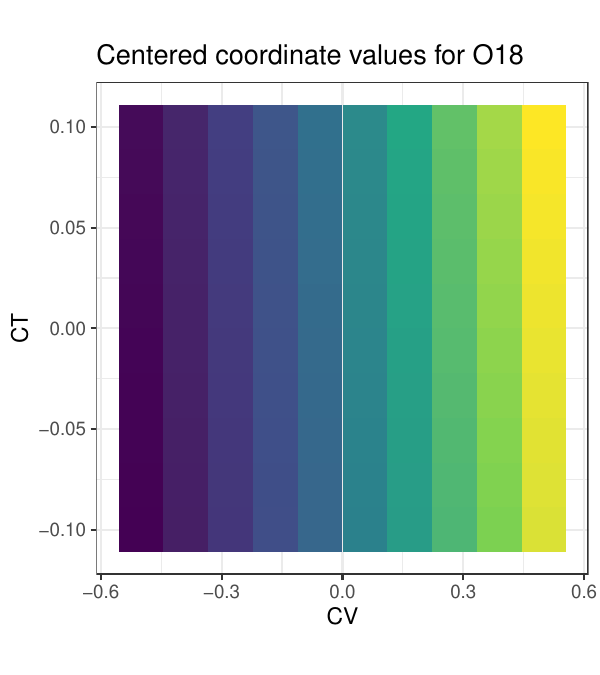}}
\caption{Slices of four clusters using average linkage and Euclidean distance on the pulls in parameter space:  the $C_{VL}-C_{SL}$ plane with $C_T=-0.1$ (top),  and the $C_{VL}-C_{T}$ plane with $C_{SL}=0.5$ (bottom). The variation of predictions across parameter space for the same slices is also shown for the coordinates $O_1$ and $O_{18}$.}
\label{f:coorvarCK}
\end{figure}

The complementarity of the two distributions can be seen in Fig.~\ref{f:coorvarCK} where we show slices of the clusters in parameter space in the $C_{VL}-C_{SL}$ plane with $C_T=-0.1$ (top)  and in the $C_{VL}-C_{T}$ plane with $C_{SL}=0.5$ (bottom). The variation of predictions across parameter space for the same slices is also shown for the coordinates $O_1$ and $O_{18}$. All coordinates in the  $\cos\theta_\tau$ distribution in $B\to D^{\star} \tau^- \bar \nu_\tau$ exhibit a similar pattern to $O_1$ that can separate $C_{VL}$ better than $C_{SL}$, while most coordinates in the $q^2$ distribution of the branching ratio for $B\to D \tau \nu$ exhibit a pattern similar to $O_{18}$,  showing a correlation between  $C_{VL}$ and $C_{SL}$. The situation is opposite in the bottom panel, where the $\cos\theta_\tau$ distribution (illustrated by the behavior of $O_1$) exhibits a correlation between  $C_{VL}$ and $C_{T}$ and the $q^2$ distribution provides little resolving power along $C_T$. The dominance of the last few $q^2$ coordinates is reflected in the shape of the cluster boundaries in the respective slices.

\begin{figure}[h]
\centering{\includegraphics[scale=0.4]{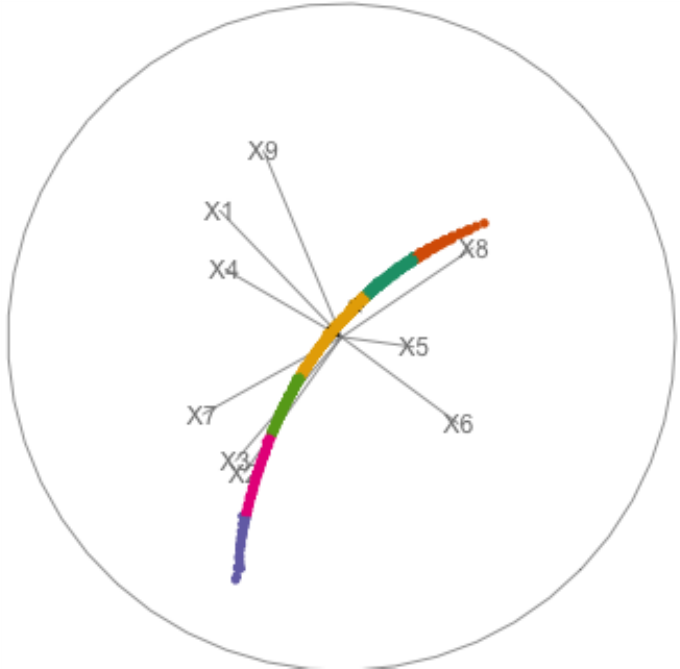}\includegraphics[scale=0.4]{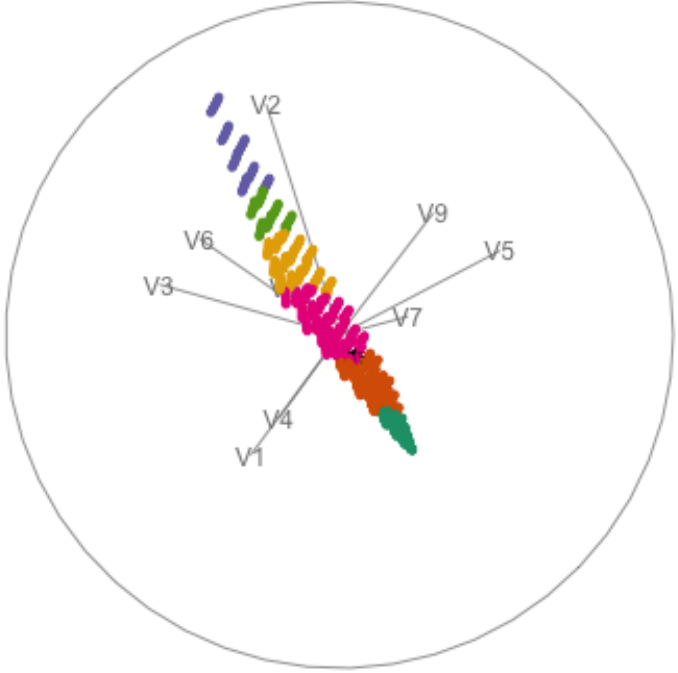}\includegraphics[scale=0.4]{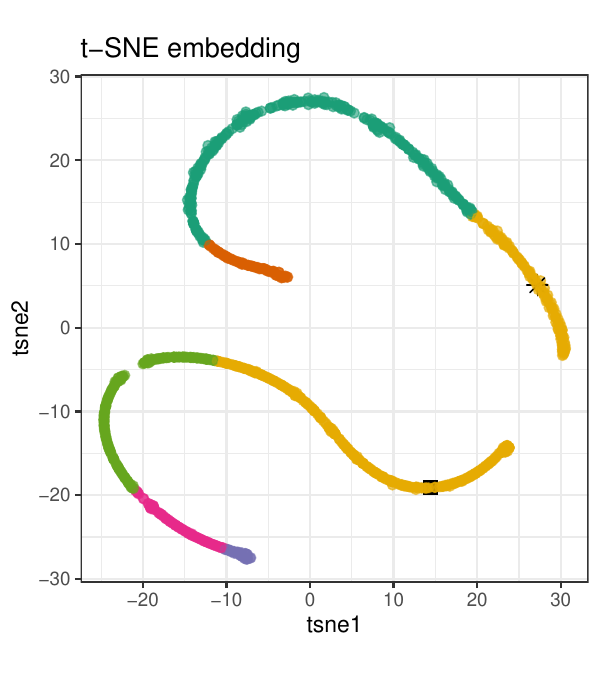}}
\caption{Still image from the tour showing six clusters with complete linkage in 9D observable space for the $\cos\theta_\tau$  distribution in $B\to D^{\star} \tau^- \bar \nu_\tau$ with user coordinates and distance to match the {\tt ClusterKing} setup (left, see animation \href{https://uschilaa.github.io/animations/pandemonium4.html}{here}) and with euclidean distance on the pulls (right, see animation \href{https://uschilaa.github.io/animations/pandemonium3.html}{here}). Right plot shows t-SNE view of the 9D observable space also illustrating that this is a 1D structure.}
\label{f:tourck}
\end{figure}

Tours of the clusters in observable space in this example also reveal the intrinsic dimensionality. With the original {\tt ClusterKing} setup there are three parameters but a constraint that requires the same normalization for all models resulting in a strip (2D surface) in nine dimensions. Using instead the pulls with Euclidean distance, without the normalization constraint, the 3D nature of the clusters in observable space is revealed. In Fig.~\ref{f:tourck} we illustrate this with two still plots from tours of six clusters in observable space for two and four parameters containing links to animated gifs. 

\subsection{Salient points}

{\tt Pandemonium} is an interactive graphical interface which provides multiple tools to study clustering. It can be used in a manner that mimics {\tt ClusterKing} but offers many ways to go beyond it. A few key differences are:

\begin{itemize}

\item {\tt Pandemonium} allows flexible coordinate input to suit a specific problem. Without additional user input, {\tt Pandemonium} relies on pulls, which are useful to compare models against each other. {\tt ClusterKing} uses coordinates appropriate to determine how likely it is that two histograms are drawn from the same underlying distribution.

\item  We have illustrated that stopping criteria are very sensitive to small numerical differences. In these scenarios our approach of visualizing a range of statistics would be preferred over the automatic stopping criterion used in {\tt ClusterKing.}

\item {\tt Pandemonium} allows the user to include multiple histograms or other observables when comparing different models. It also allows the user to select from a menu of possible linkages and distance functions and we have shown how this flexibility can be used to obtain new insights on the data.

\item We use a range of different high dimensional visualization of the observable space using parallel coordinate plots, dimension reduction and a tour display in the same interface, allowing a better understanding of the clustering outcome.

\end{itemize}

\section{Summary and Conclusions}

This paper explores the use of hierarchical clustering in combination with data visualization and an interactive user interface for applications in physics. The work has been implemented in the interactive tool {\tt pandemonium}, best suited for clustering (in observable space) multiple predictions organized as an array of coordinates normalized to uncertainty. The results of the clustering outcome can then be visualized in parameter space. We have included examples that study different problems from flavor physics, showing how clustering can be used to complement a global fit (Section~\ref{s:existingfit}) and understand the potential of future measurements (subsection~\ref{s:futurefit}). We have also included a comparison to the approach presented in {\tt ClusterKing} (Section~\ref{s:compareCK}).

By relating the clustering outcome with a set of linked displays, we illustrate how to explore relations  between parameter and observable space. This exploration benefits from the interactive environment, which enables easy comparison of the results with different settings for the clustering. The different settings can be chosen judiciously to enhance or suppress the importance of particular observables.

The applications included in this paper present insights into well studied and understood B physics problems which serve to hone the method, and we summarized our physics results in Section \ref{s:psum}. The approach and software tools can be applied to a large number of phenomenological studies in particle physics and beyond, promising new insights in particular in scenarios that are not as well understood.

Consider for example new physics models that introduce new particles, e.g. models with an extended Higgs sector or supersymmetric models, where we may wish to understand the connection between the model parameters and the physical particle masses at the weak scale. Here we might use clustering to understand the connections between these two parameter spaces (rather than with an experimental observable space).

The observable space differs from the preferred theoretical parametrization in a range of fields. Contrary to particle physics, the transfer functions between the two spaces might not be known but need to be estimated from the data as well, see \cite{hydro} for an example from hydrology. In this case the clustering approach could help understand these connections better.
More generally speaking, this approach can be useful whenever we aim to better understand a complex connection between two representations of the same points. A very general example is the understanding of statistical models with multiple latent variables that are inferred from the observed ones (potentially using complex black-box models).

We have made the software available on GitHub in the form of an R package, \url{https://github.com/uschiLaa/pandemonium}.

\section*{Acknowledgments} This work was supported in part by the Australian Government through the Australian Research Council. We are grateful to Dianne Cook for discussions and to David Straub for help with {\tt flavio} questions.

\appendix

\section{Classical tools for the interpretation of clustering outcome}

\subsection{Cluster statistics}\label{s:clustat}

A large number of cluster statistics (or internal validation measures) has been developed to help analysts decide the preferred number of clusters for a given dataset, or to compare two different clustering outcomes. Internal cluster statistics typically compare the dissimilarities within each cluster to the dissimilarity between clusters. In our work we use the R package \texttt{fpc}~\cite{fpc} to compute a number of statistics for cluster validation:

 \begin{itemize}
 \item Cluster SS: sum of the within cluster squared dissimilarities divided by two times the cluster size.
 \item WB ratio: the ratio of average dissimilarity within to between clusters.
 \item Normalized gamma: the correlation between distances and an indicator vector that is 0 (1) for points in the same (different) cluster~\cite{halkidi}.
 \item Dunn index: the ratio of the minimum cluster separation over the maximum cluster diameter.
 \item Calinski and Harabasz index: the ratio of average dissimilarity between clusters to within clusters, but adjusting based on the number of data points and the number of clusters~\cite{ch}.
 \end{itemize}

The first two (Cluster SS and WB ratio) always decrease when increasing the number of clusters, and we would typically look for an elbow in the index measure as a function of the number of clusters to decide that additional clusters are not relevant. For all other indices the maximum should indicate the best clustering result. However, in practice, when there is no clear answer, the different indices often disagree and we look at the combination to get a better idea of the uncertainty involved when choosing the number of clusters.

Finally, we may also wish to compare two clustering solutions and measure how similar they are. We can do this using the adjusted Rand index~\cite{rand}. The Rand index takes values between zero and one, with larger values indicating more overlap in grouping. For the adjusted Rand index a negative value is also possible in cases where the overlap is below what is expected for independent groupings. An adjusted Rand index close to zero indicates independent clustering, and a value of one means identical grouping. This index can be used, for example, to compare the hierarchical clustering results with a given set of options to partitioning based on the value of the $\chi^2$ statistic.

\subsection{Visualization of hierarchical clustering}

An effective visual summary of hierarchical clustering is obtained when combining a heatmap showing the full dissimilarity matrix with a dendrogram~\cite{WilkinsonLeland2009THot}. A dendrogram shows the tree-like structure of hierarchical clustering where the height represents the dissimilarity between observations or clusters, and each branching corresponds to the merging of clusters. The ordering of rows and columns in the heatmap display is based on the dendrogram, thus grouping points based on similarity (and reflecting the choice in distance metric and linkage). This makes it easy to spot patterns in the dissimilarity matrix.

The dendrogram can often indicate the preferred number of clusters by identifying when more dissimilar clusters are combined, as read off the heights of the nodes. A specific solution is obtained by cutting the tree structure, for example we can specify a requested number of clusters and cut the tree structure at the corresponding height. All observations that are merged below this cut are considered to be within a cluster. In our implementation of the display we indicate this solution by coloring the edges in the tree according to the clustering solution, using the R package \texttt{dendextend}~\cite{dendextend}.

This visualization is most useful if there are clearly separated clusters in the data. In our applications, where points in observable space are distributed according to continuous functions, other displays are more important. 

\section{Usage of the app}

\begin{figure}[h]
\centering{\includegraphics[scale=0.3]{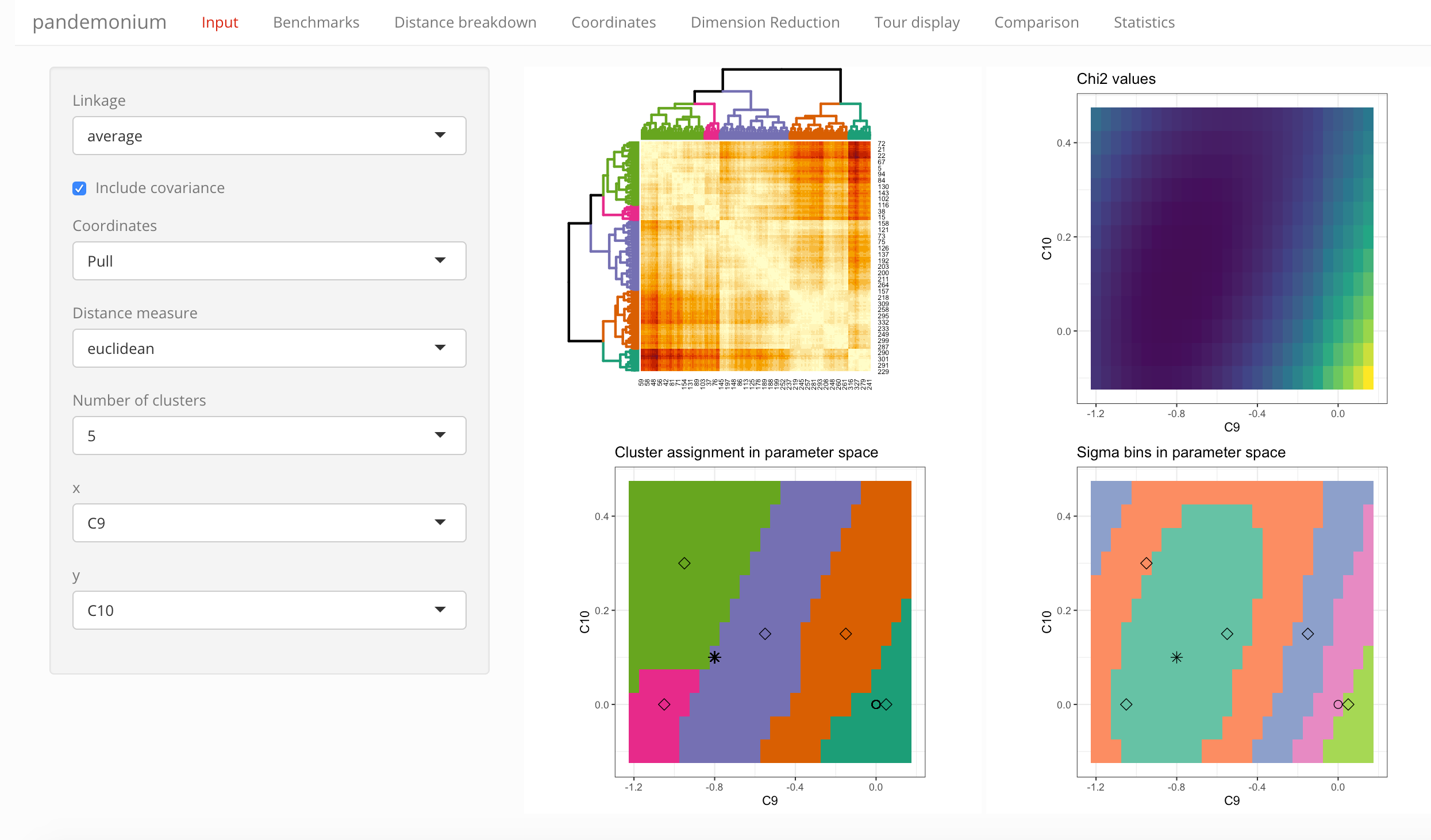}}
\caption{Screenshot of the graphical interface showing the input tab. The user can select coordinates, linkage and number of clusters and is presented with a first overview of the resulting clustering.}
\label{f:app}
\end{figure}

For documentation of the interface and usage instructions we refer the reader to the package page at \url{https://uschilaa.github.io/pandemonium/}. For illustration a screenshot of the app, open on the input tab, is shown in Figure \ref{f:app}.

\subsection{Practical considerations}

As a help to new users we now provide a few guidelines to get started with using the app.

\begin{itemize}
\item The parameter range considered should be selected with care, this can be done given the result from a global fit and selecting a region based on the $\Delta(\chi^2)$ contours.
\item The parameter space needs to be scanned on a regular grid for the visualizations to be useful. This works well in the case of few free parameters and simplifies the representation of the results in parameter space. This will be relaxed in future versions of the interface as random scans are desirable for high dimensionality.
\item The drop down menu to select coordinates also has a ``user'' option that allows direct upload of user defined coordinates that can be combined with any of the distance measures available in the app.
\item One of the choices in the drop-down menu for distances is a user defined distance which allows direct input of $d_m(k,l)$ when loading the app. This choice enables the user to compute distances using custom methods (and alternative programming languages) and to provide the matrix in numerical form.
\end{itemize}

\section{$B\to K^{(\star)}\ell^+\ell^-$ observables}\label{a:obs}

The prediction for the fourteen observables chosen for Section~\ref{s:existingfit} (with a factorization scale $\mu=4.8$~GeV), along with their average measured value and uncertainty (given by the square root of the diagonal entries of the experimental covariance matrix), was obtained  using  {\tt flavio}. The observables are listed in Table~\ref{t:obs}, the last column gives the ID that this observable had in the study of \cite{Capdevila:2018jhy}, noting that the definitions of the observables are not always identical: the sign of $P_2$ is reversed here; and in some cases different experimental measurements are being averaged. These numbers include the results from:
for $P_5^\prime$ LHCb \cite{Aaij:2020nrf}, CMS \cite{Sirunyan:2017dhj} and ATLAS \cite{Aaboud:2018krd}; $P_2$ LHCb \cite{Aaij:2020nrf}; $R_K$  LHCb \cite{Aaij:2019wad} and Belle \cite{Abdesselam:2019lab}; $R_{K^\star}$ LHCb \cite{Aaij:2017vbb} and Belle \cite{Abdesselam:2019wac}.
\begin{table}[htp]
{
 \centering
\begin{tabular}{|c|c|c|c|c|c|}  \hline
ID & Observable & Exp.  & ID in \cite{Capdevila:2018jhy} \\ \hline
1 & $P_5^\prime(B \to K^* \mu\mu) [0.1-0.98] $ &$0.52\pm 0.10$  &20 \\
2 & $P_5^\prime(B \to K^* \mu\mu) [1.1-2.5] $ & $0.36\pm 0.12$  & 28\\
3 & $P_5^\prime(B \to K^* \mu\mu) [2.5-4] $ & $-0.15\pm0.14$   & 36 \\
4 & $P_5^\prime(B \to K^* \mu\mu) [4-6] $ &$-0.39\pm0.11$ & 44 \\
5 & $P_5^\prime(B \to K^* \mu\mu) [6-8] $ & $-0.58\pm0.09$  & 52 \\
6 & $P_5^\prime(B \to K^* \mu\mu) [15-19] $ & $-0.67\pm0.06$ & 60\\
7 & $P_2(B \to K^* \mu\mu) [0.1-0.98] $ & $0\pm 0.04$ &  17\\
8 & $P_2(B \to K^* \mu\mu) [1.1-2.5] $ & $-0.44\pm0.10$  & 25\\
9  & $P_2(B \to K^* \mu\mu) [2.5-4] $ & $-0.19\pm0.12$  & 33\\
10 & $P_2(B \to K^* \mu\mu) [4-6] $ &$0.10\pm0.07$ &  41\\
11 & $P_2(B \to K^* \mu\mu) [6-8] $ & $0.21\pm0.05$ &  49\\
12  & $P_2(B \to K^* \mu\mu) [15-19] $ & $0.36\pm 0.02$ & 57 \\
13 & $R_K(B^+  \to K^+ ) [1.1-6] $ & $0.86 \pm 0.06$ & 98 \\
14 & $R_{K^*} (B^0  \to K^{0 *}) [1.1-6] $ & $0.73\pm 0.11$ &  100\\ \hline
\end{tabular} 
\caption{List of observables used in Section~\ref{s:existingfit} to cluster measurements with an underlying $b\to s\ell^+\ell^-$ quark transition.}
\label{t:obs}
}
\end{table}

A list of the 75 additional observables included in \ref{s:moreobs} is given in Table~\ref{t:75more}. The numbering of this set for Figure~\ref{f:euc89} starts at 15 and is given in parenthesis for each observable.

\begin{table}[htp]
\resizebox{1.0\textwidth}{!}{
\begin{tabular}{|l|l|l|l|l|}  \hline
(15) $Br(B^+->K\mu^+ \mu^- )$[0.1-0.98] & 
(16) $Br(B^+->K\mu^+ \mu^- )$[1.1-2] & 
(17) $Br(B^+->K\mu^+ \mu^- )$[2-3] & 
(18) $Br(B^+->K\mu^+ \mu^- )$[3-4] & 
(19) $Br(B^+->K\mu^+ \mu^- )$[4-5] \\
(20) $Br(B^+->K\mu^+ \mu^- )$[5-6] & 
(21) $Br(B^+->K\mu^+ \mu^- )$[6-7] & 
(22) $Br(B^+->K\mu^+ \mu^- )$[7-8] & 
(23) $Br(B^0->K\mu^+ \mu^- )$[0.1-2] & 
(24) $Br(B^0->K\mu^+ \mu^- )$[2-4] \\
(25) $Br(B^0->K\mu^+ \mu^- )$[4-6] & 
(26) $Br(B^0->K\mu^+ \mu^- )$[6-8] & 
(27) $Br(B^0->K\mu^+ \mu^- )$[15-22] & 
(28) $F_L(B^0->K^\star \mu^+ \mu^- )$[1.1-2.5] & 
(29) $F_L(B^0->K^\star \mu^+ \mu^- )$[2.5-4] \\ 
(30) $F_L(B^0->K^\star \mu^+ \mu^- )$[4- 6] & 
(31) $F_L(B^0->K^\star \mu^+ \mu^- )$[15-19] & 
(32) $P_1(B^0->K^\star \mu^+ \mu^- )$[0.1-0.98] & 
(33) $P_1(B^0->K^\star \mu^+ \mu^- )$[1.1-2.5] & 
(34) $P_1(B^0->K^\star \mu^+ \mu^- )$[2.5-4] \\ 
(35) $P_1(B^0->K^\star \mu^+ \mu^- )$[4- 6] & 
(36) $P_1(B^0->K^\star \mu^+ \mu^- )$[6-8] & 
(37) $P_1(B^0->K^\star \mu^+ \mu^- )$[15-19] & 
(38) $P_3(B^0->K^\star \mu^+ \mu^- )$[0.1-0.98] & 
(39) $P_3(B^0->K^\star \mu^+ \mu^- )$[1.1-2.5] \\ 
(40) $P_3(B^0->K^\star \mu^+ \mu^- )$[2.5-4] & 
(41) $P_3(B^0->K^\star \mu^+ \mu^- )$[4- 6] & 
(42) $P_3(B^0->K^\star \mu^+ \mu^- )$[6-8] & 
(43) $P_3(B^0->K^\star \mu^+ \mu^- )$[15-19] & 
(44) $P_4^\prime(B^0->K^\star \mu^+ \mu^- )$[0.1-0.98] \\
(45) $P_4^\prime(B^0->K^\star \mu^+ \mu^- )$[2.5-4] & 
(46) $P_4^\prime(B^0->K^\star \mu^+ \mu^- )$[4- 6] & 
(47) $P_4^\prime(B^0->K^\star \mu^+ \mu^- )$[6-8] & 
(48) $P_4^\prime(B^0->K^\star \mu^+ \mu^- )$[15-19] & 
(49) $P_6^\prime(B^0->K^\star \mu^+ \mu^- )$[0.1-0.98] \\ 
(50) $P_6^\prime(B^0->K^\star \mu^+ \mu^- )$[1.1-2.5] & 
(51) $P_6^\prime(B^0->K^\star \mu^+ \mu^- )$[2.5-4] & 
(52) $P_6^\prime(B^0->K^\star \mu^+ \mu^- )$[4- 6] & 
(53) $P_6^\prime(B^0->K^\star \mu^+ \mu^- )$[6-8] & 
(54) $P_6^\prime(B^0->K^\star \mu^+ \mu^- )$[15-19] \\ 
(55) $P_8^\prime(B^0->K^\star \mu^+ \mu^- )$[0.1-0.98] & 
(56) $P_8^\prime(B^0->K^\star \mu^+ \mu^- )$[1.1-2.5] & 
(57) $P_8^\prime(B^0->K^\star \mu^+ \mu^- )$[2.5-4] & 
(58) $P_8^\prime(B^0->K^\star \mu^+ \mu^- )$[4- 6] & 
(59) $P_8^\prime(B^0->K^\star \mu^+ \mu^- )$[6-8] \\ 
(60) $P_8^\prime(B^0->K^\star \mu^+ \mu^- )$[15-19] & 
(61) $Br(B^0->K^\star \mu^+ \mu^- )$[0.1-0.98] & 
(62) $Br(B^0->K^\star \mu^+ \mu^- )$[1.1-2.5] & 
(63) $Br(B^0->K^\star \mu^+ \mu^- )$[2.5-4] & 
(64) $Br(B^0->K^\star \mu^+ \mu^- )$[4-6] \\ 
(65) $Br(B^0->K^\star \mu^+ \mu^- )$[6-8] & 
(66) $Br(B^0->K^\star \mu^+ \mu^- )$[15-19] & 
(67) $Br(B^+->K^\star \mu^+ \mu^- )$[0.1-2] & 
(68) $Br(B^+->K^\star \mu^+ \mu^- )$[2-4] & 
(69) $Br(B^+->K^\star \mu^+ \mu^- )$[4-6] \\ 
(70) $Br(B^+->K^\star \mu^+ \mu^- )$[6-8] & 
(71) $Br(B^+->K^\star \mu^+ \mu^- )$[15-19] & 
(72) $F_L(Bs->\Phi \mu^+ \mu^- )$[0.1-2] & 
(73) $F_L(Bs->\Phi \mu^+ \mu^- )$[2-5] & 
(74) $F_L(Bs->\Phi \mu^+ \mu^- )$[5-8] \\ 
(75) $Br(Bs->\Phi \mu^+ \mu^- )$[0.1-2] & 
(76) $Br(Bs->\Phi \mu^+ \mu^- )$[2-5] & 
(77) $Br(Bs->\Phi \mu^+ \mu^- )$[5-8] & 
(78) $F_L(B^0->K^\star ee)$[0.0020-1.120] & 
(79) $P_1(B^0->K^\star ee)$[0.0020-1.120] \\ 
(80) $P_2(B^0->K^\star ee)$[0.0020-1.120] & 
(81) $F_L(B^0->K^\star \mu^+ \mu^- )$[0.1-0.98] & 
(82) $F_L(B^0->K^\star \mu^+ \mu^- )$[6-8] & 
(83) $Br(B^+->K\mu^+ \mu^- )$[15-22] & 
(84) $P_4^\prime(B^0->K^\star \mu^+ \mu^- )$[1.1-2.5] \\ 
(85) $BR(Bs->\Phi \gamma)$ & 
(86) $BR(Bs->\mu^+ \mu^- )$ & 
(87) $BR(B^0->K^\star \gamma)$ & 
(88) $BR(B^+->K^\star \gamma)$ & 
(89) $BR(B->X_s \gamma)$ \\  \hline
\end{tabular} }
\caption{List of observables used in Section~\ref{s:moreobs}.}
\label{t:75more}
\end{table}%

\subsection{Future observables and uncertainty estimates}

In Table~\ref{t:futobs} we list the observables used in this case along with their predicted values within the SM.

\begin{table}[htp]{
\centering
\begin{tabular}{|c|c|c|c|}  \hline
ID & Observable  & ID in \cite{Capdevila:2018jhy}& SM prediction from {\tt flavio} \\ \hline
1 & $Q_2(B \to K^*) [1.1, 2.5]$ & 20&$-(6.83\pm 0.32)\times 10^{-3}$\\
2 & $Q_2(B \to K^*) [2.5, 4]$ & 26&$ -(0.16\pm0.71)\times 10^{-3} $\\
3 & $Q_2(B \to K^*) [4, 6]$ & 32 &$(1.37\pm 0.36)\times 10^{-3}$\\
4 & $Q_5(B \to K^*) [1.1, 2.5]$ & 22 &$(0.72\pm 1.3)\times 10^{-3}$\\
5 & $Q_5(B \to K^*) [2.5, 4]$ & 28&$-(3.72\pm 0.67)\times 10^{-3} $\\
6 & $Q_5(B \to K^*) [4, 6]$  & 34&$-(3.45\pm 0.30)\times 10^{-3}$
 \\ \hline
\end{tabular} 
\caption{List of observables used in Section~\ref{s:futurefit} to cluster future measurements to test lepton flavor universality in $b\to s\ell^+\ell^-$ quark transition. The theory errors at the SM point are presented for comparison to the projected statistical uncertainty in Eq.~\ref{guesscov}. The theory predictions within the parameter range considered can reach values that exceed  the SM by two orders of magnitude.}
\label{t:futobs}
}
\end{table}%

In this example, we estimate experimental statistical uncertainties from the sensitivity projections for Belle II \cite{Kou:2018nap}. We can also model what correlations might look like, for example, by taking their known values from LHCb \cite{Aaij:2020nrf} for the corresponding {\it muon} observables. These choices can serve as a plausible experimental covariance matrix, perhaps as a future benchmark. The resulting covariance matrix is:
\begin{eqnarray}
\Sigma_{exp}= \begin{pmatrix}
6.4 & 0 & 0 & -1.9 & 0 & 0 \\ 0 & 5.2 & 0 & 0 & -0.36 & 
  0 \\ 0 & 0 & 3.4 & 0 & 0 & -0.42 \\ -1.9 & 0 & 0 & 12 & 0 &
   0 \\ 0 & -0.36 & 0 & 0 & 10 & 0 \\ 0 & 0 & -0.42 & 0 & 0 & 6.4
\end{pmatrix}\times 10^{-3}
\label{guesscov}
\end{eqnarray}
Under the assumption that experimental uncertainty will ultimately be dominated by statistical errors, we can  scale these numbers with the luminosity. The theoretical uncertainty can also be included,  but in this case it turns out to be significantly smaller. 

\section{$B\to D^{(\star)} \tau \nu$ kinematic distributions}\label{aCK}

For the processes $B\to D \tau^- \bar \nu_\tau$ and $B\to D^\star \tau^- \bar \nu_\tau$  the following kinematic distributions are considered: 
\begin{itemize}
\item $\cos\theta_\tau$: the angle between the $\tau$ and $B$ meson in the dilepton mass frame distribution in $B\to D^{\star} \tau^- \bar \nu_\tau$.
\item $q^2$: the lepton invariant mass squared distribution in $B\to D^0\tau\nu$.  
\end{itemize}
They are divided into bins that are treated as separate observables as per Table~\ref{t:compCK}. 

The parameters that encode the unknown physics in this example are the Wilson coefficients in the effective low energy Hamiltonian responsible for the corresponding quark level transition $b\to c \tau^- \bar\nu_\tau$:
\begin{eqnarray}
{\cal H}_{\rm eff} = \frac{4G_F}{\sqrt{2}}V_{cb}\sum C_i{\cal O}_i.
\end{eqnarray}
Three operators are included in this study:
\begin{eqnarray}
{\cal O}_{VL}=(\bar c \gamma_\mu P_L b)(\bar \tau \gamma^\mu P_L \nu_\tau),~
{\cal O}_{SL}=(\bar c  P_L b)(\bar \tau P_L \nu_\tau),~
{\cal O}_{T}=(\bar c  \sigma_{\mu\nu}P_L b)(\bar \tau \sigma^{\mu\nu}P_L \nu_\tau)
\end{eqnarray}
The factorization scale used as input to {\tt flavio} is again taken to be 4.8 GeV, other operators are ignored, and the parameters are assumed to be real. In this case there is only one coefficient that does not vanish in the  standard model, $C_{VL}=1$. As in the previous section we use $C_{i}=C_{i}^{SM}+C_{i}^{NP}$, the parameters varied in the example correspond  to the deviations from the SM. 

\begin{table}[htp]{
\centering
\begin{tabular}{|c|c|c|c|c|c|c|c|c|c|c|}  \hline
ID & \multicolumn{10}{c|}{Observable}   \\ \hline
1-9 &\multicolumn{10}{c|} {$\cos\theta_\tau$ distribution of $BR(B\to D^{\star} \tau \nu)$, nine bins}\\ \hline
SM & 0.0079 &  0.008 &  0.008 &  0.008 &  0.008 &  0.0079 &  0.0077 &  0.0075 &  0.0073 & \\ \hline
$\sigma_{SM}$ & 0.0012 &  0.0012 &  0.0012 &  0.0012 &  0.0012 &  0.0012 &  0.0014 &  0.001 &  0.0011 & \\ \hline
$\sigma_{eff}$ & 0.0011 &  0.0011 &  0.0011 &  0.0011 &  0.0011 &  0.0011 &  0.0011 &  0.001 &  0.001 & \\ \hline
10-19 & \multicolumn{10}{c|}{$q^2$  distribution of $BR(B\to D \tau \nu)$, ten bins}\\ \hline
${\rm SM} \times 10^3$ & 0.002 &  0.5 &  1.0 &  1.3 &  1.4 &  1.3 &  1.2 &  0.9 &  0.57 &  0.11   \\ \hline
$\sigma_{SM} \times 10^4$ & 0.014 &  2.5 &  4.3 &  3.8 &  3.6 &  2.5 &  1.7 &  0.84 &  0.36 &  0.046   \\ \hline
$\sigma_{eff} \times 10^4$ & 0.041 &  0.82 &  1.4 &  1.7 &  1.8 &  1.7 &  1.5 &  1.3 &  0.9 &  0.33 
 \\ \hline
\end{tabular} 
\caption{List of observables used in Section~\ref{s:compareCK}. We show the SM predictions along with the corresponding theoretical uncertainty, $\sigma_{SM}$,  as given by {\tt flavio}. The rows labeled $\sigma_{eff}$ show the relative error that results for the SM point assuming a yield of 1000 events and a 10\% systematic error as described in the text.}
\label{t:compCK}
}
\end{table}%

Early experimental results from BaBar \cite{Lees:2013uzd} exist but they have very large errors and are not used here. The future uncertainties for this example are estimated following the assumptions in  \cite{Aebischer:2019zoe}: 
a yield of 1000 events is assumed for each process which are distributed across the bins according to the model predictions; a Poisson error is then assigned to each bin based on the expected bin count; a relative systematic error of 10\% is added to each bin (the two errors are combined in quadrature).

The user defined coordinates for this section are determined as follows. A set of Wilson coefficients is used to predict the total rate for $BR(B\to D^{\star} \tau \nu)$ and its distribution into the nine bins. The bin counts are then determined by scaling the prediction to a yield of 1000 events. The coordinate representing each bin is the ratio of this yield to the error determined as above.

\section{Additional Plots}

\begin{figure}
\includegraphics[scale=0.4]{fig_5_3_3.pdf}\includegraphics[scale=0.6]{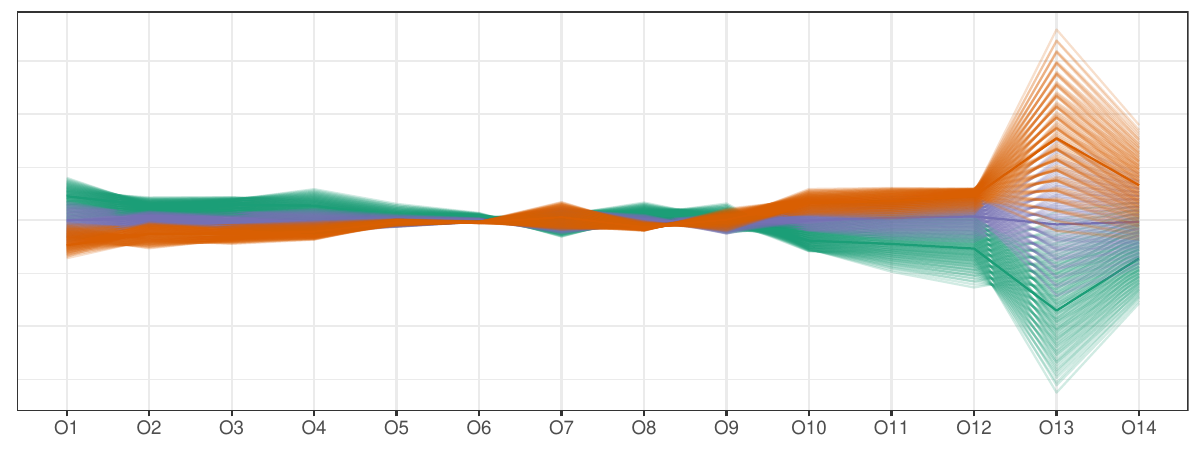}
\includegraphics[scale=0.4]{fig_5_3_4.pdf}\includegraphics[scale=0.6]{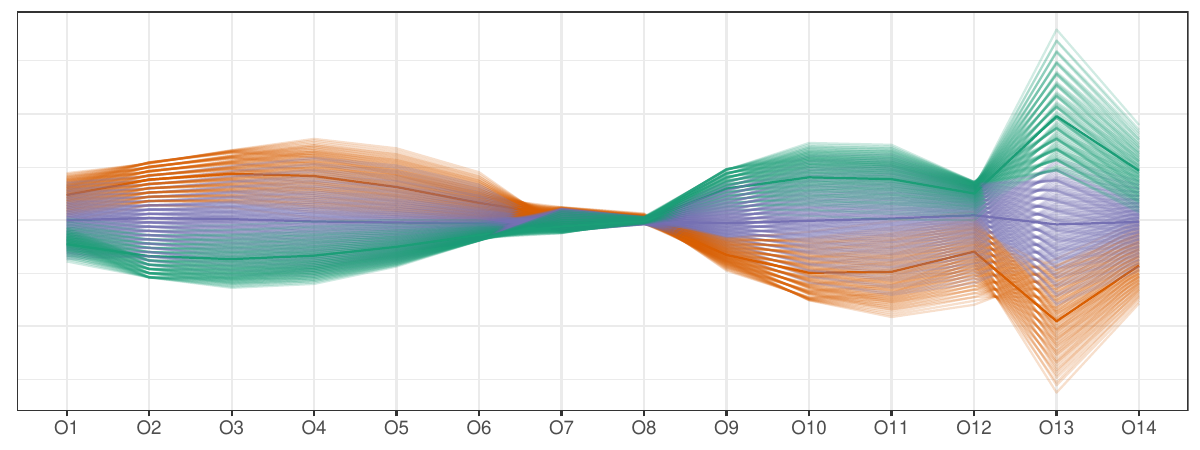}
\includegraphics[scale=0.4]{fig_5_3_5.pdf}\includegraphics[scale=0.6]{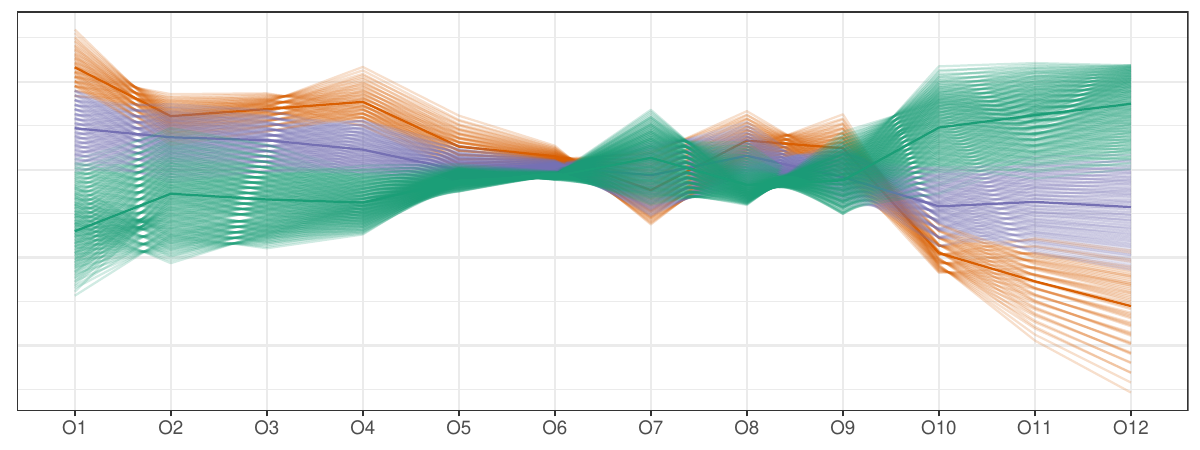}
\caption{The results of three clusters using average linkage and pull coordinates with different distance metrics: Manhattan distance with covariance (top panel), Maximum distance without covariance (middle panel) and Euclidean distance with correlations, but dropping  $R_K$ and $R_{K^\star}$ (bottom panel). The left panel is reproducing the results in parameter space, as already shown in Fig.~\ref{f:3cl}, the right panel is showing the corresponding parallel coordinate plots.}
\label{f:3clwpc}
\end{figure}

Figure~\ref{f:3clwpc} shows the clustering results seen in Fig.~\ref{f:3cl} together with the corresponding parallel coordinate plots in the right panel. 

\begin{figure}[h]
\centering{
\includegraphics[scale=0.5]{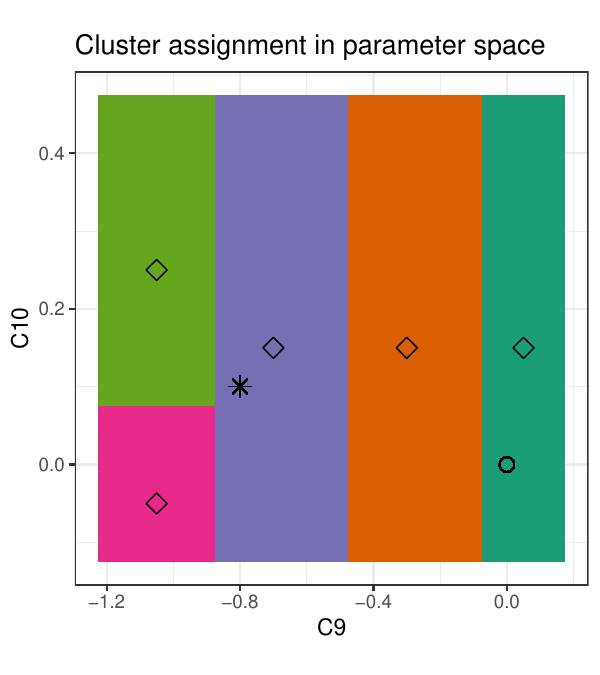}\includegraphics[scale=0.5]{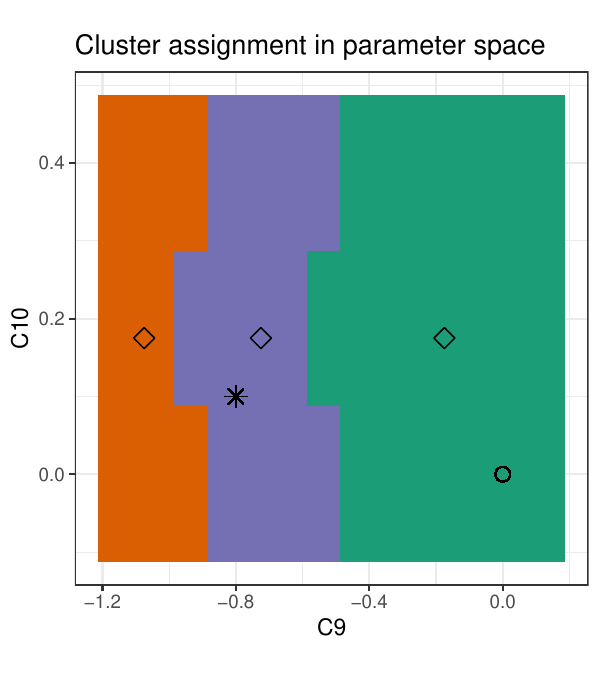}}
\caption{Partitioning with average linkage, Euclidean distance and no correlations showing the weaker dependence on $C_{10}$. Left panel: original grid with 5 clusters; right panel: finer grid with three clusters.}
\label{f:new_depc10}
\end{figure}

Figure~\ref{f:new_depc10} investigates the dependence on $C_{10}$ when no correlations are used in the coordinate definition. In this case the partitioning shows that, without correlations, this dependence is weaker than that on $C_9$. At a finer level of detail in the partitioning this dependence will appear. For example, the left panel of Figure~\ref{f:new_depc10} indicates that this dependence emerges in the fifth partitioning. The observability of finer details depends on additional factors such as the coarseness of the grid in parameter space. Doubling the number of points in each direction, for example, produces the three cluster partitioning shown in the right panel of Figure~\ref{f:new_depc10}. The main point is the same in all three cases: ignoring the correlations increases the importance of the angular observables over $R_{K^{(*)}}$. It is important to keep in mind that there is an intrinsic precision in the data affecting any result from the tool, at some point finer details will correlate with numerical uncertainty in the data itself.

\begin{figure}[h]
\centering{
\includegraphics[scale=0.5]{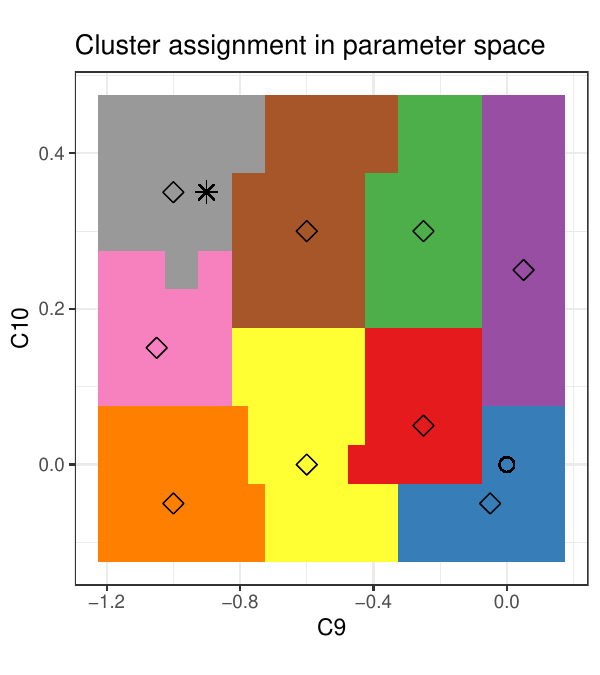}}
\caption{The resolution with 89 observables increases to the nine clusters shown here for average linkage and Euclidean distance.}
\label{f:new_9clu}
\end{figure}

Figure~\ref{f:new_9clu} shows the partitioning into nine clusters as found when using 89 observables.

\newpage

\bibliography{biblio}

\end{document}